\documentclass[useAMS,usenatbib]{mn2e}

\usepackage{natbib}

\usepackage{latexsym,graphicx}
\usepackage{color}
\usepackage{fixltx2e}
\usepackage{verbatim}
\usepackage{float}
\usepackage{amsmath,amssymb}
\usepackage{times}
\usepackage{soul}
\usepackage{natbib}

\usepackage{setspace}

\newcommand\hii{H\,{\sc ii} \,}

\def\apgt{\ {\raise-.5ex\hbox{$\buildrel>\over\sim$}}\ }
\def\aplt{\ {\raise-.5ex\hbox{$\buildrel<\over\sim$}}\ }

\newcommand{\degree}{\ensuremath{^\circ}}

\def\HII{\rm{H}\,{\textsc {ii}}}

\title[Bow shock nebulae of hot massive stars in a magnetized medium]{Bow shock nebulae of hot massive stars in a magnetized medium}

\author[D. M.-A.~Meyer et al.]
       {D. M.-A.~Meyer,$^{1}$\thanks{E-mail: dominique.meyer@uni-tuebingen.de} A. Mignone,$^{2}$ R. Kuiper,$^{1}$ A. Raga$^{3}$ and W. Kley$^{1}$ 
       \\
       $^{1}$Institut f\" ur Astronomie und Astrophysik, Universit\" at T\" ubingen,  Auf der Morgenstelle 10, 72076 T\" ubingen, Germany \\
       $^{2}$Dipartimento di Fisica Generale Facolt\`a di Scienze M.F.N., Universit\`a degli Studi di Torino, Via Pietro Giuria 1, 10125 Torino, Italy \\ 
       $^{3}$Instituto de Ciencias Nucleares, Universidad Nacional Aut\' onoma de M\' exico, Ap. 70-543, 04510 D.F., M\' exico \\
       }

\begin{document}

\date{Received 2015; accepted 2015}

\maketitle

\label{firstpage}

\begin{abstract}

\textcolor{black}{
A significant fraction of OB-type, main-sequence massive stars are classified 
as {\it runaway} and move supersonically through the interstellar medium (ISM). 
Their strong stellar winds interact with their surroundings \textcolor{black}{where the} 
typical strength of the local ISM magnetic field is about $3.5$-$7\, \mu \rm G$, which 
can result in the formation of bow shock nebulae. 
We investigate the effects of such magnetic field\textcolor{black}{s}, aligned with the motion of 
the flow, on the formation and emission properties of these circumstellar structures. 
}
\textcolor{black}{
Our axisymmetric, magneto-hydrodynamical simulations with optically-thin radiative 
cooling, heating and anisotropic thermal conduction show that the  
presence of the background ISM magnetic field affects the projected optical emission our bow 
shocks at H$\alpha$ and [O{\sc iii}] $\lambda \, 5007$ 
which become fainter by about $1$-$2$ orders of magnitude, respectively. 
Radiative transfer calculations against dust opacity indicate that the magnetic 
field slightly diminishes their projected infrared emission and that our bow 
shocks emit brightly at $60\, \mu \rm m$. 
} 
This may explain why the bow shocks generated 
by ionizing runaway massive stars are often difficult to identify. 
Finally, we discuss our results in the context of the bow shock of $\zeta$ Ophiuchi and 
we support the interpretation of its imperfect morphology as an evidence of the presence 
of an ISM magnetic field not aligned with the motion of its driving star. 

\end{abstract}

\begin{keywords}
methods: numerical -- MHD  -- circumstellar matter -- stars: massive.
\end{keywords}


\section{Introduction}
\label{sect:introduction}

Massive star formation is a \textcolor{black}{rare} event that strongly impacts the whole
Galactic machinery. These stars can release strong winds and ionizing radiation
which \textcolor{black}{shape their close surroundings into} beautiful billows of swept-up and
irradiated interstellar gas, that, in the case of a static or a slowly-moving
star, can produce structures such as the Bubble Nebula (NGC 7635) in the
constellation of Orion~\citep{moore_aj_124_2002}. The detailed study of the
circumstellar medium of these massive stars provides us an insight into
their internal physics~\citep{langer_araa_50_2012}, \textcolor{black}{it provides information on} their
intrinsic rotation~\citep{langer_ApJ_520_1999}, their envelope's
(in)stability~\citep{yoon_apj_717_2012} and allows us to understand the
properties of their close surroundings throughout their 
evolution~\citep{vanmarle_aa_460_2006,chita_aa_488_2008} and after their 
death~\citep{orlando_apj_678_2008,chiotellis_aa_537_2012}. \textcolor{black}{This
information is relevant for evaluating their feedback}, i.e. the amount of energy,
momentum and metals that massive stars inject into the interstellar medium (ISM)
of the Galaxy~\citep{vink_asp_353_2006}.

\textcolor{black}{In particular}, the bow shocks that develop around some fast-moving massive stars
ejected from their parent stellar clusters provide an  
opportunity to constrain both their wind and local ISM
properties~\citep{huthoff_aa_383_2002,meyer_mnras_439_2014}. Over the past decades,
stellar wind bow shocks have first been serendipitously noticed as bright [O{\sc
iii}] $\lambda \, 5007$ spectral line arc-like shapes and/or distorted
bubbles surrounding some massive stars having a particularly large space
velocity with respect to their ambient medium.
As a textbook example of such a bow shock, we refer the reader, e.g. 
to $\zeta$ Ophiuchi~\citep[][see Fig.~\ref{fig:obs} below]{gull_apj_230_1979}.
Further infrared observations, e.g. with the {\it Infrared Astronomical
Satellite}~\citep[{\it IRAS},][]{neugebauer_278_apj_1984} and 
the {\it Wide-Field Infrared Satellite Explorer}~\citep[ {\it WISE}, ][]{wright_aj_140_2010} facilities have made possible the compilation 
of catalogues of dozens of these bow shock nebulae~\citep{buren_apj_329_1988,vanburen_aj_110_1995,
noriegacrespo_aj_113_1997} and have motivated early numerical simulations
devoted to the parsec-scale circumstellar medium of moving
stars~\citep{brighenti_mnras_277_1995, brighenti_mnras_273_1995}. Recently,
modern facilities led to the construction of multi-wavelengths databases, see e.g.
the E-BOSS catalog~\citep{peri_aa_538_2012,2015arXiv150404264P} 
\textcolor{black}{ or the recent study of~\citet{kobulnicky_2016}}. Moreover, a
connection with high-energy astrophysics has been established, showing that
stellar wind bow shocks produce cosmic rays in the same way as the expanding shock waves of
growing supernova remnants do~\citep{valle_mnras_448_2015}.

It is the \textcolor{black}{discovery of bow shocks} around the historical stars
Betelgeuse~\citep{noriegacrespo_aj_114_1997} and
Vela-X1~\citep{kaper_apj_475_1997} that revived the interest of the scientific
community \textcolor{black}{regarding such circumstellar} structures generated by massive stars. The
fundamental study of~\citet{comeron_aa_338_1998} demonstrates that complex
morphologies can arise from massive stars' wind-ISM interactions. Bow shocks are subject to a 
wide range of shear-like and non-linear instabilities~\citep{blondin_na_57_1998} 
producing severe distortions of their overall forms, \textcolor{black}{which can only be 
analytically approximated}~\citep{wilkin_459_apj_1996} in the particular
situations of either a star moving in a relatively dense 
ISM~\citep{comeron_aa_338_1998} or a high-mass star hypersonically moving 
through the Galactic plane~\citep[][hereafter Paper~I]{meyer_2014b}. Tayloring 
numerical models to runaway red supergiant stars allows us to constrain 
the mass loss and local ISM properties of 
Betelgeuse~\citep{vanmarle_apj_734_2011,cox_aa_537_2012,mackey_apjlett_751_2012} or
IRC$-$10414~\citep{Gvaramadze_2013,meyer_mnras_439_2014}. \textcolor{black}{For} the sake of simplicity, these
models neglect the magnetisation of the ISM.

\textcolor{black}{
However, magnetic fields are an essential component of the ISM of the Galaxy, 
e.g. its large scale component has a tendency to be aligned with the galactic 
spiral arms~\citep{gaensler_apj_493_1998}. If the strength of the ISM magnetic field 
can reach up to several tenths of Gauss in the center of our 
Galaxy~\citep[see][]{rand_apj_343_1989, 
ohno_mnras_262_1993,opher_natur_462_2009,
shabala_mnras_405_2010}, it can be even stronger in the cold 
phase of the ISM~\citep{crutcher_apj_515_1999}. In particular, radio polarization 
measures of the magnetic field in the context of Galactic ionized supershells are 
reported to be $2$-$6\, \mu \rm G$ in~\citet{harvey_apj_736_2011}. This value is 
in accordance with previous estimates of the field strength in the warm phase of the 
ISM~\citep{troland_301_apj_1986} and was supported by hydrodynamical 
simulations~\citep{fiedler_apj_415_1993}.  
Such a background magnetic field should therefore be included in realistic models of 
circumstellar nebulae around massive stars. 
}

Numerical studies of magneto-hydrodynamical flows around an obstacle is approximated 
in the \textcolor{black}{plane-parallel} approach in~\citet{sterck_phpl_1998,
sterck_aa_343_1999}. A significant number of circumstellar structures, such as
the vicinity of our Sun~\citep{pogolerov_aa_321_1997}, 
planetary nebulae developing in the vicinity of intermediate-mass
stars~\citep{heiligman_mnras_191_1980} or supernova
remnants~\citep{rozyczka_274_MNRAS_1995} have been studied in such a two-dimensional
approach~\citep[see also][]{soker_apj_484_1997,pogolerov_aa_354_2000}. The 
presence of a weak magnetic field can inhibit the \textcolor{black}{growth rate of
shear} instabilities in the bow shocks around cool stars such as the runaway 
red supergiant Betelgeuse in the constellation of
Orion~\citep{vanmarle_aa_561_2014}. \textcolor{black}{We place our work} in this context, focusing 
on bow shocks generated by hot, fast winds of main-sequence massive stars.

In this study, we continue our investigation of the circumstellar medium of
runaway massive stars moving within the plane of the Milky 
way~\citep[Paper~I,][]{meyer_mnras_450_2015,meyer_mnras_459_2016}. As a
logical extension of them, we present magneto-hydrodynamical
models of a sample of some \textcolor{black}{of} the most common main-sequence, runaway massive
stars~\citep{kroupa_mnras_322_2001} moving at the most probable space
velocities~\citep{eldridge_mnras_414_2011}. We ignore any intrinsic inhomogenity
or turbulence in the ISM. Particularly, we assume an axisymmetric magnetisation
of the ISM surrounding the bow shocks in the spirit of~\citet{vanmarle_aa_561_2014}. We concentrate our
efforts on an initially $20\, \rm M_{\odot}$ star, however, we also consider bow
shocks generated by lower and higher initial mass stars. This project
principally differs from Paper~I because \textcolor{black}{of} (i) the inclusion of an ISM background
\textcolor{black}{magnetic field leads to anisotropic heat conduction}~\citep[see, e.g.][]{balsara_mnras_386_2008} and (ii) our study does not
concentrate on the \textcolor{black}{secular} stellar wind evolution of our bow-shock-producing stars. Note that our
study introduces a reduced number of representative models due to the high
numerical cost of the magneto-hydrodynamical simulations.  
\textcolor{black}{
Following~\citet{acreman_mnras_456_2016}, we additionally appreciate the effects of the ISM magnetic field on the 
bow shocks with the help of radiative transfer calculations of dust continuum emission. 
}

This paper is organised as follows. We start in Section 2 with a review of the
physics included in our models for both the stellar wind and the ISM. We also
recall the adopted numerical methods. Our models of bow shocks generated by
main-sequence, runaway massive stars moving in a magnetised medium are presented
together with a discussion of their morphology and internal structure in Section
3. We detail the emission properties of our bow shocks and discuss their observational 
implications in Section 4. Finally, we formulate our conclusions in Section 5.


\section{Method}
\label{sect:methods}

In the present section, we briefly summarise the numerical methods and
microphysics utilised to produce magneto-hydrodynamical bow shock models of
the circumstellar medium surrounding hot, runaway massive stars.

\subsection{Governing equations}
\label{sect:eq}

We consider a magnetised flow past a source of hot, \textcolor{black}{ionized and magnetized} stellar wind. 
The dynamics are described by the ideal equations of 
magneto-hydrodynamics and the dissipative character of the 
thermodynamics originates from the treatment of the gas with heating and losses 
by optically-thin radiation together with electronic heat conduction. 
These equations are,
\begin{equation}
	   \frac{\partial \rho}{\partial t}  + 
	   \bmath{\nabla}  \cdot \big(\rho\bmath{v}) =   0,
\label{eq:mhdeq_1}
\end{equation}
\begin{equation}
	   \frac{\partial \bmath{m} }{\partial t}  + 
           \bmath{\nabla} \cdot \Big( \bmath{m} \otimes \bmath{v}  + \bmath{B} \otimes \bmath{B} + \bmath{\hat{I}}p_{\rm t} \Big) 
            =   \bmath{0},
\label{eq:mhdeq_2}
\end{equation}
\begin{equation}
	  \frac{\partial E }{\partial t}   + 
	  \bmath{\nabla} \cdot \Big( (E+p_{\rm t})\bmath{v}-\bmath{B}(\bmath{v}\cdot\bmath{B}) \Big)  = \zeta(T,\rho,\mu),
\label{eq:mhdeq_3}
\end{equation}
and,
\begin{equation}
	  \frac{\partial \bmath{B} }{\partial t}   + 
	  \bmath{\nabla} \cdot \Big( \bmath{v}  \textcolor{black}{\otimes} \bmath{B} - \bmath{B} \textcolor{black}{\otimes} \bmath{v} \Big)  =
	  \bmath{0},
\label{eq:mhdeq_4}
\end{equation}
where $\rho$ and $\bmath{v}$ are the mass density and the velocity of the
plasma. In the relation of momentum conservation Eq.~(\ref{eq:mhdeq_2}), the
quantity $\bmath{m}=\rho\bmath{v}$ is the linear momentum of a gas element,
$\bmath{B}$ the magnetic field, $\bmath{\hat{I}}$ the \textcolor{black}{identity matrix} and,  
\begin{equation}
	  p_{\rm t}  =	p + \frac{ \bmath{B} \cdot \bmath{B} }{2},
\label{eq:pres}
\end{equation}
is the total pressure of the gas, i.e. the sum of its thermal component $p$ 
and its magnetic contribution $(\bmath{B} \cdot \bmath{B})/2$, respectively. 
Eq.~(\ref{eq:mhdeq_3}) describes the conservation of the total energy of the gas, 
\begin{equation}
	E = \frac{p}{(\gamma - 1)} + \frac{ \bmath{m} \cdot \bmath{m} }{2\rho} + \frac{ \bmath{B} \cdot \bmath{B} }{2},
\label{eq:energy}
\end{equation}
where $\gamma$ is the adiabatic index, \textcolor{black}{which is taken to be $5/3$}, i.e. we assume 
an ideal gas. The right-hand source term $\zeta(T,\rho,\mu)$ in
Eq.~(\ref{eq:mhdeq_3}) represents (i) the heating and the losses by
optically-thin radiative processes and (ii) the heat transfers by anisotropic
electronic thermal conduction (see Section~\ref{ssect:microphysics}). Finally,
Eq.~(\ref{eq:mhdeq_4}) is the induction equation and governs the time evolution
of the vector magnetic field $\bmath{B}$. The relation, 
\begin{equation}
	  c_{\rm s} = \sqrt{ \frac{\gamma p}{\rho} },
\label{eq:cs}
\end{equation}
closes the system Eq.(\ref{eq:mhdeq_1})$-$(\ref{eq:mhdeq_4}), where $c_{\rm
s}$ denotes the adiabatic speed of sound.

\subsection{Boundary conditions and numerical scheme}
\label{sect:scheme}

We solve the above described system of equations
Eqs.~(\ref{eq:mhdeq_1})$-$(\ref{eq:cs}) using the open-source {\sc pluto}
code\footnote{http://plutocode.ph.unito.it/}~\citep{mignone_apj_170_2007,
migmone_apjs_198_2012} on a uniform two-dimensional grid covering a rectangular
computational domain in a cylindrical frame of reference $(O; R, z)$ of origin
$O$ and symmetry axis about $R=0$. The grid $[O;R_{\rm max}]\times[-z_{\rm
min};z_{\rm max}]$ where $R_{\rm max}$, $-z_{\rm min}$ and $z_{\rm max}$ are the
upper and lower limits of the $OR$ and $Oz$ directions, respectively, which are
discretised with $N_{\rm R}=2N_{\rm z}=1000$ cells such that the grid resolution
is $\Delta_R=\Delta_z=R_{\rm max}/N_{\rm R}$. Learning from previous bow shock
models~\citep{comeron_aa_338_1998,vanmarle_aa_460_2006}, we impose inflow
boundary conditions corresponding to the stellar motion at $z=z_{\rm max}$
whereas outflow boundaries are set at $R=R_{\rm max}$ and $z=-z_{\rm min}$.
Moreover, the stellar wind is modelled setting inflow boundaries conditions
centered around the origin (see Section~\ref{sect:ism}).

We integrate the system of partial differential equations within the eight-wave
formulation of the magneto-hydrodynamical Eqs.~(\ref{eq:mhdeq_1})$-$(\ref{eq:cs}),
using a cell-centered representation consisting in evaluating $\rho$,
$\bmath{m}$, $E$ and $\bmath{B}$ using the barycenter of the cells (see
section 2 of Paper~I). This formulation, used together with the \textcolor{black}{Harten-Lax-van
Leer approximate Riemann solver~\citep{hll_ref}}, conserves the divergence-free condition
$\bmath{\nabla} \cdot \bmath{B} = \bmath{0}$. The method is a second order,
unsplit, time-marching algorithm scheme controlled by the Courant-Friedrich-Levy
parameter initially set to $C_{\rm cfl}=0.1$. The gas cooling and heating
rates are linearly interpolated from tabulated cooling curves (see
Section~\ref{ssect:microphysics}) and the corresponding rate of change is 
subtracted from the \textcolor{black}{total} energy $E$. The parabolic term of
heat conduction is integrated with the Super-Time-Stepping
algorithm~\citep{alexiades_cnme_12_1996}.


\subsection{Gas microphysics}
\label{ssect:microphysics}

The source term $\zeta(T,\rho,\mu)$ in Eq.~(\ref{eq:mhdeq_3}) represents the non-ideal 
thermodynamics processes that we take into account, and reads, 
\begin{equation}
      \zeta(T,\rho,\mu) = \itl{\Phi}(T,\rho) + \bmath{\nabla} \cdot \bmath{{F}_{\rm c}}
\label{eq:terms}
\end{equation}
where $\itl{\Phi}(T,\rho)$ is a function that stands for the processes by 
optically-thin radiation where, 
\begin{equation}
	T =  \mu \frac{ m_{\mathrm{H}} }{ k_{\rm{B}} } \frac{p}{\rho},
\label{eq:temperature}
\end{equation}
is the gas temperature, with $\mu=0.61$ the mean molecular 
weight of the gas, $k_{\rm{B}}$ the Boltzmann constant 
and $m_{\rm H}$ the proton mass, respectively. The gain and losses by optically-thin 
radiative processes are taken into account via the following law,
\begin{equation}  
	 \itl \Phi(T,\rho)  =  n_{\mathrm{H}}\itl{\Gamma}(T)   
		   		 -  n^{2}_{\mathrm{H}}\itl{\Lambda}(T),
\label{eq:dissipation}
\end{equation}
where $\itl{\Lambda}(T)$ and $\itl{\Gamma}(T)$ are the rate of change of the gas 
internal energy induced by heating and cooling as a function of $T$, respectively, 
and where $n_{\mathrm{H}}=\rho/\mu (1+\chi_{\rm He,Z}) m_{\mathrm{H}}$ is the 
hydrogen number density with $\chi_{\rm He,Z}$ the mass fraction of the coolants heavier 
than hydrogen. Details about the processes included into the cooling $\Lambda(T)$ 
and heating $\Gamma(T)$ laws are given in section 2 of Paper~I.

The divergence term in the source function in Eq.~(\ref{eq:terms}) represents 
the anisotropic heat flux, 
\begin{equation}
      \bmath{{F}_{\rm c}} = \kappa_{||} \hat{\bmath{b}} \Big(  \hat{\bmath{b}} \cdot \bmath{ \nabla } T  \Big) + \kappa_{\perp} \Big( \bmath{ \nabla } T  -  \hat{\bmath{b}} \cdot \bmath{ \nabla } T  \Big),
\label{eq:tc}
\end{equation}
where $\hat{\bmath{b}}=\bmath{B}/||\textcolor{black}{\bmath{B}}||$ is the magnetic field unit vector. It is
calculated through the interface of the nearest neighbouring cells in the whole
computational domain according to the temperature difference $\Delta T$ and to
the local field orientation $\hat{\bmath{b}}$~\citep[\textcolor{black}{see appendix of}][]{migmone_apjs_198_2012}.
The coefficients $\kappa_{||}$ and $\kappa_{\perp}$ are the heat coefficients
along the directions parallel and normal to the local magnetic field streamline,
respectively.   Along the direction of the local magnetic field, 
\begin{equation}
      \kappa_{||}=K_{||} T^{5/2},
      \label{fig:eq_kappa_par}
\end{equation}
with,
\begin{equation}
      K_{||} =  \frac{ 1.84 \times 10^{-5} }{ \ln( \mathcal{L} ) }\, \rm erg\, \rm s^{-1}\, \rm K^{-1}\, \rm cm^{-1},
      \label{fig:eq_kappa}
\end{equation}
where $\ln( \mathcal{L} ) = 29.7 + \ln ( T / 10^{6}\sqrt{n}) $ is the Coulomb
logarithm, with $n$ the gas total number density~\citep{spitzer_1962}. The heat
conduction coefficients satisfy $\kappa_{\perp}/\kappa_{||} \approx 10^{-16} \ll 1$ for the
densities that we
consider~\citep{parker_1963_book,velazquez_apj_601_2004,balsara_mnras_386_2008,
orlando_apj_678_2008}. The value of $\textcolor{black}{\bmath{{F}_{\rm c}}}$ varies between the
classical flux in Eq.~(\ref{eq:tc}) and the saturated conduction
regime~\citep{balsara_mnras_386_2008} which limits the heat flux to, 
\begin{equation}
    F_{\rm sat} = 5 \phi \rho c_{\rm iso}^{3},
    \label{fig:eq_sat}
\end{equation}
for very large temperature gradients (\textcolor{black}{$\ge\, 10^{6}\, \rm K\, \rm pc^{-1}$}), with 
$ c_{\rm iso} = p/\rho$ the isothermal speed of sound and $\phi<1$ a 
free parameter that we set to the typical value of $0.3$~\citep{cowie_apj_211_1977}.

\begin{table*}
	\centering
	\caption{
	\textcolor{black}{
	 Stellar wind parameters at the beginning of the simulations, at a time $t_{\rm start}$ 
	 after the beginning of the zero-age main-sequences of the star.
	 Parameter $M_{\star}$ (in $\rm M_{\odot}$) is the initial mass of the star, $L_{\star}$ 
	 the stellar luminosity (in $\rm L_{\odot}$), $\dot{M}$ 
	 its mass loss and $v_{\rm w}$ the wind velocity, see also table~1 of~\citet{meyer_mnras_459_2016}.
	 }   
	 }
	\begin{tabular}{ccccccc}
	\hline
	$M_{\star}\, (\rm M_{\odot})$  &  $t_{\mathrm{ start}}\, (\rm Myr)$
				   &   $\textcolor{black}{\log(L_{\star}/\rm L_{\odot})}$                              
			           &   $\log(\dot{M}/\rm M_{\odot}\, \rm yr^{-1})$
			           &   $v_{\rm w}\, (\mathrm{km}\, \mathrm{s}^{-1})$			           
			           &   $T_{\rm eff}\, (\mathrm{K})$
			\\ \hline   
	$10$ &  $5.0$  &  $3.80$ 		    &  $-9.52$   & $1082$ & $25200$         \\        
	$20$ &  $3.0$  &  $4.74$                    &  $-7.38$   & $1167$ & $33900$         \\        
  	$40$ &  $0.0$  &  $5.34$  	            &  $-6.29$   & $1451$ & $42500$         \\ 
	\hline 
	\end{tabular}
\label{tab:wind_para}
\end{table*}


\subsection{Setting up the stellar wind}
\label{sect:ism}

We impose the stellar wind at the surface of a sphere of 
radius $20\Delta z\, \rm pc$ centered into the origin $O$ with 
wind material. Its density is, 
\begin{equation}
	\rho_{w} = \frac{ \dot{M} }{ 4\pi r^{2} v_{\rm w} },
\label{eq:wind}
\end{equation}
where $\dot{M}$ is the star's mass-loss rate and $r$ the distance to the origin 
$O$. We interpolate the wind parameters from stellar evolution models of 
non-rotating massive stars with solar metallicity that we used for previous 
studies, see Paper~I. 
\textcolor{black}{
Our stellar wind models are have been generated with the stellar evolution code 
described in~\citet{heger_apj_626_2005} and subsequently updated 
by~\citet{yoon_443_aa_2005,petrovic_aa_435_2005} and~\citet{brott_aa_530_2011a}.
It utilises the mass-loss prescriptions 
of~\citet{kudritzki_aa_219_1989} for the main-sequence phase of our massive 
stars and of~\citet{dejager_aas_72_1988} for the red supergiant phase. 
Despite of the fact that our 
wind models report the marginal evolution of 
main-sequence winds, see Paper~I, they remain quasi-constant during the part of 
the stellar evolution that we follow. 
We refer the reader 
interested in a graphical representation of the utilised wind models to the fig~3 
of Paper~I, while we report the wind properties at the beginning of our simulations 
in our Table~\ref{tab:wind_para}.  
}
\textcolor{black}{
Note that our adopted values for the stellar wind velocity belong to the lower 
limit of the range of validity for stellar winds of OB stars (see below in 
Section~\ref{sect:physics33}).
}

Since we assume a spherically symmetric stellar wind density, thermal pressure and 
velocity profiles, we use the Parker prescription~\citep{parker_paj_128_1958} to model 
the magnetic field in the stellar wind. It consists of a radial component of the field, 
\begin{equation}
	B_{r} = B_{\star} \Big( \frac{R_{\star}}{r} \Big)^{2},
\label{eq:Bstellarfield_1}
\end{equation}
where $B_{\star}$ and $R_{\star}$ are the stellar surface magnetic field and the stellar radius, 
respectively, and of a toroidal component, which, in the case of a non-rotating star, this 
reduces to $B_{\phi} = 0$. The $\propto 1/r^{2}$ radial dependence of
Eq.~(\ref{eq:Bstellarfield_1}) makes the strength
of the stellar magnetic field almost negligible at the wind termination shock
that is typically about a few tenths of $\rm pc$ from the star that we study 
(Paper~I). However, imposing a null
magnetic field in the stellar wind region would let the direction of the heat
flux $\bmath{{F}_{\rm c}}$ undetermined in the region of (un)shocked wind 
material of the bow shock, see magnetic field unit vector $\hat{\bmath{b}}$ 
in the right-hand side of Eq.~(\ref{eq:tc}). Note that, given their analogous radial 
dependance on $r$, stellar wind and stellar magnetic field are 
similarly implemented into our axisymmetric simulations. 
In these simulations the stellar surface magnetic 
field is set to $B_{\star} \simeq 1.0\,\rm kG$~\citep{donati_mnras_333_2002} 
at $R_{\star}=3.66\, R_{\odot}$~\citep{brott_aa_530_2011a} where $R_{\odot}$ 
is the solar radius.

\subsection{Setting up the ISM}
\label{sect:medium}

Our runaway stars are moving through the warm ionised phase of the ISM, i.e.
we assume that they run in their own $\HII$ region inside \textcolor{black}{which the gas is considered 
as} homogeneous, laminar and fully ionised fluid. The ISM composition 
assumes solar metalicity~\citep{lodders_apj_591_2003}, with 
$n_{\rm H}=0.57\, \mathrm{cm}^{-3}$~\citep{wolfire_apj_587_2003} and
with $T_{\rm ISM}\approx 8000\, \rm K$, initially. 
The model is a moving star within an ISM at rest. We solve the equations 
of motion \textcolor{black}{in} the frame in which the star is at rest and, hence, 
the ISM moves with $v_{\mathrm{ISM}}=-v_{\star}$, 
where $v_{\star}$ is the bulk motion of the star. 
The gas in the computational domain is 
evaluated with the cooling curve for photoionised gas described in fig.~4a of Paper~I. 
\textcolor{black}{In particular}, our initial conditions neglect the possibility that a bow shock might trap the ionising front 
of the $\HII$ region (see section~2.4 of Paper~I for an extended discussion of 
the assumptions underlying our method for modelling bow shocks from hot massive stars). 
Additionally, an axisymmetric magnetic field $\bmath{B}=-B_{\rm ISM}\, \bmath{\hat{z}}$ field 
is imposed over the whole computational domain, with $B_{\rm ISM}>0$ its 
strength and $\bmath{\hat{z}}$ the unit vector along the $Oz$ direction. Finally, 
our simulations trace the respective proportions of ISM gas with respect to the 
wind material using a passive scalar tracer according to the advection equation,
\begin{equation}
	\frac{\partial (\rho Q) }{\partial t } +  \bmath{ \nabla } \cdot  ( \bmath{v} \rho Q) = 0,
\label{eq:tracer}
\end{equation}
where $Q$ is a passive tracer which initial value is $Q(\bmath{r})=1$ for the wind material and 
$Q(\bmath{r})=0$ for the ISM gas, respectively.

\subsection{Simulation ranges}
\label{sect:models}

We first focus on a baseline bow shock generated by an initially 
$20\, \rm M_{\odot}$ star moving with a velocity $v_{\star}=40\, \rm km\, \rm s^{-1}$ in
the Galactic plane of the Milky Way \textcolor{black}{whose magnetic field} is assumed to be $B_{\rm
ISM}=7\, \mu \rm G$~\citep{draine_piim_bool_2011}. Then, we consider models with velocity $v_{\star}=20$ to 
$70\, \rm km\, \rm s^{-1}$, \textcolor{black}{explore the effects of a magnetisation
of $B_{\rm ISM}=3.5\, \mu \rm G$}, and carry out simulations 
of initially $10$ and $40\, \rm M_{\odot}$ stars 
moving at velocities $v_{\star}=40$ and $70\, \rm km\, \rm s^{-1}$, respectively. 
We investigate the effects of the ISM magnetic field carrying out a couple of 
\textcolor{black}{additional purely hydrodynamical simulations}, as comparison runs. All our 
simulations are started at a time about $4.5\, \rm Myr$ 
after the zero-age main-sequence phase of our stars and are run at least 
four crossing times $|z_{\rm max}-z_{\rm min}|/v_{\star}$ of the gas through 
the computational domain, such that the system \textcolor{black}{reaches} a steady or quasi-stationary   
state in the case of a stable or unstable bow shock, respectively.

\textcolor{black}{We label our magneto-hydrodynamical simulations concatenating 
the values of the initial mass $M_{\star}$ of the moving
star (in $\rm M_{\odot}$), its bulk motion} $v_{\star}$ (in $\rm km\,
\rm s^{-1}$) and the included physics ``Ideal'' for dissipativeless simulations,
``Cool" if the model includes heating and losses by optically-thin radiative
processes, ``Heat" for heat conduction and ``All" if cooling, heating and heat 
conduction are taken into account together). Finally, the labels inform about the
strength of the ISM magnetic field. We distinguish our magneto-hydrodynamical runs
from our previously published hydrodynamical studies (Paper~I) adding the prefix 
``HD" and ``MHD" to the simulations labels of our hydrodynamical and 
magneto-hydrodynamical simulations, respectively. All the informations 
relative to our models are summarised in Table~\ref{tab:ms}.

\begin{table*}
	\centering
	\caption{
	 Nomenclature and grid parameters used in our (magneto-)hydrodynamical
simulations. The quantities $M_{\star}$ (in $M_{\odot}$) and
$v_{\star}$ (in $\mathrm{km}\, \mathrm{{s}^{-1}}$) are the initial mass of the
stars and their space velocity, respectively, whereas $B_{\rm ISM}$ (in $\mu \rm G$) is 
the strength of the ISM magnetic field. Parameters $\it \Delta$,
$z_{\mathrm{min}}$ and $R_{\mathrm{max}}$ are the resolution of the uniform grid
(in $\mathrm{pc}\, \mathrm{{cell}^{-1}}$) and the lower and upper limits of the 
domain along the $\rm R$-axis and $\rm z$-axis (in $\mathrm{pc}$), respectively.  The
last column contains the physics included in each simulation. Heat conduction (HC) 
refers to isotropic thermal conduction in the case of an hydrodynamical (HD) simulation and 
to anisotropic thermal conduction in the case of an magneto-hydrodynamical (MHD) simulation, 
respectively. 
	 }
	\begin{tabular}{lccccccc}
	\hline
	${\rm {Model}}$ &   $M_{\star}\, (M_{\odot})$                              
 			&   $v_{\star}\, (\mathrm{km}\, \mathrm{s}^{-1})$
			&   $B_{\rm ISM}\, (\mu \rm G)$ 
			&   $\itl{\Delta}\, ( 10^{-3}\, \mathrm{pc}\, \mathrm{cell}^{-1})$ 
			&   $z_{\mathrm{min}}\, (\mathrm{pc})$ 
			&   $R_{\mathrm{max}}\, (\mathrm{pc})$
			&   $\rm Included\, microphysics$
			\\ \hline   
	HD2040Ideal      &  $20$  &  $40$ & $-$   &  $8.0$   &  $-2.0$     &  $~8.0$      & HD, adiabatic  \\             
	HD2040Cool       &  $20$  &  $40$ & $-$   &  $8.0$   &  $-2.0$     &  $~8.0$      & HD, cooling, heating  \\        
	HD2040Heat       &  $20$  &  $40$ & $-$   &  $8.0$   &  $-2.0$     &  $~8.0$      & HD, HC  \\ 
	HD2040All        &  $20$  &  $40$ & $-$   &  $8.0$   &  $-2.0$     &  $~8.0$      & HD, cooling, heating, HC  \\             
	MHD2040IdealB7   &  $20$  &  $40$ & $7.0$   &  $8.0$   &  $-2.0$     &  $~8.0$   & MHD   \\        
	MHD2040CoolB7    &  $20$  &  $40$ & $7.0$   &  $8.0$   &  $-2.0$     &  $~8.0$   & MHD, cooling, heating  \\  
	MHD2040HeatB7    &  $20$  &  $40$ & $7.0$   &  $8.0$   &  $-2.0$     &  $~8.0$   & MHD, HC  \\         
	MHD2040AllB7     &  $20$  &  $40$ & $7.0$   &  $8.0$   &  $-2.0$     &  $~8.0$   & MHD, cooling, heating, HC  \\     
	MHD1040AllB7     &  $10$  &  $40$ & $7.0$   &  $3.0$   &  $-2.0$     &  $~6.0$   & MHD, cooling, heating, HC  \\         	
	MHD2020AllB7     &  $20$  &  $20$ & $7.0$   &  $6.0$   &  $-3.0$     &  $12.0$  & MHD, cooling, heating, HC  \\         
	MHD2040AllB3.5   &  $20$  &  $40$ & $3.5$   &  $8.0$   &  $-2.0$     &  $~8.0$   & MHD, cooling, heating, HC  \\  
	MHD2070AllB7     &  $20$  &  $70$ & $7.0$   &  $1.2$   &  $-1.0$     &  $~3.0$   & MHD, cooling, heating, HC  \\     
	MHD4070AllB7     &  $40$  &  $70$ & $7.0$   &  $1.6$   &  $-4.0$     &  $16.0$  & MHD, cooling, heating, HC  \\
	\hline    
	\end{tabular}
\label{tab:ms}
\end{table*}


\section{Results and discussion}
\label{sect:results}

This section presents the magneto-hydrodynamical simulations carried out in the
context of our Galactic, ionizing, runaway massive stars. We detail the
effects of the included microphysics on a baseline bow shock model,  
we discuss the morphological differences between our hydrodynamical and 
magneto-hydrodynamical simulations and we consider the 
effects of the adopted stellar wind models. 
Finally, review the limitations of the model.

\subsection{Bow shock thermodynamics}
\label{sect:physics1}

\subsubsection{Effects of the included physics: hydrodynamics}
\label{sect:physics1}

\begin{figure*}
	\begin{minipage}[b]{ 0.48\textwidth}
		\includegraphics[width=1.0\textwidth]{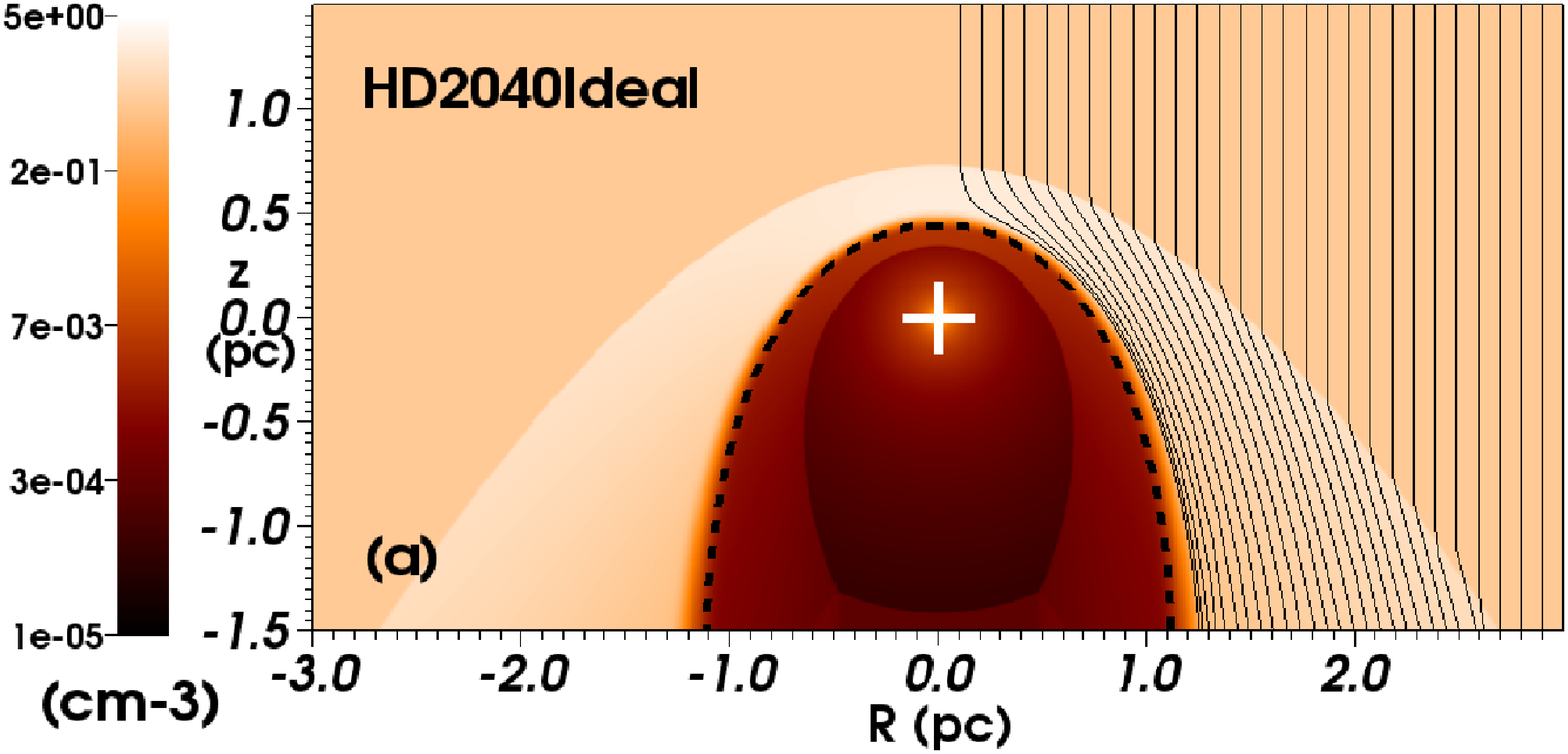}
	\end{minipage}
	\begin{minipage}[b]{ 0.48\textwidth}
		\includegraphics[width=1.0\textwidth]{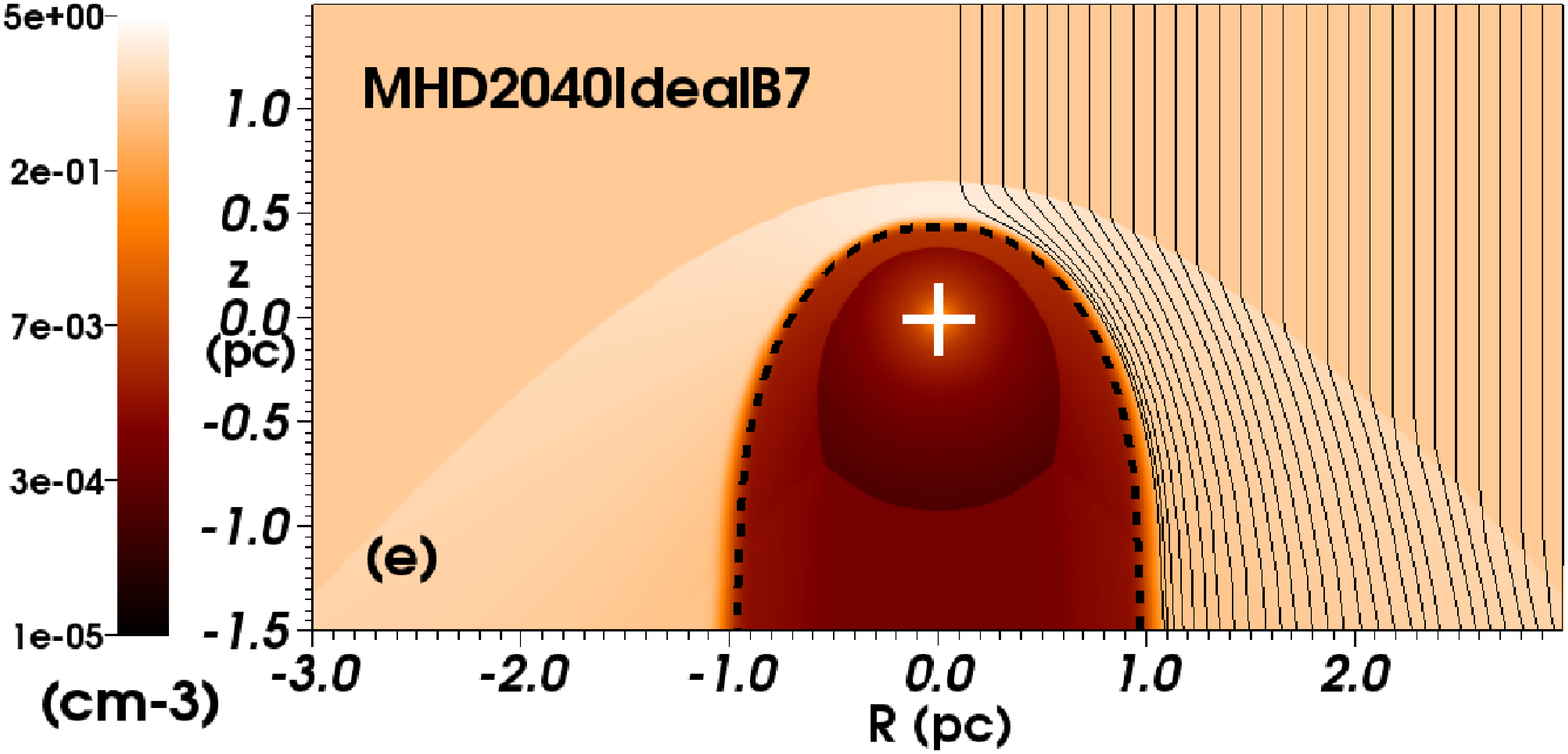}
	\end{minipage} \\
	\begin{minipage}[b]{ 0.48\textwidth}
		\includegraphics[width=1.0\textwidth]{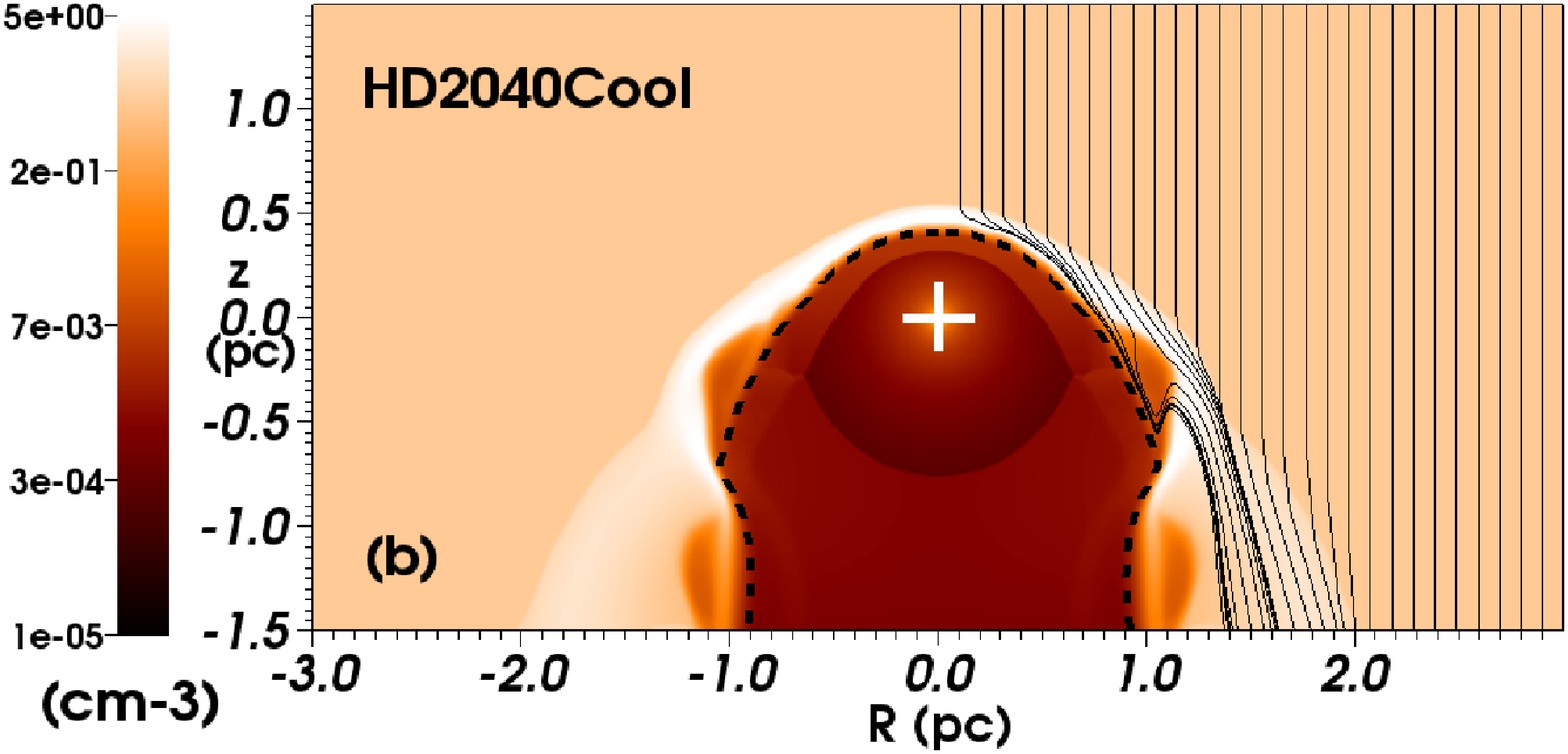}
	\end{minipage}
	\begin{minipage}[b]{ 0.48\textwidth}
		\includegraphics[width=1.0\textwidth]{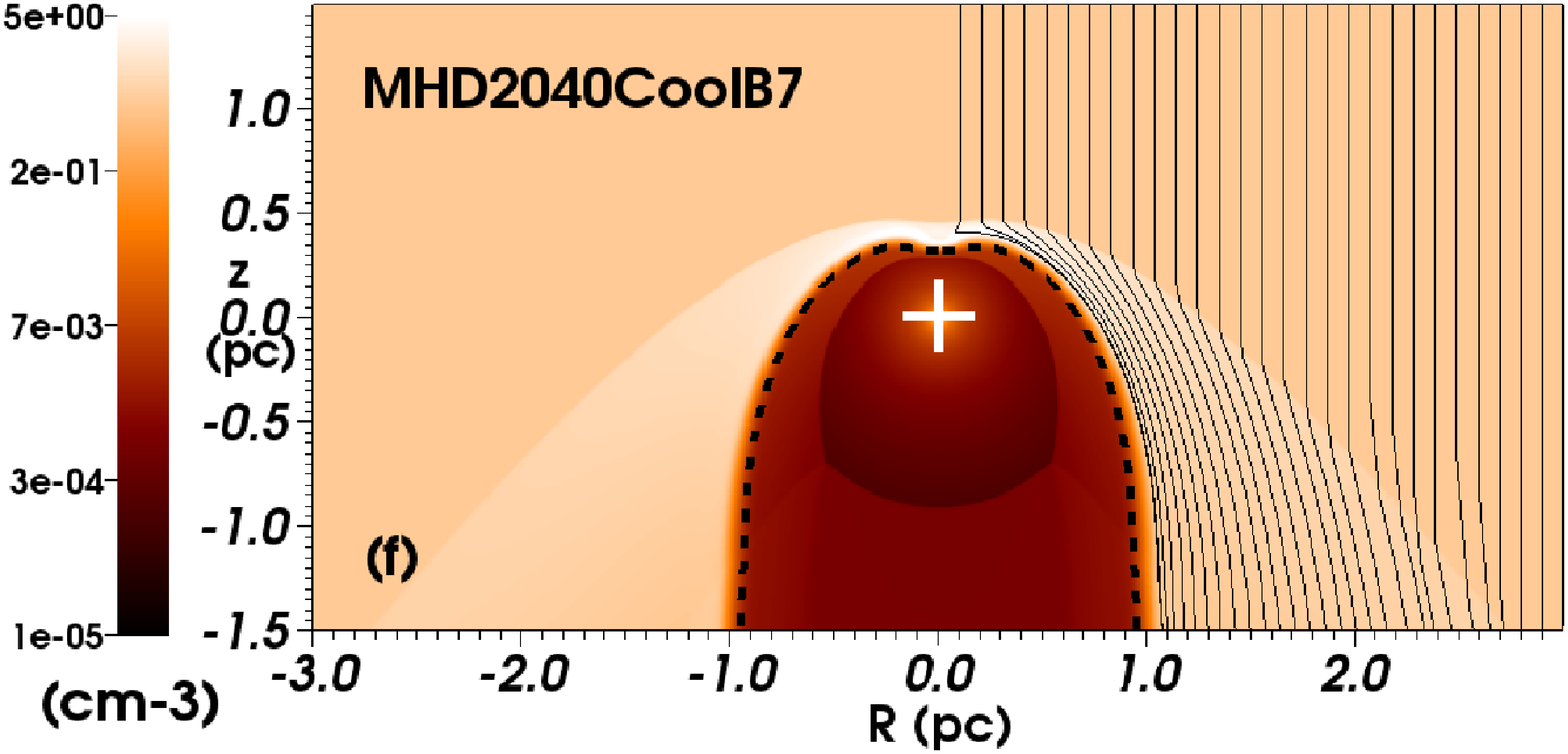}
	\end{minipage} \\
	\begin{minipage}[b]{ 0.48\textwidth}
		\includegraphics[width=1.0\textwidth]{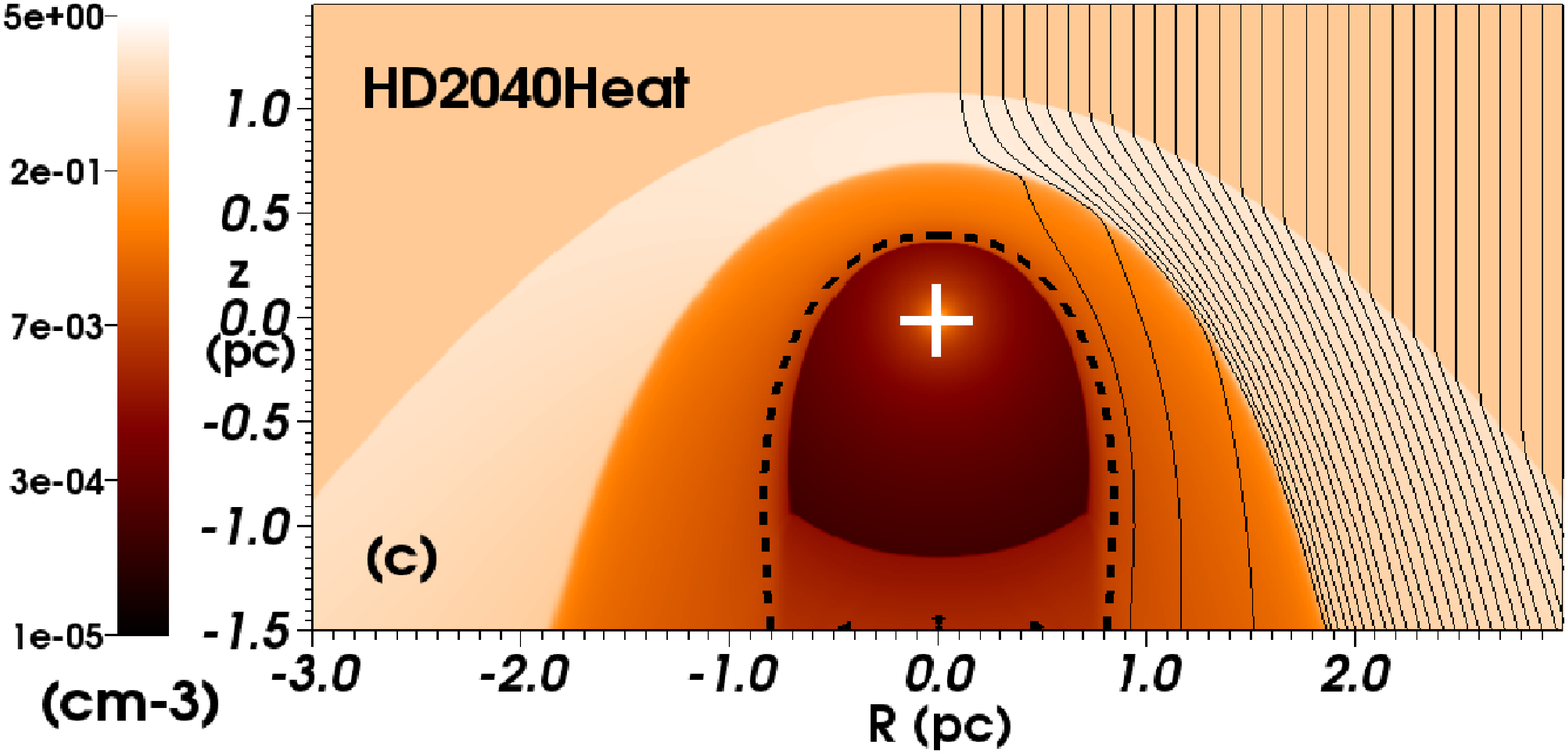}
	\end{minipage}
	\begin{minipage}[b]{ 0.48\textwidth}
		\includegraphics[width=1.0\textwidth]{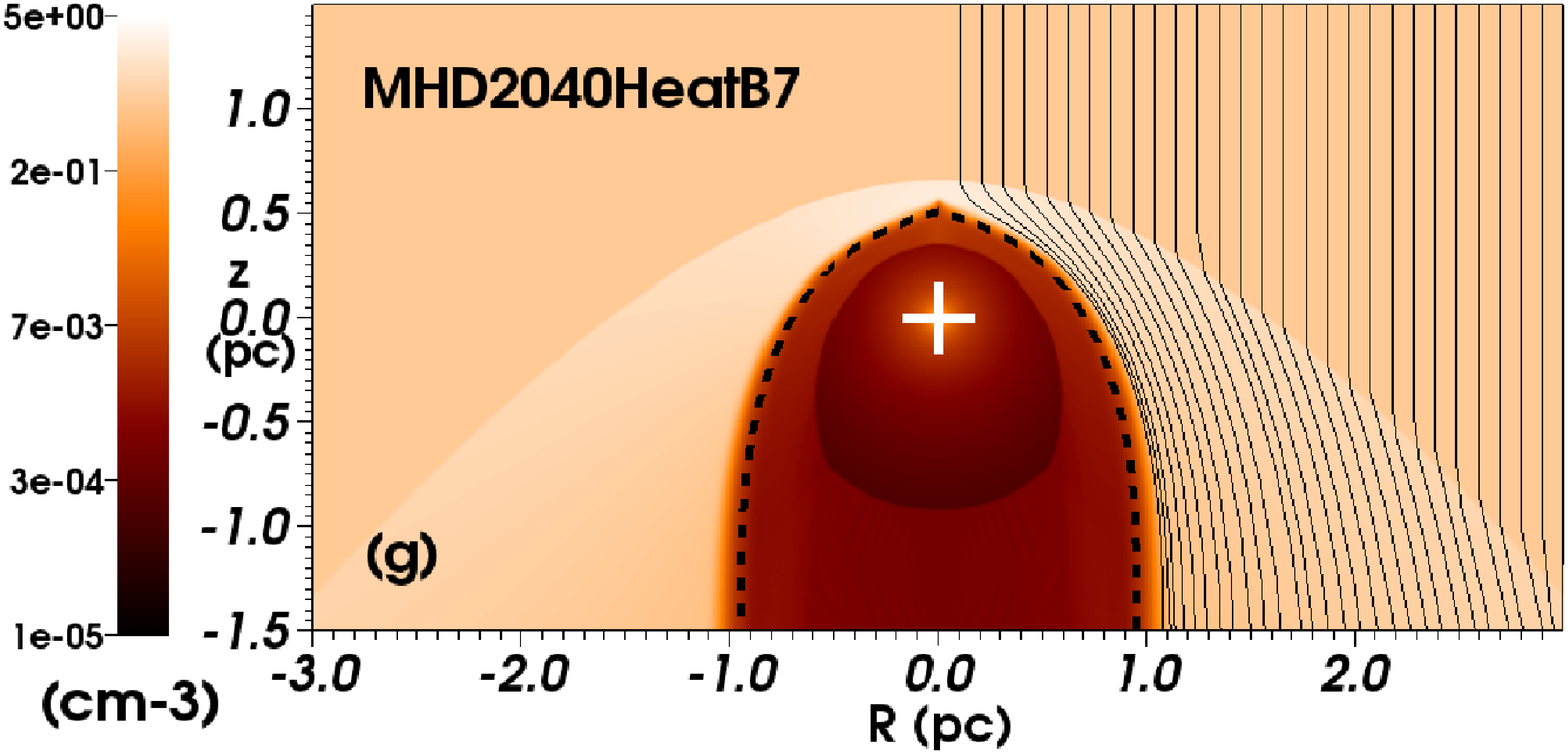}
	\end{minipage} \\ 
	\begin{minipage}[b]{ 0.48\textwidth}
		\includegraphics[width=1.0\textwidth]{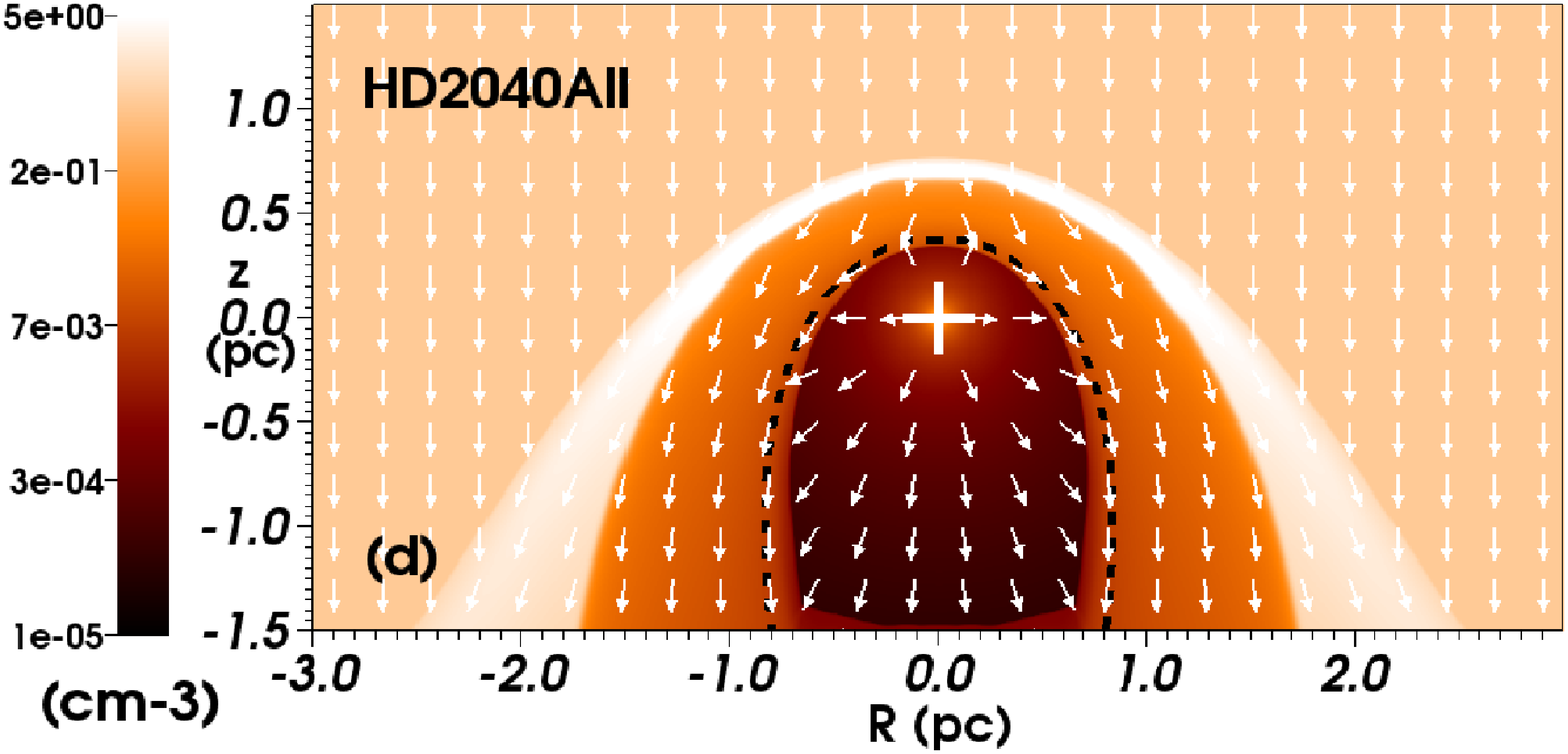}
	\end{minipage}
	\begin{minipage}[b]{ 0.48\textwidth}
 		\includegraphics[width=1.0\textwidth]{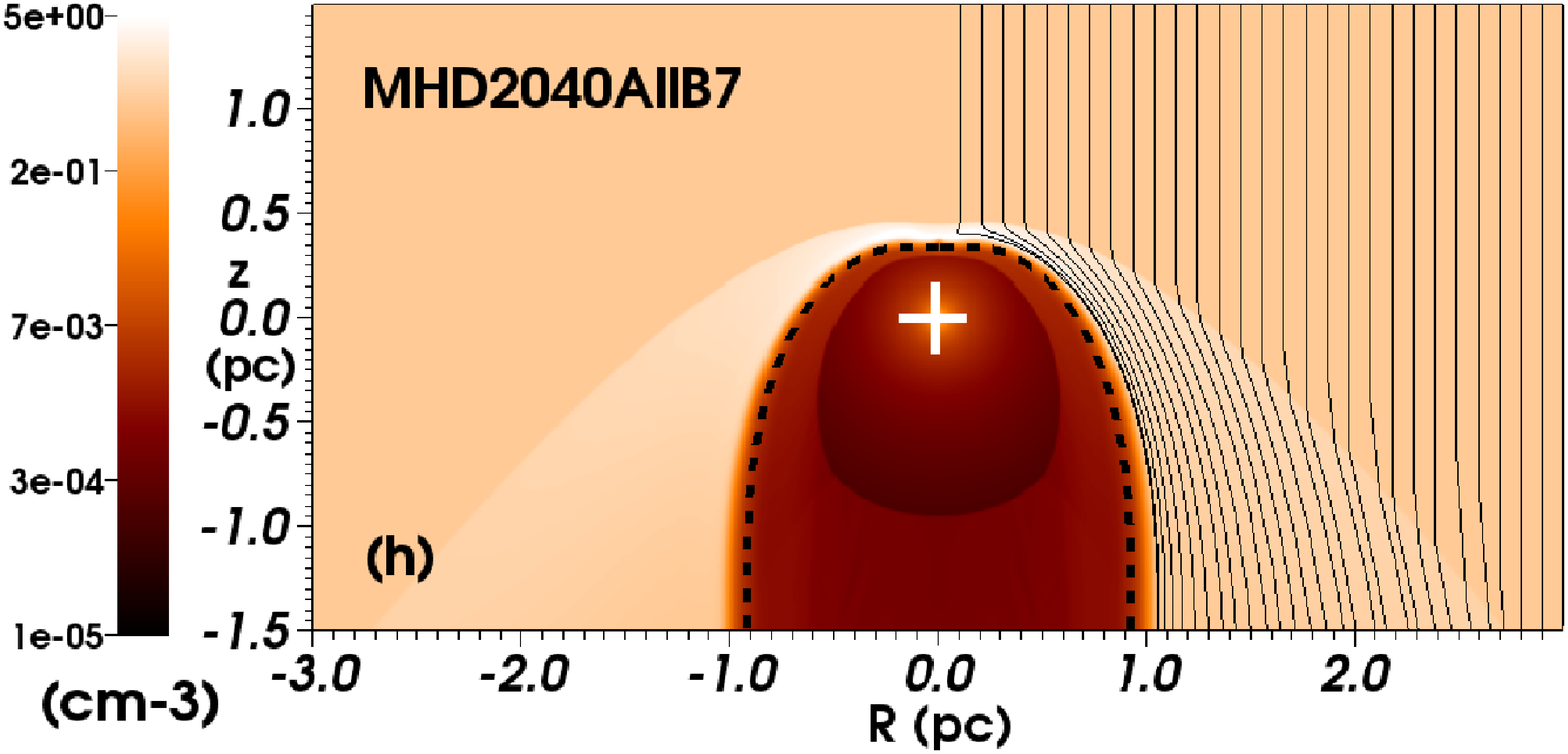}
	\end{minipage} \\ 	
	\caption{   
	Changes in the morphology of a stellar wind bow shock with variation of
the included physics. Figures show gas number density plotted with a density
range from $10^{-5}$ to $5\, \rm cm^{-3}$ in the logarithmic scale for an
initially $20\, \rm M_{\odot}$ star moving with velocity $40\, \rm km\, \rm
s^{-1}$. Left-hand panels are the hydrodynamical models whereas right-hand
panels are the magneto-hydrodynamical models with $B_{\rm ISM}=7\, \rm \mu G$.
The first line of panels shows adiabatic (a) and ideal magneto-hydrodynamical
(e) models, respectively. The second line of panels plots models with optically-thin 
radiative processes (b,f), the third line shows models including thermal (an-)isotropic
conduction (c,g) and the last line plots models models including cooling,
heating and (an-)isotropic thermal conduction (d,h). The nomenclature of the
models follows our Table~\ref{tab:ms}. For
each figure the dotted thick line traces the material discontinuity, i.e. 
the interface of the wind/ISM regions, $Q(\bmath{r})=1/2$. The right part of each figure overplots ISM flow
streamlines, except panel (d) which explicitly plots the velocity 
field as white arrows over the whole computational domain. 
The crosses mark the position of the star. 
\textcolor{black}{ The R-axis represents the radial direction and the z-axis 
the direction of stellar motion (in $\rm pc$).}
Only a fraction of the computational domain is shown.
		 }
	\label{fig:physics}  
\end{figure*}

In Fig.~\ref{fig:physics}, we show the gas density field in a series of bow
shock models of our initially $20\, \rm M_{\odot}$ star moving with velocity
$40\, \rm km\, \rm s^{-1}$ through a medium of ISM background density $n_{\rm
H}=0.59\, \rm cm^{-3}$ and of magnetic field strength $B_{\rm ISM}=7\, \mu \rm
G$. The crosses indicate the position of the moving star. 
The figures correspond to a time about $5\, \rm Myr$ after the beginning of
the main-sequence phase of the star. The stellar wind and ISM properties are
the same for all figures, only the included physics is different for each models
(our Table~\ref{tab:ms}). Left-hand panels are hydrodynamical simulations whereas  
right-hand panels are magneto-hydrodynamical simulations, respectively. 
From top to bottom, the included thermodynamic processes are adiabatic (a), 
take into account optically-thin radiative processes of the gas (b), heat 
transfers (c) or both (d). The black dotted lines are the contours 
$Q(\bmath{r})=1/2$ which trace the discontinuity between the stellar wind 
and the ISM gas. The streamlines (a-c) and vector velocity field (d) 
highlight the penetration of the ISM gas into the different layers 
of the bow shock.

The internal structure of the bow shocks can be understood \textcolor{black}{by} comparing the timescales
associated to the different physical processes at work. The dynamical 
timescale represents the time interval it takes the gas to advect  
through a given layer of our bow shocks, i.e. the region of shocked 
ISM or the layer of shocked wind. It is defined as, 
\begin{equation}
t_{\mathrm{dyn}}=\frac{ \Delta l }{ v }, 
\label{eq:time_dyn}
\end{equation}
where $\Delta l$ is the characteristic lengthscale of the region of the bow shock 
measured along the $Oz$ direction and where $v$ is the gas velocity in the post-shock 
region of the considered layers. According to the Rankine-Hugoniot relations and taking 
into account the non-ideal character of our model, we should have 
$v \simeq v_{\star}/4$ in the shocked ISM and $v \simeq v_{\rm w}/4$ in the post-region at the 
forward shock and at the reverse shock, respectively.

The cooling timescale is defined as,  
\begin{equation}
t_{\mathrm{cool}} =  \frac{E_{\rm int}}{\dot{E_{\rm int}}}  = \frac{p}{(\gamma-1)\itl{\Lambda}(T) n_{\rm H}^{2}},  
\label{eq:time_cool}
\end{equation}
where $\dot{E_{\rm int}}$ is the rate of change of internal energy $E_{\rm int}$~\citep{orlando_aa_444_2005}. 
The heat conduction timescale measures the rapidity of heat transfer into the 
bow shock, and is given by, 
\begin{equation}
  t_{\rm heat} = \frac{7pl^{2}}{2(\gamma-1)\kappa(T)T}, 
\label{eq:time_heat}
\end{equation}
where $l$ is a characteristic length of the bow shock along which heat transfers take 
place. Measuring the density, 
pressure and velocity fields in our simulations, we evaluate and compare those quantities 
defined in Eqs.~(\ref{eq:time_dyn})-(\ref{eq:time_heat}) at both 
the post-shock regions at the forward and reverse shocks. 
Results for both the layers of shocked wind and shocked 
ISM material are given in Table~\ref{tab:ts}.

\begin{table}
	\centering
	\caption{
	Characteristics dynamical timescale $t_{\rm dyn}$, cooling timescale $t_{\rm cool}$ and thermal
conduction timescale $t_{\rm heat}$ (in $\rm Myr$) measured along the $Oz$ direction from our simulations
of our initially $20\, \rm M_{\odot}$ star moving velocity $40\, \mathrm{km}\,
\mathrm{s}^{-1}$ (see Fig.~\ref{fig:physics}a-h). We estimate the various
timescales in both the post-shock region at the forward shock (FS) and the
reverse shock (RS) of our bow shocks. The black hyphen indicate that the
corresponding physical process is not included in the models (our
Table~\ref{tab:ms}).  
	}
	\begin{tabular}{lccccccccc}
	\hline
	${\rm {Model}}$ &   $t_{\rm dyn}\, (\rm Myr)$                              
 			 &   $t_{\rm cool}\, (\rm Myr)$
			 &   $t_{\rm heat}\, (\rm Myr)$ 
			\\ \hline     
	HD2040Ideal (FS)     &  $2.5\times 10^{-2}$  &  $-$  &  $-$    \\             
	HD2040Ideal (RS)     &  $4.7\times 10^{-3}$  &  $-$  &  $-$    \\             
	HD2040Cool (FS)      &  $1.0\times 10^{-2}$  &  $4.5\times 10^{-3}$  &  $-$    \\        
	HD2040Cool (RS)      &  $3.7\times 10^{-3}$  &  $3.5\times 10^{+3}$  &  $-$    \\        
	HD2040Heat (FS)      &  $6.5\times 10^{-2}$  &  $-$                  &  $1.2\times 10^{+3}$    \\ 
	HD2040Heat (RS)      &  $1.3\times 10^{-2}$  &  $-$                  &  $1.2\times 10^{-4}$    \\ 
	HD2020All (FS)       &  $9.0\times 10^{-3}$  &  $5.1\times 10^{-3}$  &  $8.7\times 10^{+5}$    \\             
	HD2020All (RS)       &  $1.1\times 10^{-2}$  &  $2.5\times 10^{+1}$  &  $4.0\times 10^{-3}$    \\             
	MHD2040IdealB7 (FS)  &  $1.8\times 10^{-2}$  &  $-$  		       &  $-$    \\        
	MHD2040IdealB7 (RS)  &  $4.3\times 10^{-3}$  &  $-$                  &  $-$    \\        
	MHD2040CoolB7 (FS)   &  $1.1\times 10^{-1}$  &  $4.3\times 10^{-2}$  &  $-$    \\ 
	MHD2040CoolB7 (RS)   &  $2.0\times 10^{-3}$  &  $1.0\times 10^{+3}$  &  $-$    \\  	
	MHD2040HeatB7 (FS)   &  $1.0\times 10^{-2}$  &  $-$                  &  $1.1\times 10^{+25}$    \\         
	MHD2040HeatB7 (RS)   &  $6.7\times 10^{-3}$  &  $-$                  &  $5.5\times 10^{+9}$    \\         
	MHD2040AllB7 (FS)    &  $3.0\times 10^{-1}$  &  $3.4\times 10^{-3}$  &  $1.4\times 10^{+18}$    \\   
	MHD2040AllB7 (RS)    &  $2.3\times 10^{-3}$  &  $3.0\times 10^{+4}$  &  $5.4\times 10^{+7}$    \\ 	
	\hline    
	\end{tabular}
\label{tab:ts}
\end{table}

Our hydrodynamical, \textcolor{black}{dissipation-free} bow shock model HD2040Ideal has a morphology
governed by the gas dynamics only (Fig.~\ref{fig:physics}a). It has a contact
discontinuity separating the outer region of cold shocked ISM from the inner 
region of hot shocked stellar wind, which are themselves bordered by the forward and reverse shocks,
respectively. There is no advection of ISM material into the wind region (see
the ISM gas streamlines in Fig.~\ref{fig:physics}a). The model HD2040Cool including 
cooling by optically-thin radiation has a considerably reduced layer of dense, 
shocked ISM gas caused by the rapid losses of internal energy 
($t_{\rm dyn} \gg t_{\rm cool}$, see timescales in our 
Table~\ref{tab:ts}). Its thinness favours 
the growth of Kelvin-Helmholtz instabilities and allows large eddies
to develop in the shocked regions (Fig.~\ref{fig:physics}b). 
The layer of hot gas is isothermal because the regular wind 
momentum input at the reverse shock prevents it from cooling and 
it therefore conserves its hot temperature ($t_{\rm cool} \gg t_{\rm dyn}$) 
whereas the distance between the star and the contact discontinuity, 
\begin{equation}
	R(0) = \sqrt{ \frac{\dot{M}v_{\mathrm{w}}}{4\pi\rho_{\mathrm{ISM}}v_{\star}^{2} } },
\label{eq:Ro}
\end{equation}
does not evolve~\citep{wilkin_459_apj_1996}.

\textcolor{black}{
The model HD2040Heat takes into account thermal conduction which is isotropic in the 
case of the absence of magnetic field. The heat flux reads,} 
\begin{equation}
     \bmath{{F}_{\rm c}} = \kappa \bmath{ \nabla } T, 
\label{eq:tciso}
\end{equation}
and transports internal energy from the reverse shock to the contact
discontinuity ($t_{\rm dyn} \gg t_{\rm heat}$) which in its turn splits 
the dense region into a hot ($t_{\rm dyn} \gg t_{\rm heat}$) and a cold layer of shocked ISM
gas ($t_{\rm dyn} \ll t_{\rm heat}$), respectively. 
This modifies the penetration of ISM gas into the bow shock and
causes the region of shocked wind to shrink to a narrow layer of material close to the
\textcolor{black}{reverse shock} (Fig.~\ref{fig:physics}c). Not surprisingly, the
model with both cooling and conduction HD2040All (Fig.~\ref{fig:physics}d)
presents both the thermally split region of shocked ISM ($t_{\rm dyn} \ll t_{\rm heat}$, $t_{\rm dyn} \gg t_{\rm cool}$, 
$t_{\rm cool} \ll t_{\rm heat}$) and a reduced layer
of shocked wind material ($t_{\rm dyn} \gg t_{\rm heat}$, 
$t_{\rm dyn} \ll t_{\rm cool}$, $t_{\rm cool} \gg t_{\rm heat}$) that reorganises the internal structure of the bow
shock together with a dense shell of cool ISM gas (see the also discussion in Paper~I).  
For the sake of clarity Fig.~\ref{fig:physics}d overplots the gas velocity fields as white arrows 
which illustrate the penetration of ISM gas into the hot layer of the bow shock.


\subsubsection{Effects of the included physics: magneto-hydrodynamics}
\label{sect:physics2}

We plot in the right-hand panels of Fig.~\ref{fig:physics} the ideal
magneto-hydrodynamical simulation of our initially $20\, \rm M_{\odot}$ star
moving with $v_{\star}=40\, \rm km\, \rm s^{-1}$ through a medium where the strength
of the magnetic field is $B_{\rm ISM}=7\, \mu \rm G$ (e) together with models
including cooling and heating by optically-thin radiation (f), anisotropic heat
conduction (g) and both (h). Despite of the fact that the overall morphology of
our magneto-hydrodynamical bow shock models is globally similar to the models
with $B_{\rm ISM}=0\, \mu \rm G$, a given number of significant changes relative
to both their shape and internal structure arise. Note that in the context of
our magneto-hydrodynamical models, $t_{\rm heat}$ represents the heat transfer
timescale normal to the fields lines.

\textcolor{black}{Our ideal magneto-hydrodynamical model has the typical structure 
of a stellar wind bow shock, with a region of shocked ISM gas surrounding the 
one of shocked wind gas. The contact discontinuity acts as a border between 
the two kind of material (Fig.~\ref{fig:physics}e). The model with cooling 
MHD2040CoolB7 has reduced but denser layer of ISM gas (Fig.~\ref{fig:physics}f) 
due to the rapid cooling time ($t_{\rm cool} \ll t_{\rm dyn}$). The magneto-hydrodynamical 
model with thermal conduction is similar to our model MHD2040IdealB7 since, due to  
his anisotropic character, heat transport are canceled across the magnetic field lines 
($t_{\rm heat} \ggg t_{\rm dyn}$). Note the boundary effect close to the apex 
along the $Oz$ direction as a result of the heat conduction along the 
direction of the ISM magnetic field lines (Fig.~\ref{fig:physics}g).  
Finally, our model with both processes has its dynamics governed by the cooling in the 
region of shocked ISM ($t_{\rm heat} \ggg t_{\rm dyn}$, $t_{\rm heat} \ggg t_{\rm cool}$, 
$t_{\rm cool} \ll t_{\rm dyn}$) and by the wind momentum in the region of shocked wind 
($t_{\rm heat} \ggg t_{\rm dyn}$, $t_{\rm heat} \ggg t_{\rm cool}$, $t_{\rm dyn} \ll t_{\rm cool}$).  
}


\subsubsection{Effects of the boundary conditions: stellar wind models}
\label{sect:physics33}

\textcolor{black}{
The shape of the bow shock generated around a runaway massive star in the warm 
phase of the ISM is a function of the respective strength of both the ISM ram 
pressure $\rho_{\rm ISM}v_{\star}^{2}$ and the stellar wind ram pressure 
$\rho_{\rm w}v_{\rm w}^{2}$, as seen in the frame of reference of the moving 
object~\citep[see explanations in][]{mohamed_aa_541_2012}. According to 
Eq.~(\ref{eq:wind}), $\rho_{w}=\dot{M} / 4\pi r^{2} v_{\rm w}$ which implies 
that $\rho_{\rm w}v_{\rm w}^{2} \propto \dot{M} v_{\rm w}$. In other words, in a 
given ambient medium and at a given peculiar velocity, the governing quantity 
in the shaping of such bow shock is $\dot{M}\rm v_{w}$ and its stand-off 
distance $R(0)$ goes as $\sqrt{ \dot{M} v_{w} }$, see Eq.~(\ref{eq:Ro}). 
Nevertheless, if the production of stellar evolution models depends on specific 
prescriptions relative to $\dot{M}$ that are consistently used through the 
calculations~\citep[in our case the recipe of][]{kudritzki_aa_219_1989}, 
the estimate of the wind velocity is posterior to the calculation of the 
stellar structure and it does not influence $\dot{M}$, $T_{\rm eff}$ or 
$L_{\star}$. 
}

\begin{figure}
	\begin{minipage}[b]{ 0.47\textwidth}
		\includegraphics[width=1.0\textwidth]{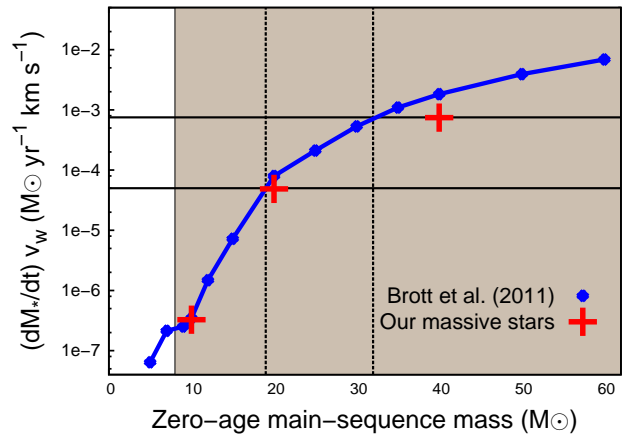}
	\end{minipage} \\ 	
	\caption{   
	\textcolor{black}{
	Comparison of the quantity $\dot{M}\rm v_{w}$ between our weak-winded 
	stars and the non-rotating Galactic models of~\citet{brott_aa_530_2011a}. 
	The grey zone of the plot corresponds to the mass regime of massive stars ($M_{\star}\ge8\,\rm M_{\odot}$). 
	Solid and dotted lines are lines of constant $\dot{M} v_{w}$ and $M_{\star}$, respectively.
	}
		 }
	\label{fig:brot}  
\end{figure}

\textcolor{black}{
The manner to calculate $\rm v_{w}$ is not unique~\citep{castor_apj_195_1975,kudritzki_aa_219_1989,kudritzki_aa_38_2000, 
eldridge_mnras_367_2006} and it can also be assumed to characteristic 
values for the concerned stars~\citep{comeron_aa_338_1998,vanmarle_aa_561_2014,vanmarle_2015,
acreman_mnras_456_2016}. In our study, the wind velocities are in the lower 
limit of the range of validity for the main-sequence massive stars that we 
consider, nonetheless, they still remain within the order 
of magnitude of, e.g. late O stars~\citep{martins_aa_468_2007} or weak-winded 
stars~\citep{comeron_aa_338_1998}. Furthermore, the evolution of massive stars 
are governed by physical mechanisms strongly influencing their feedback such 
as the presence of low-mass companions~\citep{sana_sci_337_2012}, \textcolor{black}{which 
are neglected in our stellar evolution models. Produced before their zero-age 
main-sequence phase, e.g. by fragmentation of the accretion disk that surrounds 
massive protostars~\citep{2016arXiv160903402M}, those dwarf stars entirely 
modify the evolution of massive stars and consequently affect their wind 
properties~\citep{demink_aa_467_2007,demink_aa_497_2009,paxton_apjs_192_2011,marchant_aa_588_2016}}. 
}

\textcolor{black}{
Using wind velocities faster by a factor $\alpha$ would 
enlarge the bow shocks by a factor $\sqrt{\alpha}$ and, eventually, in the 
hydrodynamical case, favorise the growth of \textcolor{black}{instabilities} (cf.~Fig.~\ref{fig:physics}b). 
However, the results of our numerical study would be similar in the sense 
that the presence of the field essentially stabilises the nebulae and inhibits 
the effects of heat conduction (cf.~Fig.~\ref{fig:physics}a,h), 
reduces their size (Section~\ref{sect:titi}) and modifies, e.g.  
their infrared emission accordingly (see Section~\ref{sect:infrared}). 
In Fig.~\ref{fig:brot}, we compare our values of $\dot{M} v_{w}$ 
(Table~\ref{tab:wind_para}) with the non-rotating stellar evolutionary models 
published in~\citet{brott_aa_530_2011a}. We conclude that the 
bow shocks generated with our initially $10$, $20$ and $40\, \rm M_{\odot}$ 
weak-winded stellar models correspond to nebulae produced by 
initially $\approx10$, $\approx18$ and $\approx32\, \rm M_{\odot}$ standard massive stars 
at Galactic metallicity, respectively. Therefore, our models have full validity 
for this study of magnetized bow shock nebulae, albeit of lower zero-age 
main-sequence mass in the case of our heaviest runaway star. 
}


\subsection{Hydrodynamics versus magneto-hydrodynamics}
\label{sect:physics3}

\subsubsection{The effects of the magnetic pressure}
\label{sect:titi}

The ISM magnetic pressure, \textcolor{black}{proportional to $\vec{B}_{\rm ISM}^{2}$}, 
dynamically compresses the region of shocked ISM gas such that the density in the
post-shock region at the forward shock slightly increases. Similarly, the shape
of the bow shock's wings of shocked ISM are displaced sidewards compare to our
model with $B_{\rm ISM}=0\, \mu \rm G$ (Fig.~\ref{fig:physics}a,e). The size
of the layer of ISM gas diminishes along the direction of motion of the moving star and the position of
the termination shock \textcolor{black}{sets} at a distance from the star where the wind ram
pressure equals the ISM total pressure decreases as measured along the $Oz$
axis. The effects of the cooling is standard in the sense that it makes the
region of shocked ISM thinner and denser, i.e. the position of the forward shock
decreases, together with the bow shock volume. The effects of heat conduction
are canceled ($t_{\rm dyn} \ll t_{\rm heat}$) in the direction \textcolor{black}{perpendicular} to the
field lines, i.e. in the direction perpendicular to the streamline collinear 
to both the reverse shock and the contact discontinuity.

\subsubsection{Stagnation point morphology and discussion in the context of plasma physics studies}
\label{sect:}

The topology at the apex of our magneto-hydrodynamical bow shock
(Fig.~\ref{fig:physics}h) is different from the traditional single-front bow
shock morphology (Fig.~\ref{fig:physics}d). \textcolor{black}{This can be discussed at the light 
of plasma physics studies~\citep{sterck_phpl_1998,sterck_aa_343_1999}. These 
works explore the formation of exotic shocks and discontinuities that affect 
the particularly dimpled apex of bow shocks generated by field-aligned flows around a conducting 
cylinder~\citep{sterck_phpl_1998}. They extended this result to bow shocks 
produced around a conducting sphere and showed} that the inflow 
parameter space leading to such structures is similar to plasma $\beta$ and 
\textcolor{black}{Alfv\' enic} Mach number values allowing the formation of so-called switch-on 
shocks~\citep{sterck_aa_343_1999}.

\begin{figure}
	\begin{minipage}[b]{ 0.48\textwidth}
		\includegraphics[width=1.0\textwidth]{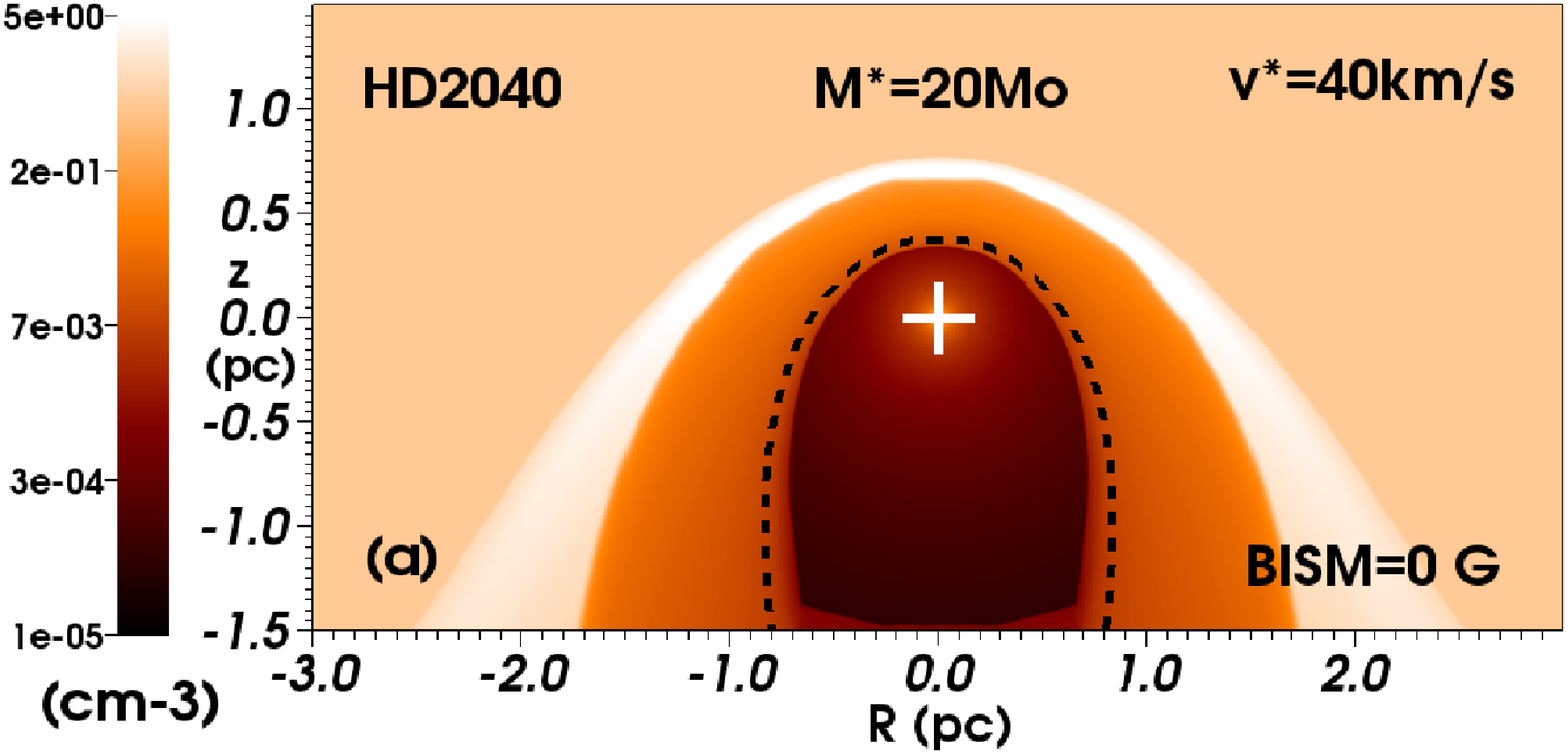}
	\end{minipage} \\
	\begin{minipage}[b]{ 0.48\textwidth}
		\includegraphics[width=1.0\textwidth]{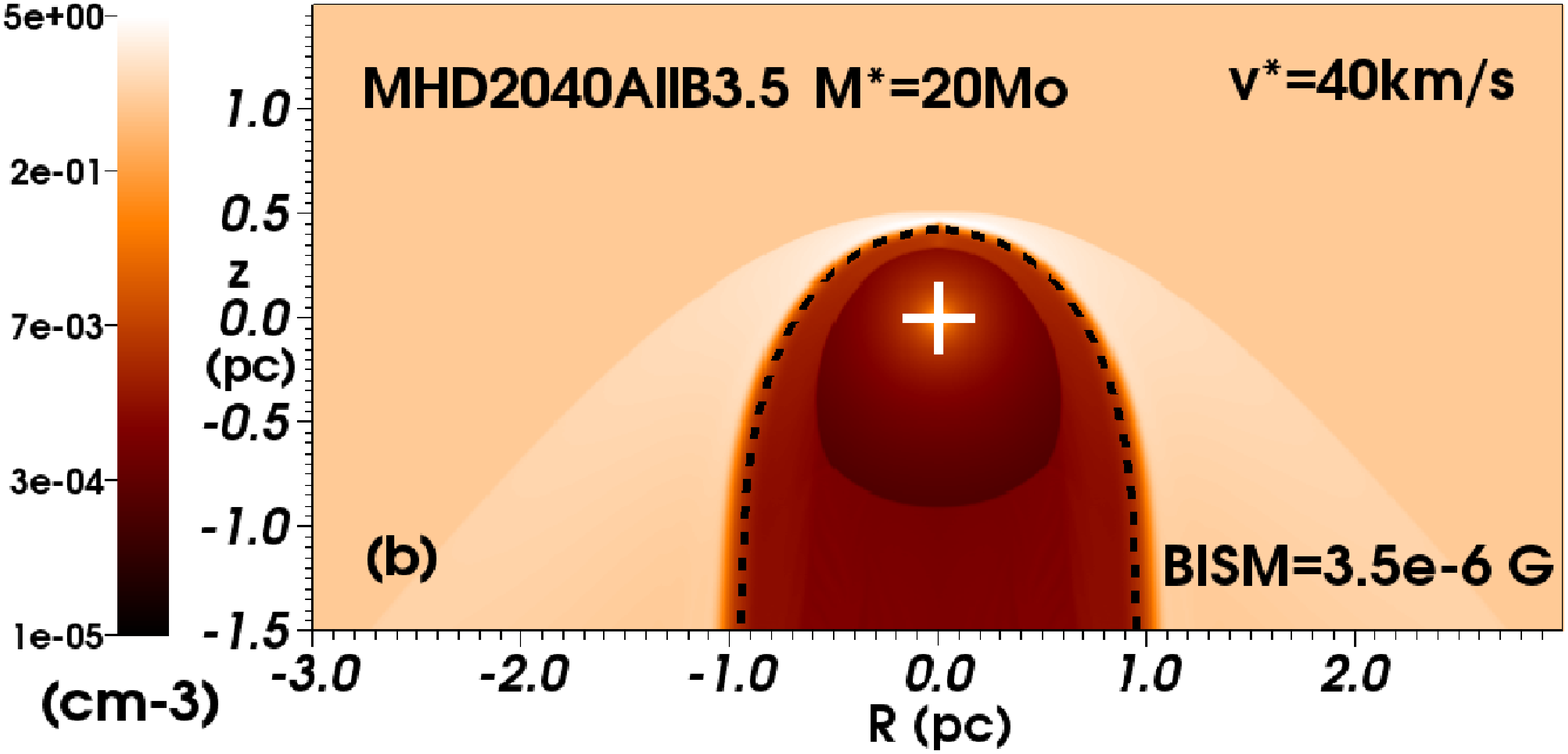}
	\end{minipage} \\
	\begin{minipage}[b]{ 0.48\textwidth}
		\includegraphics[width=1.0\textwidth]{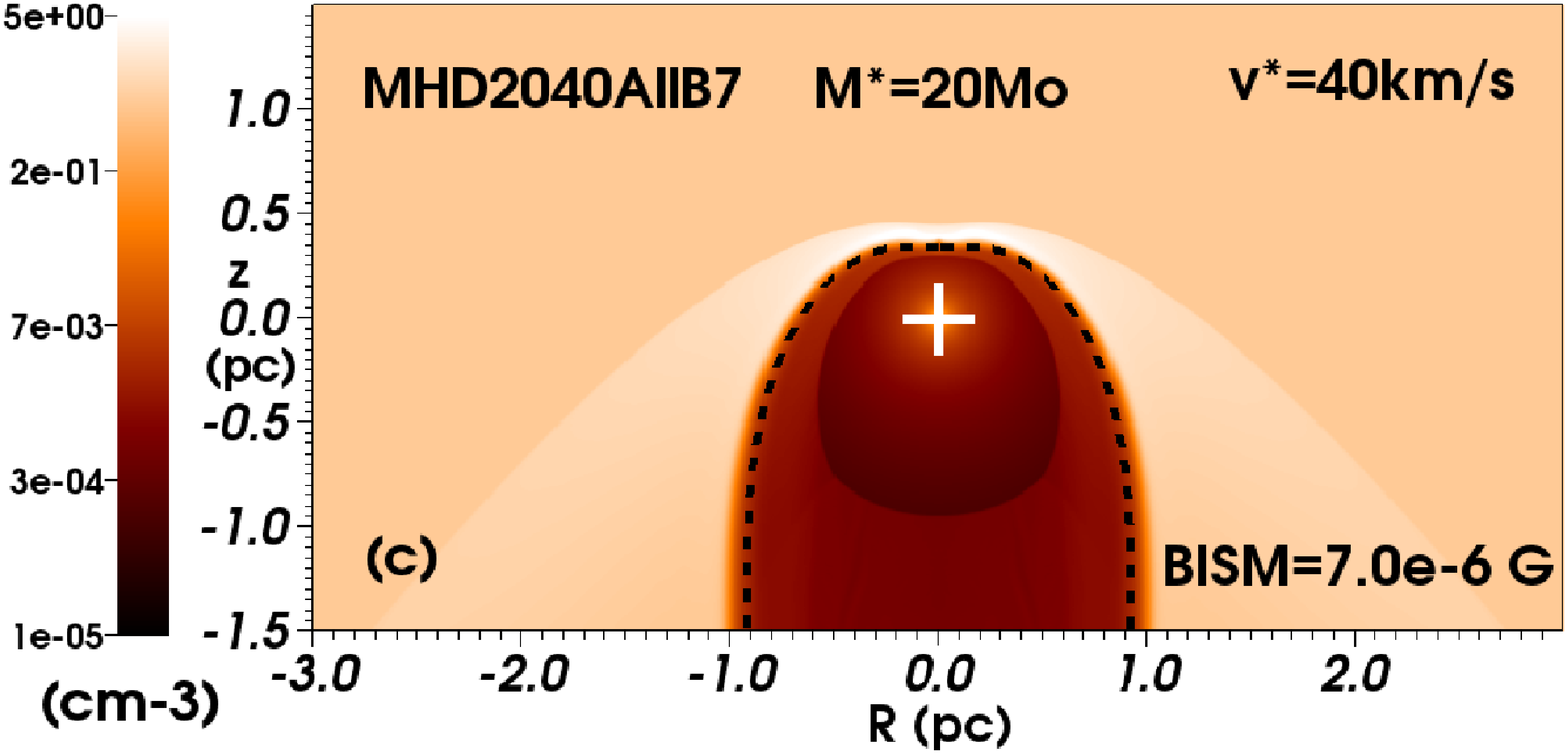}
	\end{minipage} \\  	
	\caption{
		\textcolor{black}{Models of} stellar wind bow shocks of our initially $20\, \rm
M_{\odot}$ star moving with velocity $v_{\star}=40\, \rm km\, \rm s^{-1}$ represented as a function of its ISM 
magnetic field strength, with $B_{\rm ISM}=0$ (a), $3.5$ (b) and $7.0\, \rm \mu G$ (c).  
		 }
	\label{fig:bfieldstrength}  
\end{figure}

Switch-on shocks are allowed when plasma $\beta$ of the inflowing material,
i.e. the ratio of the gas and magnetic pressures, which read,
\begin{equation}
	\beta = \frac{8\pi n k_{\rm B} T }{\bmath{B}_{\rm ISM} \cdot \bmath{B}_{\rm ISM} },
\label{eq:beta}
\end{equation}
\textcolor{black}{and its \textcolor{black}{Alfv\' enic} Mach number,
\begin{equation}
	M_{\rm A} = \frac{ v  }{ v_{\rm A} },
\label{eq:Ma}
\end{equation}
where,
\begin{equation}
	v_{\rm A} = \frac{ |\bmath{B}_{\rm ISM}|  }{ \sqrt{ 4 \pi n m_{\rm H} } },
\label{eq:va}
\end{equation}
is the Alfv\' enic velocity, satisfy some particular conditions. 
Note that in Eq.~(\ref{eq:Ma}) the velocities are taken along the shock normal. }
On the one hand, the plasma beta must be such that,
\begin{equation}
	\beta < \frac{2}{\gamma},
\label{eq:cond1}
\end{equation}
whereas on the other hand, the \textcolor{black}{Alfv\' enic} Mach number verifies the following order 
relation,
\begin{equation}
	1 < M_{\rm A} < \sqrt{ \frac{\gamma(1-\beta)+1 }{\gamma-1} },
\label{eq:cond2}
\end{equation}
where $\gamma$ is the adiabatic index, \textcolor{black}{see Eq.~1 in~\citet{pogolerov_aa_354_2000}}. Numbers from our simulations indicate
that the ISM thermal pressure $n_{\rm ISM} k_{\rm B} T_{\rm ISM}\approx 8.62
\times 10^{-13}\, \rm dyne\, \rm s^{-2}$, \textcolor{black}{therefore we find 
$\beta > 2/\gamma \approx 1.2$ for $B_{\rm ISM} \le 3.5\, \mu \rm G$ 
(see bow shocks with normal morphologies in Fig.~\ref{fig:bfieldstrength}a,b)
but $\beta \approx
\textcolor{black}{0.44}< 2/\gamma \approx 1.2$ for $B_{\rm ISM}=7\, \mu \rm G$ 
(see dimpled bow shock in Fig.~\ref{fig:bfieldstrength}c).
The \textcolor{black}{Alfv\' enic} Mach number $M_{\rm A} =
v/v_{\rm A} \approx 40.0\, \rm  km\, \rm s^{-1}/17.2\, \rm  km\, \rm
s^{-1} \approx 2.33$ which is outside the range 
$1<M_{\rm A}<((\gamma(1-\beta)+1)/(\gamma-1))^{1/2} \approx \textcolor{black}{1.70}$. 
Similarly, the model with $v_{\star}=70\, \rm  km\, \rm s^{-1}$ is such that 
$M_{\rm A}>((\gamma(1-\beta)+1)/(\gamma-1))^{1/2}$ whereas our slower model with 
$v_{\star}=20\, \rm  km\, \rm s^{-1}$ gives 
$1<M_{\rm A}\approx 1.16<((\gamma(1-\beta)+1)/(\gamma-1))^{1/2}$, which 
is inside the range in Eq.~(\ref{eq:cond2}).  
We conclude that the upstream ISM conditions in our magneto-hydrodynamical 
simulations producing dimpled bow shocks have values consistent with the 
existence of switch-on shocks}, see also sketch of the ($\beta$,$M_{\rm A}$) 
plane in Fig.~3 of~\citet{sterck_aa_343_1999}.

\textcolor{black}{However, we can not affirm that the dimpled apex topology of our 
magneto-hydrodynamical bow shocks models is of origin similar to the ones 
in~\citet{sterck_phpl_1998,sterck_aa_343_1999}. Only their particular 
concave-inward form that differs from the classical shape of hydrodynamical 
bow shocks (Fig.~\ref{fig:physics}e) authorizes a comparison between the 
two studies. Nevertheless, we notice that our bow shocks are also generated 
in an ambient medium in which the plasma beta and the \textcolor{black}{Alfv\' enic} Mach number 
have parameter values consistent with the formation of switch-on shocks, 
which has been showed to be similar to the parameter values producing dimpled 
bow shocks around charged obstacles~\citep[see][and references therein]
{sterck_phpl_1998,sterck_aa_343_1999}.} Additional investigations, left for future studies, 
are required to assess the question of \textcolor{black}{the exact nature the 
various discontinuities affecting magneto-hydrodynamical bow shocks of OB stars. }

\begin{figure*}
	\begin{minipage}[b]{ 0.65\textwidth}
		\includegraphics[width=1.0\textwidth]{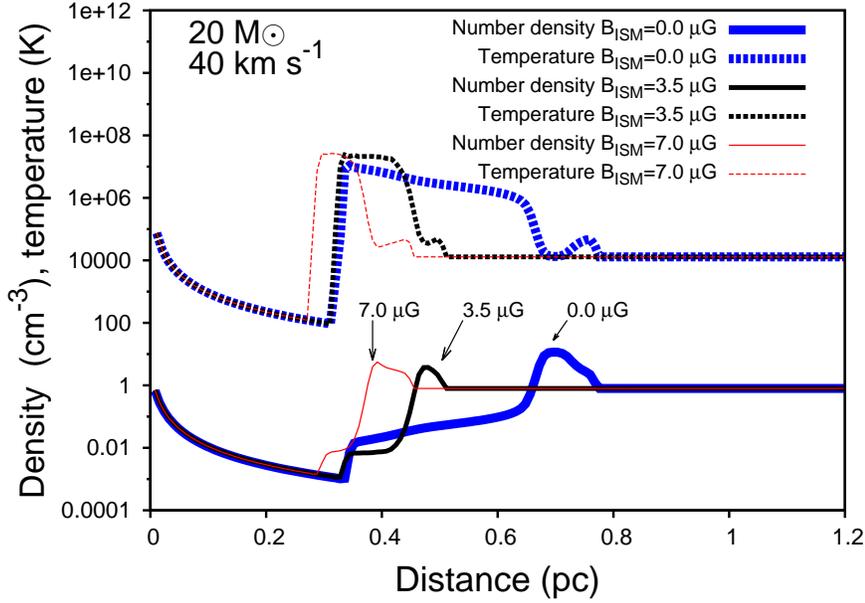}
	\end{minipage} \\  	
	\caption{
	         Number density (solid lines, in $\rm cm^{-3}$) and temperature 
	         (dotted lines, in $\rm K$) profiles in our hydrodynamical (thick blue lines) 
	         and magneto-hydrodynamical (thin red lines) 
	         bow shocks models of an initial $20\, \rm M_{\odot}$ star moving with velocity 
	         $v_{\star}=40\, \mathrm{km}\, \mathrm{s}^{-1}$. The profiles are measured along the 
	         symmetry axis Oz. 
		 }
	\label{fig:cuts}  
\end{figure*}

\subsubsection{Effects of the magnetic field strength}
\label{sect:}

Fig.~\ref{fig:bfieldstrength} is similar to Fig.~\ref{fig:physics} and displays
the effects of the ISM magnetic field strength $B_{\rm ISM}=0$ (a), $3.5$ (b)
and $7.0\, \rm \mu G$ (c) on the shape of the bow shocks produced by our
initially $20\, \rm M_{\odot}$ star moving with velocity $v_{\star}=40\, \rm
km\, \rm s^{-1}$. In Fig.~\ref{fig:cuts} we show density (solid lines) and
temperature (dotted lines) profiles from our hydrodynamical simulation (thick
blue lines) and magneto-hydrodynamical model (thin red lines) of the bow shocks
in Fig.~\ref{fig:bfieldstrength}. The profiles are taken along the symmetry axis
of the computational domain. The global structure of the bow shock is similar
for both simulations, i.e. it consists of a hot bubble ($T\approx\, 10^{7}\, \rm
K$) surrounded by a shell of dense ($n\approx\, 10\, \rm cm^{-3}$) shocked ISM
gas. The profiles in Fig.~\ref{fig:cuts} highlights the progressive compression 
of the bow shocks by the the ISM total pressure which magnetic component increases 
as $B_{\rm ISM}$ is larger. \textcolor{black}{Several mechanisms at work might be 
responsible for such discrepancy:
\begin{enumerate}
\item The magnetic pressure in the ISM. If one neglects the thermal pressures $nk_{\rm B}T$ 
      in both the supersonic stellar wind and the inflowing ISM, and omits the magnetic 
      pressure $\propto B_{\star}^{2} \propto r^{-4}$ at the wind termination shock,  
      then the pressure balance between ISM and stellar wind gas reads, 
\begin{equation}
	\rho_{\rm w} v_{\mathrm{w}}^{2} = \rho_{\rm ISM} v_{\star}^{2} + \frac{B_{\rm ISM}^{2}}{8\pi},
\label{eq:pressures}
\end{equation}
from which one can derive the bow shock stand-off distance in a planar-aligned field bow shock,
\begin{equation}
	R(0) = \Bigg( \frac{2\dot{M}v_{\mathrm{w}}}{ B_{\rm ISM}^{2} + 8\pi \rho_{\mathrm{ISM}}v_{\star}^{2} }\Bigg)^{1/2},
	\label{eq:Romag}
\end{equation}
that is slightly smaller from the one derived in a purely hydrodynamical context~\citep{wilkin_459_apj_1996}.  
\item The cooling by optically-thin radiative processes. Changes in the density at the 
      post-shock region at the forward shock influence the temperature in the shocked ISM gas, 
      which in their turn modify the cooling rate of the gas, itself affecting its thermal 
      pressure. This results in an increase of the density of the shell of ISM gas but also a decrease 
      of the temperature in the hot region of shocked stellar wind material that shrinks in order 
      to maintain its total pressure equal to $\rho_{\rm ISM} v_{\star}^{2}+B_{\rm ISM}^{2}/8\pi$. 
\item The magnetic field field lines inside the bow shock. The compression of the layer of shocked ISM gas 
      modifies the arrangement of the field lines in the post-shock region at the forward shock. 
      Thus, the term $B_{\rm ISM}^{2}/8\pi$ corresponding to the magnetic pressure increases and modifies 
      the effects of radiative cooling in the simulations (see above). 
\item Symmetry effects. The solution may also be affected by the intrinsic two-dimensional nature of 
      our simulations, which may develop numerical artifices close to the symmetry axis. In the case 
      of magneto-hydrodynamical simulations of objects moving supersonically along the direction 
      of the ISM magnetic field, such effects are more complex than a simple accumulation of material 
      at the apex of the nebula, but might present artificial shocks, see also Section~\ref{sect:physics4}.       
\end{enumerate}
}

\textcolor{black}{
Appreciating in detail which of the above cited processes dominates the solution would 
require three-dimensional numerical simulations which are beyond the scope of this work. 
Moreover, establishing an analytic theory of the position of the contact 
discontinuity of a magnetized bow shock is a non-trivial task since the 
thin-shell limit~\citep{wilkin_459_apj_1996} is not applicable. 
}
In particular, the hot bubble loses about three quarter of its size along
the $Oz$ direction when the ISM magnetic field strength increases up to $B_{\rm
ISM}=7\, \mu \rm G$ (Fig.~\ref{fig:bfieldstrength}a,c). This modifies 
the volume of hot shocked ISM gas advected thanks to heat transfers towards the
inner part of the bow shock of our model HD2040All, reducing it to a narrow
layer made of shocked wind material since anisotropic thermal conduction forbids
the penetration of ISM gas in the hot region. The effects of the ISM
magnetisation on our optical and infrared bow shocks' emission properties are 
further discussed in Section~\ref{sect:discussion}.


All of our magneto-hydrodynamical simulations have a stable density field
(Fig.~\ref{fig:physics}e,f,g,h). \textcolor{black}{The simulations} with cooling but
without heat transfer (Fig.~\ref{fig:physics}b) show that the presence of the
magnetic field inhibits the growth of Kelvin-Helmholtz instabilities
(Fig.~\ref{fig:physics}f) that typically develops within the contact
discontinuity of the bow shocks because they are the interface of two plasma
moving in opposite
directions~\citep[][Paper~I]{comeron_aa_338_1998,vanmarle_aa_469_2007}. 
\textcolor{black}{The solution does} not change performing the simulation MHD2040AllB7 at double and
quadruple spatial resolution, and conclude that our results are consistent with both
numerical studies devoted to the growth and saturation of these instabilities in
the presence of a planar magnetic field~\citep[see,
e.g.][]{keppens_jplph_61_1999} and with results obtained for slow-winded, cool
runaway stars moving in a planar-aligned magnetic
field~\citep{vanmarle_aa_561_2014}. Note that detailed numerical studies
demonstrating the suppression of shear instabilities by the presence of a
background magnetic field also exist in the context of jets from
protostars~\citep{viallet_473_aa_2007}.

\subsection{Effects of the star's bulk motion}
\label{sect:physics3}


\begin{figure}
	\begin{minipage}[b]{ 0.48\textwidth}
		\includegraphics[width=1.0\textwidth]{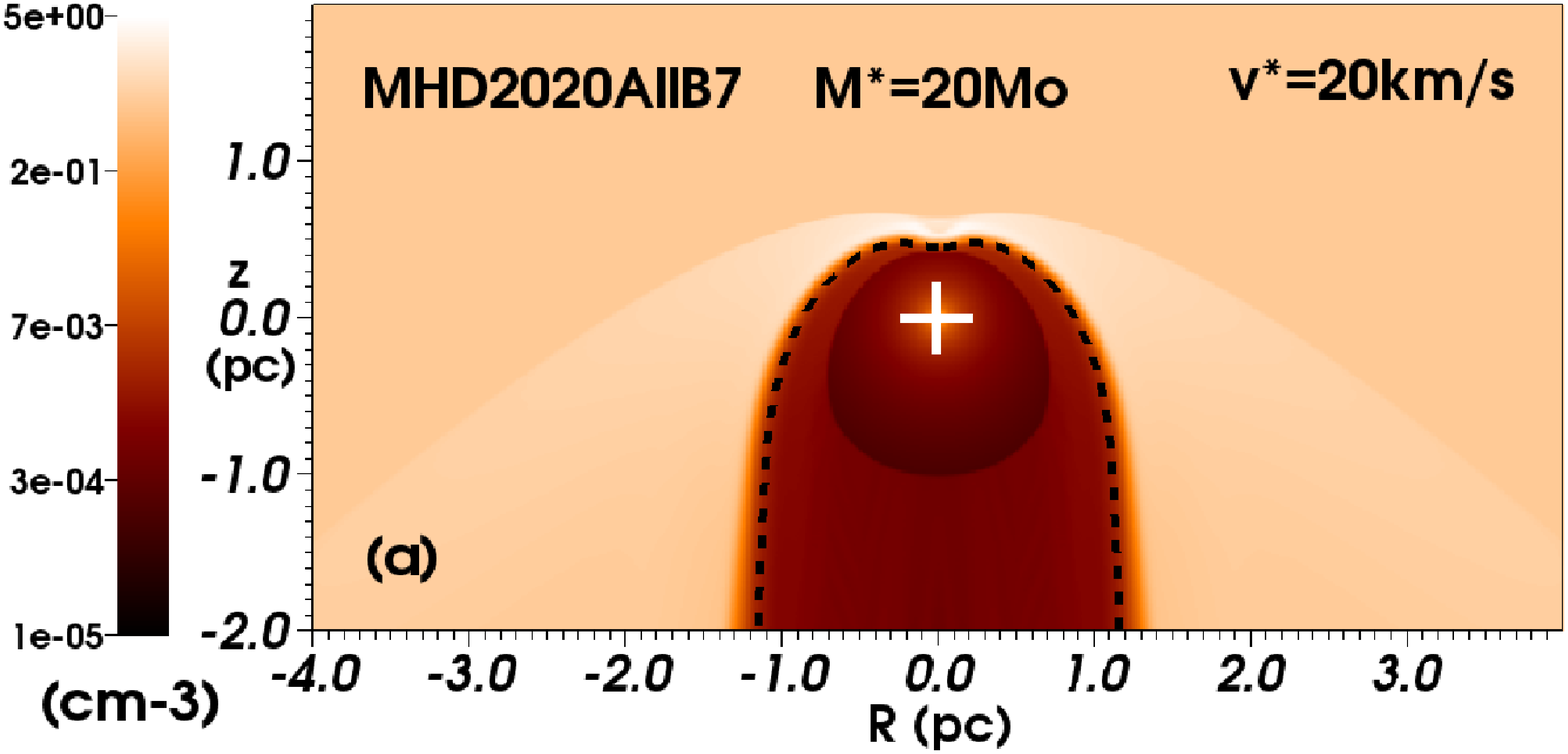}
	\end{minipage} \\
	\begin{minipage}[b]{ 0.48\textwidth}
		\includegraphics[width=1.0\textwidth]{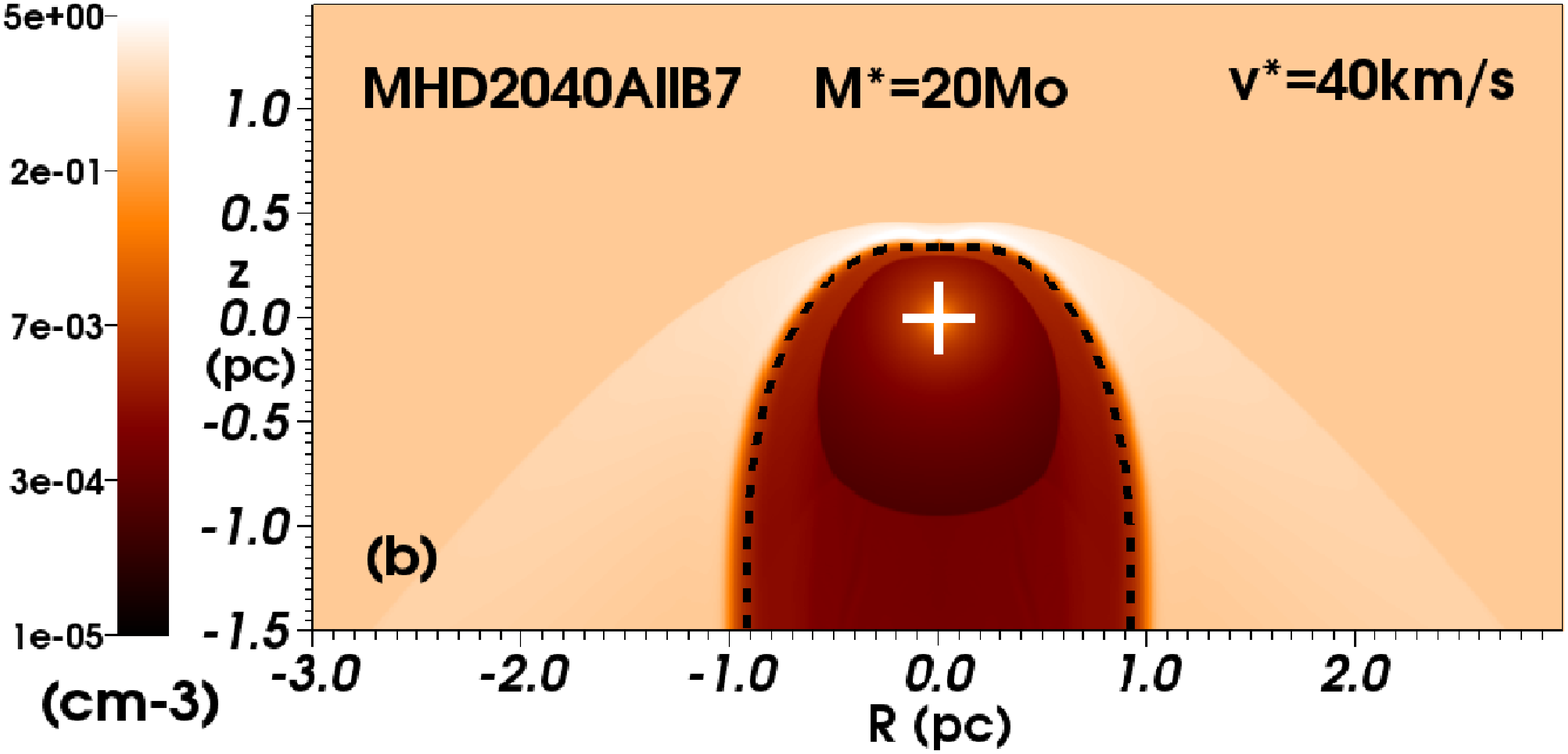}
	\end{minipage} \\
	\begin{minipage}[b]{ 0.48\textwidth}
		\includegraphics[width=1.0\textwidth]{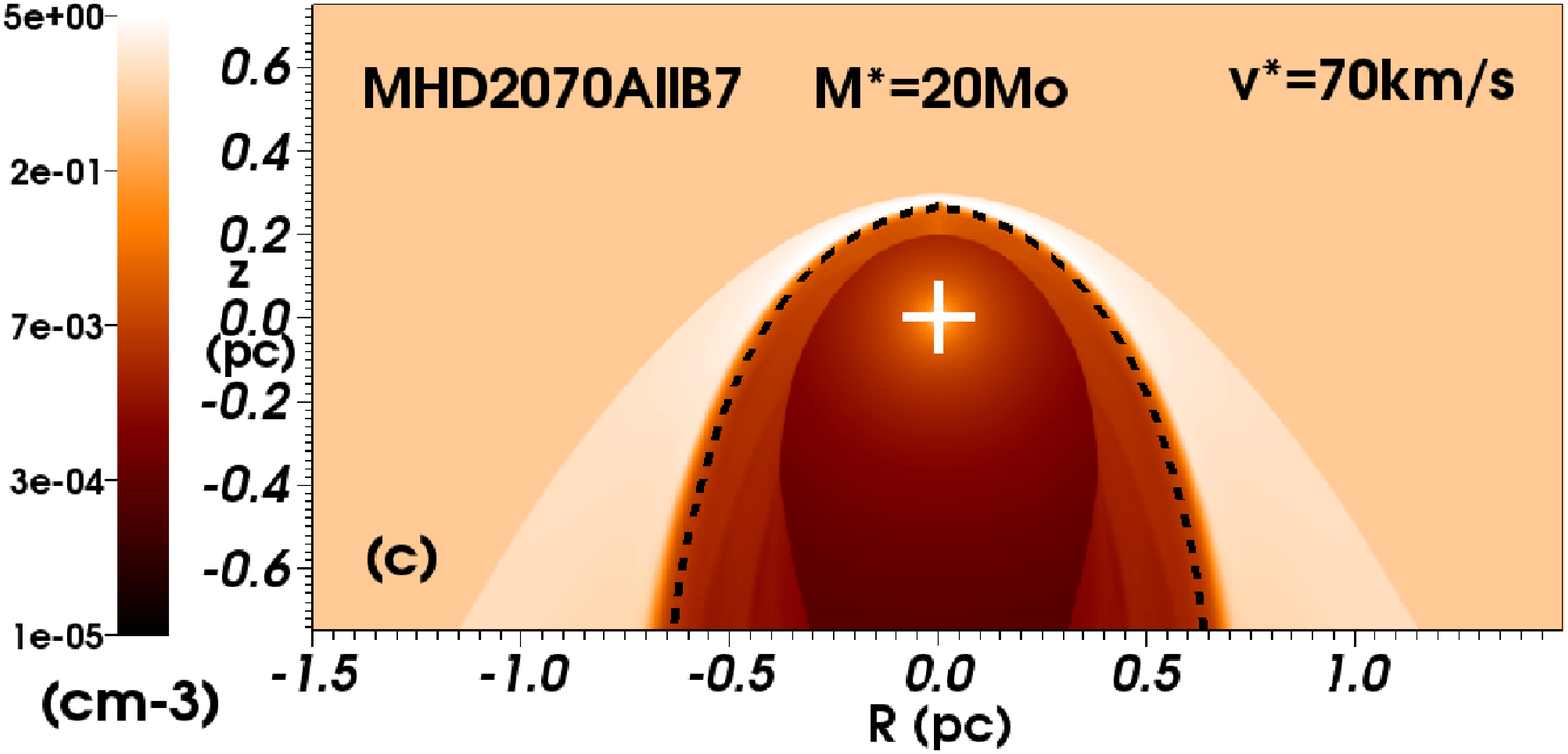}
	\end{minipage} \\  	
	\caption{
		Grid of stellar wind bow shocks from our initially $20\, \rm
M_{\odot}$ star represented as a function of its space velocity with respect to
the ISM, with velocity $v_{\star}=20$ (a), $40$ (b) and $70\, \mathrm{km}\,
\mathrm{s}^{-1}$ (c). The nomenclature of the models follows Table~\ref{tab:ms}.
The bow shocks are shown at about $5\, \rm Myr$ after the beginning of the
main-sequence phase of the central star's evolution. All our MHD models assume a
strength of the ISM magnetic field $B_{\rm ISM}=7\, \mu \rm G$. The gas number density
is shown with a density range from $10^{-5}$ to $5.0\, \mathrm{cm}^{-3}$ in the
logarithmic scale. The crosses mark the position of the star. The solid
black contour traces the boundary between wind and ISM material
$Q_{1}(\bmath{r})=1/2$. The
$R$-axis represents the radial direction and the $z$-axis the direction of
stellar motion (in $\mathrm{pc}$). Only part of the computational domain is
shown in the figures.  
		 }
	\label{fig:motion}  
\end{figure}

Fig.~\ref{fig:motion} is similar to Fig.~\ref{fig:bfieldstrength} and plots a grid of
density field of our initially $20\, \rm M_{\odot}$ star moving with velocity
$v_{\star}=20$ (a), $40$ (b), $70\, \mathrm{km}\, \mathrm{s}^{-1}$ (c). The
scaling effect of the bulk motion of the star on the bow shocks morphology is \textcolor{black}{similar
to} our hydrodynamical study (Paper~I). At a given strength of the ISM magnetic
field, the compression of the forward shock increases as the \textcolor{black}{spatial}
motion of the star increases because the ambient medium ram pressure is larger.
The relative thickness of the layers of ISM and wind behaves similarly as
described in Paper~I. Our model with $v_{\star}=20\, \rm km\, \rm s^{-1}$ has a
layer of shocked ISM larger than the layer of shocked wind because the
relatively small ISM ram pressure induces a weak forward shock
(Fig.~\ref{fig:motion}a). The shell of shocked ISM is thinner in our
simulation with $v_{\star}=70\, \rm km\, \rm s^{-1}$ because the strong forward
shock has a high post-shock temperature which allows an efficient cooling of the
plasma (Fig.~\ref{fig:motion}c).

\textcolor{black}{
The density field in our models with ISM inflow 
velocity similar to the \textcolor{black}{Alfv\' enic} speed ($v_{\star}=20 
~\simeq v_{\rm A}\approx 17.2\, \rm km\, \rm s^{-1}$) has the dimpled shape 
of its apex of the bow shock (Fig.~\ref{fig:motion}a). 
The model with $v_{\star}=70\, \rm km\, \rm s^{-1}$ has inflow ISM velocity 
larger than the \textcolor{black}{Alfv\' enic} speed and presents the classical single-front 
morphology (Fig.~\ref{fig:motion}c) typically produced by stellar wind bow 
shocks~\citep{brighenti_mnras_273_1995,brighenti_mnras_277_1995,comeron_aa_338_1998,meyer_mnras_459_2016}. 
A similar effect of the \textcolor{black}{Alfv\' enic} speed is discussed in, e.g. fig.4 of~\citet{sterck_aa_343_1999}. 
Again, exploring in detail whether the formation mechanisms of our dimpled bow shocks 
is identical to the ones obtained in calculations of 
bow shock flow over a conducting sphere is far beyond the scope of this work. }
Note the absence of instabilities in our magneto-hydrodynamical bow shocks simulations 
compare to our hydrodynamical models.

\subsection{Model limitation}
\label{sect:physics4}

\textcolor{black}{
First and above, our models suffer from their two-dimensional nature. If 
carrying out axi-symmetric models is advantageous in order to decrease the 
amount of computational ressources necessary to perform the simulations, however, 
it forbids the bow shocks from generating a structure which apex would be 
totally unaffected by symmetry-axis related phenomenons, common in this case of 
calculations~\citep{meyer_mnras_459_2016}. This prevents our simulations from being able 
to assess, e.g. the question of the relation between the seeds of the non-linear 
thin-shell instability at the tip of the structure and the growth of 
Kelvin-Helmholtz instabilities occuring later in the wings of the bow shocks. 
Only full 3D models of the same bow shocks could fix such problems and 
allow us to further discuss in detail the instability of bow shocks from OB stars.
We refer the reader to~\citet{vanmarle_2015} for a discussion of the dimension-dependence 
of numerical solutions concerning the interaction of magnetic fields with hydrodynamical 
instabilities. 
}

\textcolor{black}{
In particular, the selection of admissible shocks which is generally treated 
using artificial viscosity in purely hydrodynamical simulations is more 
complex in our magneto-hydrodynamical context~\citep[see 
discussion in][]{ pogolerov_aa_354_2000}. This can lead to additional 
fragilities of the solution, especially close to the symmetry axis of our 
cylindrically-symmetric models. Although the stability of these kinds of shocks 
is still under debate~\citep{desterck_aipc_537_2000,desterck_2001}, we will try 
to address these issues in future three-dimensional simulations. 
Moreover, such models would (i) allow us to explore the effects of a non-aligned 
ISM magnetic field on the morphology of the bow shocks and (ii) will make 
subsequent radiative transfer calculations meaningful, e.g. considering 
polarization maps using full anisotropic scattering of the photons on the dust 
particles in the bow shocks. The space of parameters investigated in our study is 
also limited, especially in terms of the explored range of space velocity 
$v_{\star}$ and ISM density $n_{\rm ISM}$ and will be extended in a follow-up 
project. Finally, other physical processes such as the presence of a 
surrounding $\HII$ region or the intrinsic viscous, granulous and turbulent 
character of the ISM are also neglected and deserve additional investigations. 
}


\begin{table*}
	\centering
	\caption{
	Maximum optical surface brightness of our magneto-hydrodynamical
simulations with $B_{\rm ISM}=7\, \mu G$. The second and third columns are the
quantities $\Sigma_{[\rm H\alpha]}^{\rm max}$ and $\Sigma_{[\rm O{\sc III}]}^{\rm
max}$ representing the maximum projected emission in [O{\sc iii}] $\lambda \,
5007$ and H$\alpha$ (in $\rm erg\, \rm cm^{-2}\, \rm s^{-1}\, \rm arcsec^{-2}$),
respectively. Models consisting of the hydrodynamical counterpart of our bow shocks 
models have their labels in italic in the first column (see description in Table~1 
in Paper~I). The surface brightnesses are measured along the direction of motion of 
the star at the apex of our bow shocks, close to the symmetry axis $Oz$. 
	}
	\begin{tabular}{cccc}
	\hline
	${\rm {Model}}$ &   $\Sigma_{[\rm H\alpha]}^{\rm max}\, (\rm erg\, \rm cm^{-2}\, \rm s^{-1}\, \rm arcsec^{-2})$                              
			&   $\Sigma_{[\rm O{\sc III}]}^{\rm max}\, (\rm erg\, \rm cm^{-2}\, \rm s^{-1}\, \rm arcsec^{-2})$    
			&   $\Sigma_{[\rm O{\sc III}]}^{\rm max}/\Sigma_{[\rm H\alpha]}^{\rm max}$
			\\ \hline   
	MHD1040AllB7    &  $2.5\times 10^{-19}$  &  $7.0\times 10^{-18}$ & $28.0$   \\ 
	{\it MS1040 }   &  $1.0\times 10^{-18}$  &  $2.5\times 10^{-17}$ & $25.0$   \\     		
	MHD2020AllB7    &  $1.7\times 10^{-17}$  &  $6.8\times 10^{-17}$ & $4.0$   \\   
	{\it MS2020 }   &  $6.0\times 10^{-17}$  &  $7.2\times 10^{-17}$ & $1.2$   \\     		
	MHD2040AllB7    &  $2.9\times 10^{-17}$  &  $1.6\times 10^{-16}$ & $5.5$   \\    
	MHD2040AllB3.5  &  $1.0\times 10^{-16}$  &  $3.2\times 10^{-16}$ & $3.2$   \\    	
	{\it MS2040 }   &  $1.2\times 10^{-16}$  &  $2.5\times 10^{-16}$ & $2.1$   \\    
	MHD2070AllB7    &  $8.0\times 10^{-18}$  &  $2.0\times 10^{-16}$ & $25.0$   \\  
	{\it MS2070 }   &  $1.5\times 10^{-16}$  &  $8.5\times 10^{-16}$ & $5.7$   \\  	
	MHD4070AllB7    &  $1.2\times 10^{-17}$  &  $5.5\times 10^{-16}$ & $45.8$   \\   
	{\it MS4070 }   &  $4.0\times 10^{-16}$  &  $1.0\times 10^{-15}$ & $2.5$   \\     			
	\hline    
	\end{tabular}
\label{tab:sigma}
\end{table*}

\section{Comparison with observations and implications of our results}
\label{sect:discussion}

\begin{figure}
	\begin{minipage}[b]{ 0.48\textwidth}
		\includegraphics[width=1.0\textwidth]{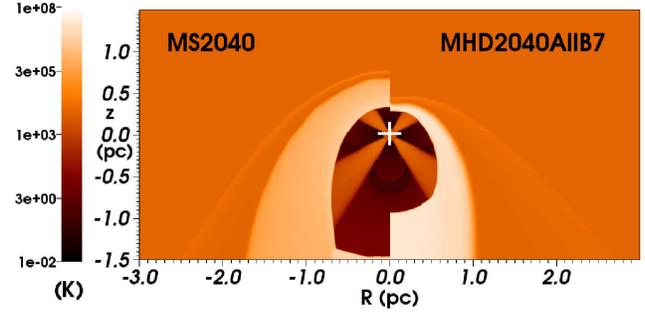}
	\end{minipage} \\  	
	\caption{
	         \textcolor{black}{Temperature field (in $\rm K$) in the models MS2040 and MHD2040AllB7. 
	         The cross-like structure in the  central region of freely-expanding 
	         stellar wind is a boundary effects caused by the pressure
	         }}
	\label{fig:temperature}  
\end{figure}

In this section, we extract observables from our simulations, compare them to observations 
and discuss their astrophysical implications. 
We first recall the used post-processing methods and then compare 
the emission by optically-thin radiation 
of our magneto-hydrodynamical bow shocks with hydrodynamical
models of the same star moving at the same \textcolor{black}{velocity}.
\textcolor{black}{Given the high temperature generated by collisional heating (Fig.~\ref{fig:temperature}), 
we particularly focus on the H$\alpha$ and [O{\sc iii}] $\lambda \, 5007$ optical emission.   
Moreover, stellar wind bow shocks from massive stars have been first detected at these 
spectral lines and hence constitute a natural observable. } 
\textcolor{black}{
We complete our analysis with infrared radiative transfer calculations and 
comment the observability of our bow shock nebulae. Last, we discuss our findings in the
context of the runaway massive star $\zeta$ Ophiuchi.   
}

\subsection{Post-processing methods}
\label{sect:maps}

\begin{figure}
	\begin{minipage}[b]{ 0.48\textwidth}
		\includegraphics[width=1.0\textwidth]{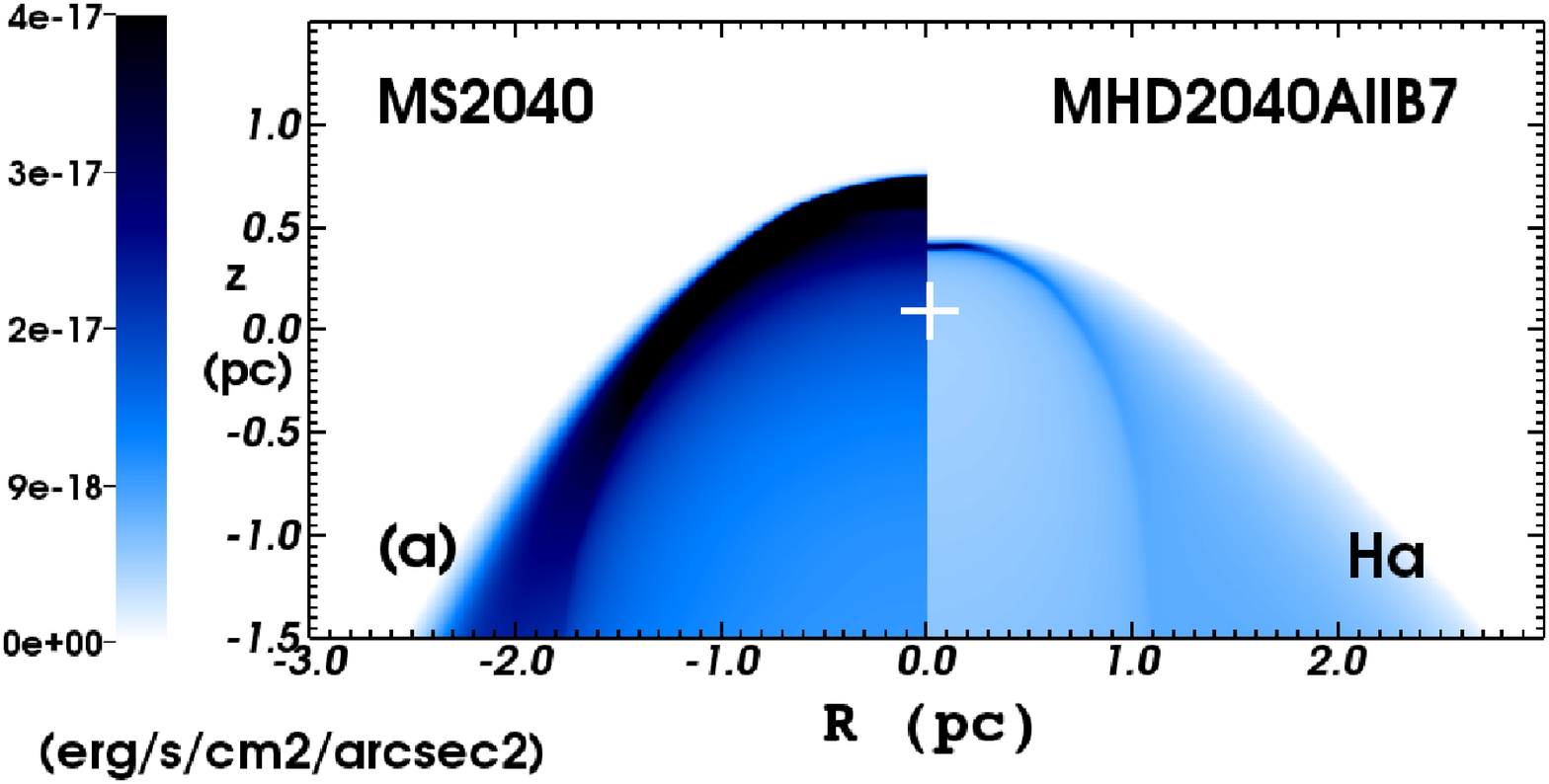}
	\end{minipage} \\
	\begin{minipage}[b]{ 0.48\textwidth}
		\includegraphics[width=1.0\textwidth]{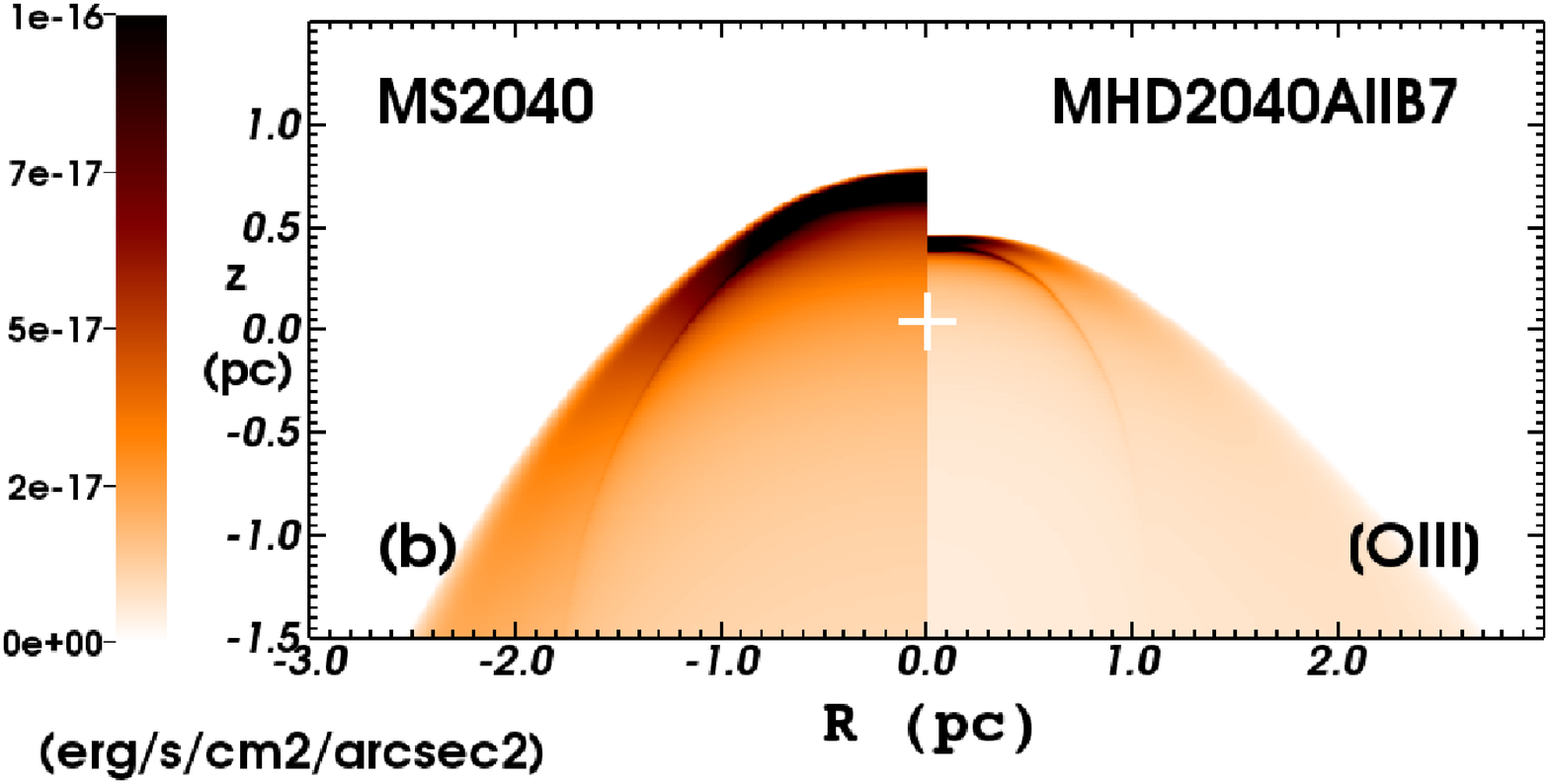}
	\end{minipage}	\\ 	
	\caption{
	         Surface brightness maps of H$\alpha$ (a), [O{\sc
iii}] (b) surface brightness (in $\mathrm{erg}\, \mathrm{s}^{-1}\,
\mathrm{cm}^{-2}\, \mathrm{arcsec}^{-2}$), respectively, of our bow shock model 
generated by our initially $20\, \rm M_{\odot}$ star moving with velocity
$v_{\star}=40\, \mathrm{km}\, \mathrm{s}^{-1}$. Quantities are calculated excluding the
undisturbed ISM and plotted in the linear scale. The left-hand part of the panels 
refers to the hydrodynamical model MS2040, the right-hand part to the 
magneto-hydrodynamical model \textcolor{black}{MHD2040AllB7}. The crosses mark the
position of the star. \textcolor{black}{For the sake of comparison, these optical maps 
are presented as in Paper~I}. 
		 }
	\label{fig:maps}  
\end{figure}

Fig.~\ref{fig:maps} plots the projected optical emission of our model of an initially
$20\, \rm M_{\odot}$ star moving at $40\, \rm km\, \rm s^{-1}$ in H$\alpha$ (a)
and [O{\sc iii}] $\lambda \, 5007$ (b) in $\mathrm{erg}\, \mathrm{s}^{-1}\, \mathrm{cm}^{-2}\,
\mathrm{arcsec}^{-2}$. Left-hand part of the panels correspond to the star
moving into an ISM with no background magnetic field (hydrodynamical model MS2040, Paper~I) whereas right-hand
parts correspond to $B_{\rm ISM}=7\, \rm \mu G$ (magneto-hydrodynamical model MHD2040AllB7). We take into
account the rotational symmetry about $R=0$ of our models and integrate the
emission rate assuming that our bow shocks lay in the plane of the sky, i.e. 
the star moves perpendicular to the observer’s line-of-sight. The
spectral lines emission coefficients are evaluated using the 
prescriptions for optical spectral line emission 
from~\citet{dopita_aa_29_1973} and~\citet{osterbrock_1989}, which read,
\begin{equation}
	j_{\rm [H\alpha]}(T) \approx 1.21\times 10^{-22} T^{-0.9} n_{\rm p}^{2}\, \rm erg\, s^{-1}\, cm^{-3}\, sr^{-1},
\label{eq:Ha}
\end{equation}
where $n_{\rm p}$ is the number of proton in the plasma, and,
\begin{equation}
	j_{\rm [O{\sc III}]}(T) \approx 3.23\times 10^{-21} \frac{ e^{-\frac{28737}{T}} }{4\pi\sqrt{T}} n_{\rm p}^{2}\, \rm erg\, s^{-1}\, cm^{-3}\, sr^{-1},
\label{eq:OIII}
\end{equation}
for the H$\alpha$ and [O{\sc iii}] $\lambda \, 5007$ spectral lines, respectively. 
Additionally, we assume solar oxygen abundances~\citep{lodders_apj_591_2003} and cease to
consider the oxygen as triply ionised at temperatures larger than $10^{6}\, \rm
K$~\citep[cf.][]{cox_mnras_250_1991}. 

\begin{figure*}
	\begin{minipage}[b]{ 0.58\textwidth}
		\includegraphics[width=1.0\textwidth]{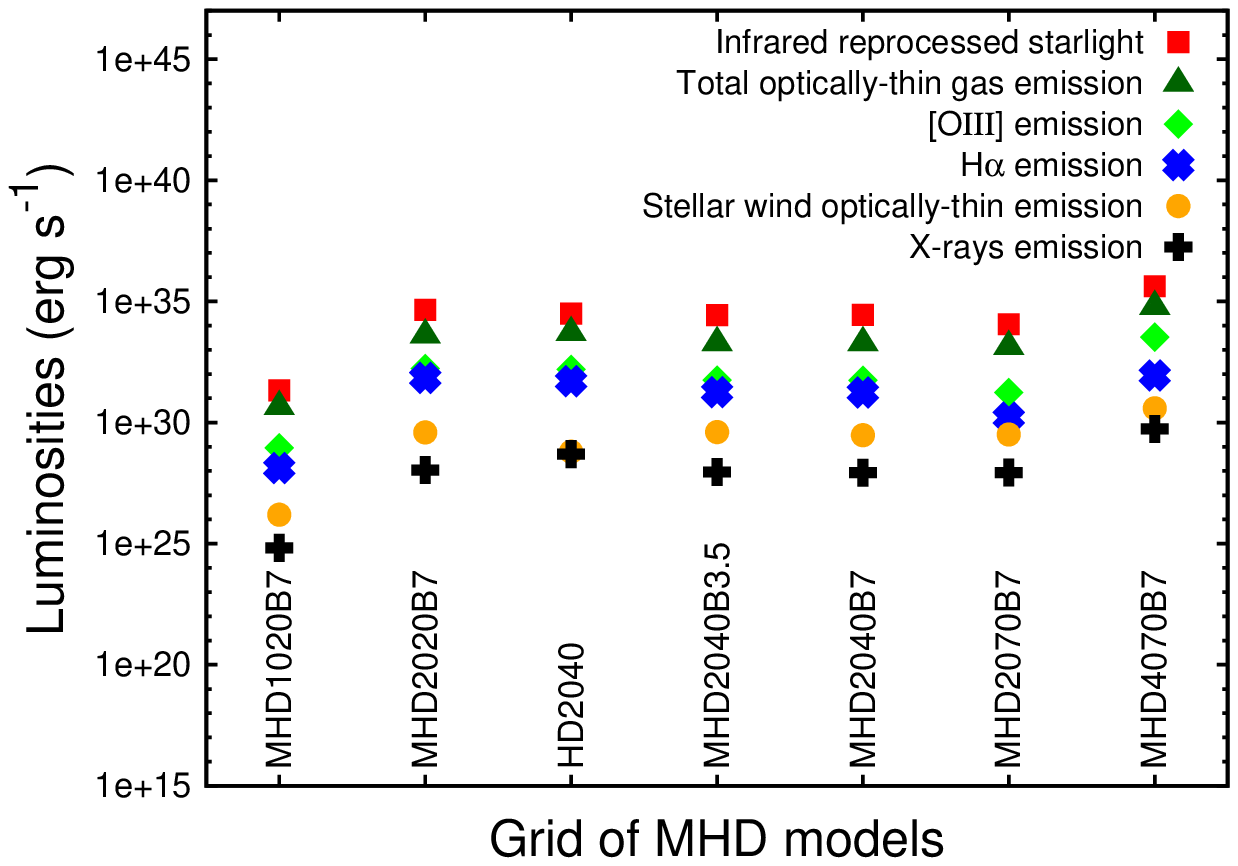}
	\end{minipage} 
	\caption{
	         \textcolor{black}{Bow shocks luminosities and feedback of our
magneto-hydrodynamical models. We separate the infrared reprocessed
starlight (red squares, in $\rm erg\, \rm s^{-1}$) and distinguish the total emission by
optically-thin radiation from the bow shock (dark-green triangles, in $\rm erg\,
\rm s^{-1}$) from the emission from the shocked wind material only (orange dots,
in $\rm erg\, \rm s^{-1}$). Additionally, we show the luminosity from $[\rm
O{\sc III}]$ $\lambda \, 5007$ emission (pale green losanges, in $\rm erg\, \rm
s^{-1}$), the luminosity from H$\alpha$ emission (blue crosses, in $\rm erg\,
\rm s^{-1}$) and the X-rays luminosity in both the soft and hard energy bands
$E>0.5\, \rm eV$ ($T>5.8\times10^{6}\, \rm K$). For the sake of comparison we
add the feedback of the hydrodynamical model HD2040 corresponding to $B_{\rm
ISM}=0\, \mu \rm G$ (originally published in Paper~I). The 
simulations labels are indicated under the corresponding values.} 
		 }
	\label{fig:feedback}  
\end{figure*}

The bow shocks luminosities $L$ are estimated integrating the emission rate,
\begin{equation}
	\itl{ L} = 2\pi \iint_{\mathcal D} \itl{\Lambda}(T) n_{\rm H}^{2} R dRdz,
\label{eq:bslum}
\end{equation}
where $\mathcal D$ represents its volume in the $z>0$ part of the 
computational domain~\citep[][Paper~I]{mohamed_aa_541_2012}. Similarly, we calulate
the momentum deposited by the bow shock \textcolor{black}{by}
subtracting the stellar motion from the ISM gas velocity field. 
We compute $L_{\rm H\alpha}$ and $L_{[\rm OIII]}$, the bow
shocks luminosity in [O{\sc iii}] $\lambda \,
5007$ and H$\alpha$, respectively. Furthermore, we
discriminate the total bow shock luminosity $L_{\rm total}$ from the shocked
wind emission $L_{\rm wind}$. For distinguishing the two kind 
of material, we make use of a passive scalar $Q$ that is advected with
the gas. We estimate the overall X-rays luminosity $L_{\rm X}$ with emission coefficients
generated with the {\sc xspec} program~\citep{arnaud_aspc_101_1996} with solar 
metalicity and chemical abundances from~\citet{asplund_araa_47_2009}. Moreover, 
the total infrared emission $L_{\rm IR}$ is estimated as a fraction of the 
starlight bolometric flux $L_{\star}$~\citep{brott_aa_530_2011a} intercepted by 
the ISM silicate dust grains in the bow shock,  
\begin{equation}
	\itl{ \Gamma}_{\star}^{\rm dust} = 
	\frac{L_{\star}}{ 4\pi d^{2} }n_{\rm d}\sigma_{\rm d}(1-A)\, \rm erg\, s^{-1}\, cm^{-3},
\label{eq:radlum}
\end{equation}
plus the collisional heating,
\begin{equation}
	\itl{\Gamma}_{\rm coll,photo}^{\rm dust}(T) = \frac{ 2^{5/2} f Q 
	n n_{\rm d} \sigma_{\rm d} }{ \sqrt{\pi m_{\rm p}} } \Big( k_{\rm B}T \Big)^{3/2} \, \rm erg\, s^{-1}\, cm^{-3},
\label{eq:dustheathot}
\end{equation}
where $a=5.0\, \rm nm$ is the dust grains radius, 
\begin{equation}
  \sigma_{\rm d}=\pi a^{2}\, \rm cm^{2},
  \label{eq:crosssection}
\end{equation}
is their geometrical cross-section, $d$ their distance from the star and 
$A=1/2$ their Albedo. Additionally,  $n_{\rm d}$ is the dust number density 
whereas $Q\, \simeq 1$ represents the grains electrical properties. More 
details regarding to the estimate of the bow shock infrared luminosity are given in 
Appendix B of Paper~I.

\textcolor{black}{
Last, infrared images are computed performing dust continuum calculations 
against dust opacity for the bow shock generated by our $20\, \rm M_{\odot}$ 
star moving with velocity $40\, \rm km\, \rm s^{-1}$,  using 
the radiative transfer code {\sc radmc-3d}\footnote{ 
\textcolor{black}{http://www.ita.uni-heidelberg.de/~dullemond/software/radmc-3d/} 
}~\citep{ dullemond_2012}. We map the dust mass density fields in our models 
onto a uniform spherical grid $[0;R_{\rm sph}]\times[0; \theta_{\rm max}]$, 
where $R_{\rm sph}=(R_{\rm max}^{2}+z_{\rm max}^{2})^{1/2}$ and $\theta_{\rm 
max}=180\degree$. We assume a dust-to-gas mass ratio of $1/200$. The 
dust density field is computed with the help of the passive 
scalar tracer $Q$ that allows us to separate the dust-free stellar wind of our 
hot OB stars with respect to the dust-enriched regions of the bow shock, made of 
shocked ISM gas. Additionally, we 
exclude the regions of ISM material that are strongly heated by the shocks 
or by electronic thermal conduction (Paper~I), and which are defined as much hotter 
than about a few $10^{4}\, \rm K$. {\sc radmc-3d} then self-consistently 
determines the dust temperature using the Monte-Carlo 
method of~\citet{bjorkman_paj_554_2001} and~\citet{lucy_aa_344_1999} that we use 
as input to the calculations of our synthetic observations.
}

\begin{figure}
	\begin{minipage}[b]{ 0.50\textwidth}
		\includegraphics[width=0.98\textwidth]{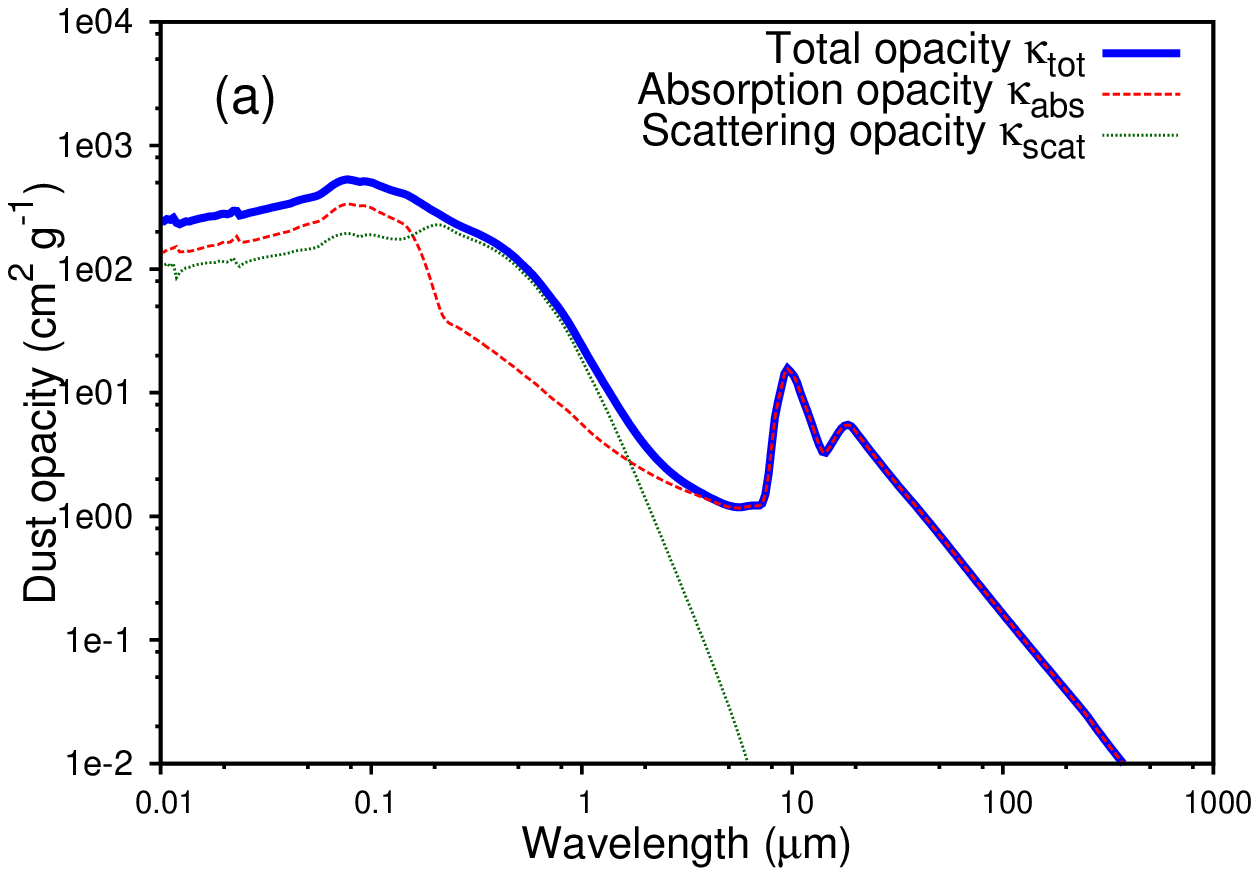}
	\end{minipage} 	\\
	\begin{minipage}[b]{ 0.50\textwidth}
		\includegraphics[width=0.98\textwidth]{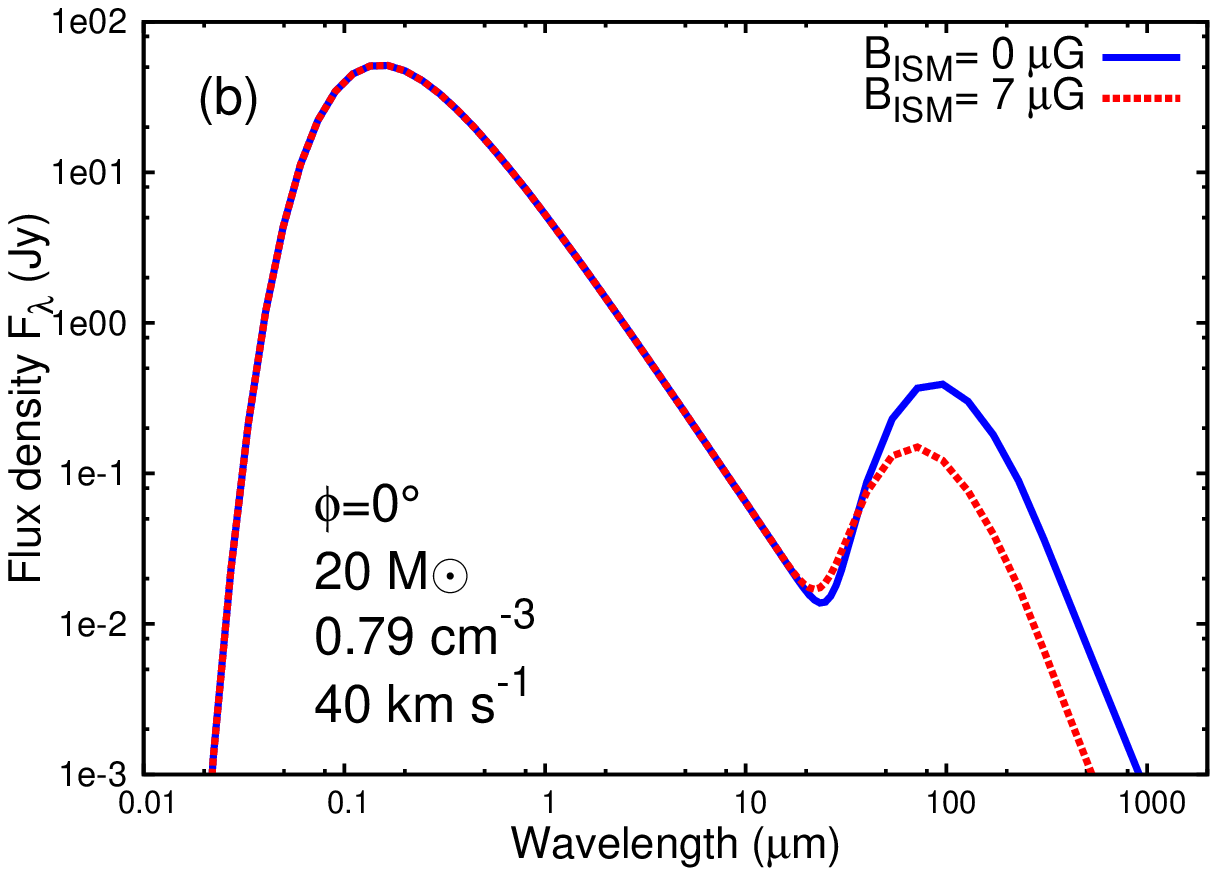}
	\end{minipage}	
	\caption{
	         \textcolor{black}{
	         Top panel: dust opacities used in this study, inspired from~\citet{acreman_mnras_456_2016}. 
	         The figure shows the total opacity 
	         $\kappa_{\rm tot}$ (blue solid thick line), the absorption opacity $\kappa_{\rm abs}$ (red dotted thin line) 
	         and the scattering opacity $\kappa_{\rm scat}$ (green dashed thin line).  
	         } 
	         \textcolor{black}{
	         Bottom panel: spectral energy distributions of our model involving a $20\, \rm M_{\odot}$ star 
	         moving with a velocity of $40\, \rm km\, \rm s^{-1}$, considered in the hydrodynamical 
	         (model MS2040 with $B_{\rm ISM}=0\, \mu \rm G$, solid blue line) and 
	         in the magneto-hydrodynamical contexts (model MHD2040AllB7 with $B_{\rm ISM}=7\, \mu \rm G$, dotted red line).
	         The plot shows the flux density $F_{\lambda}$ (in $\rm Jy$) as a function 
	         of the wavelength $\lambda$ (in $\mu$m) for an inclination angle $\phi=0\degree$ of the bow shock. 
	         }	         
		 }
	\label{fig:dust}  
\end{figure}

\textcolor{black}{
The code solves the transfer equation by ray-tracing photons packages from the
stellar atmosphere that we model as a black body point source of temperature
$T_{\rm eff}$ (see our Table~\ref{tab:wind_para}) that is located at the origin of
the spherical grid. The dust is assumed to be composed of
silicates~\citep{draine_apj_285_1984} of mass density $3.3\, \rm g\, \rm
cm^{-3}$ that follow the canonical power-law distribution $n(a)\propto a^{-q}$
with $q=-3.3$~\citep{mathis_apj_217_1977} and where $a_{\rm min}=0.005\, \mu \rm m$
and $a_{\rm max}=0.25\, \mu \rm m$ the minimal and maximal dust
sizes~\citep{vanmarle_apj_734_2011}.  We generate the corresponding {\sc
radmc-3d} input files containing the dust scattering $\kappa_{\rm scat}$ and
absorption $\kappa_{\rm abs}$ opacities such that the total opacity $\kappa_{\rm
tot}=\kappa_{\rm scat}+\kappa_{\rm abs}$ (see Fig.~\ref{fig:dust}a) on the basis
of a run of the Mie code of Bohren and Huffman~\citep{bohren_book_1983} which is
available as a module of the {\sc
hyperion}\footnote{\textcolor{black}{http://www.hyperion-rt.org/}}
package~\citep{robitaille_aa_536_2011}. Our radiative transfer calculations 
produces spectral energy distributions (SEDs) and isophotal images of the bow
shocks at a desired wavelength, which we choose to be $\lambda=24$ and $60\, \mu
\rm m$ because they corresponds to the wavelengths at which stellar wind bow
shocks are typically observed, see~\citet{sexton_mnras_446_2015}
and~\citet{buren_apj_329_1988,vanburen_aj_110_1995, noriegacrespo_aj_113_1997},
respectively. Our SEDs and images are calibrated to such that we 
consider that the objects are located at a distance $1\, \rm pc$ from 
the observer. 
}


\subsection{Results: optically-thin emission}
\label{sect:feedback}

In Table~\ref{tab:sigma} we report the maximum surface brightness measured along
the direction of motion of the stars in the synthetic emission maps build from
our models at both the H$\alpha$ and [O{\sc iii}] $\lambda \, 5007$ spectral
line emission. We find that the presence of an ISM magnetic field makes \textcolor{black}{the}
H$\alpha$ signatures fainter by about 1-2 orders of magnitudes whereas \textcolor{black}{the}
[O{\sc iii}]  $\lambda \, 5007$ emission maps are about 1 order of magnitude
fainter, respectively. \textcolor{black}{The luminosity of stellar wind bow shocks is a volume
integral (Paper~I) and this volume decreases when a large ISM magnetic pressure 
compresses the nebula} (Fig.~\ref{fig:physics}d,h). Thus, their
surface brightness is fainter despite of the fact that the density and
temperature of their shocked regions is similar (Fig.~\ref{fig:cuts}).

\begin{figure*}
		\centering
	\begin{minipage}[b]{ 0.32\textwidth}
			\centering
	        \includegraphics[width=1.0\textwidth]{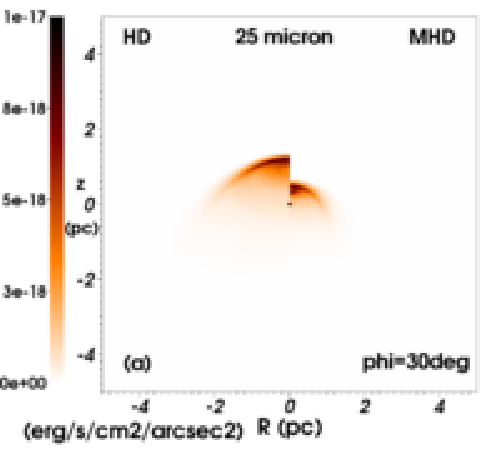}
	\end{minipage}
	\begin{minipage}[b]{ 0.32\textwidth}
			\centering
	        \includegraphics[width=1.0\textwidth]{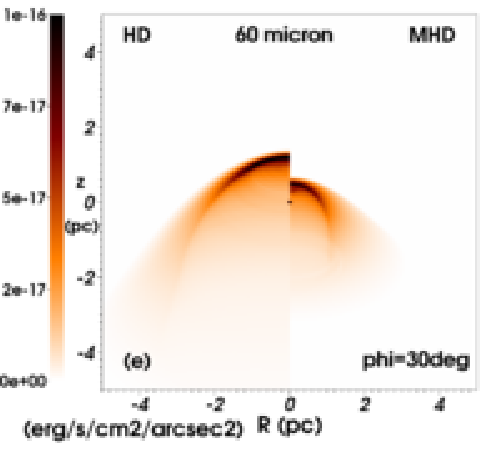}
	\end{minipage}		
	\begin{minipage}[b]{ 0.32\textwidth}
			\centering
	        \includegraphics[width=1.0\textwidth]{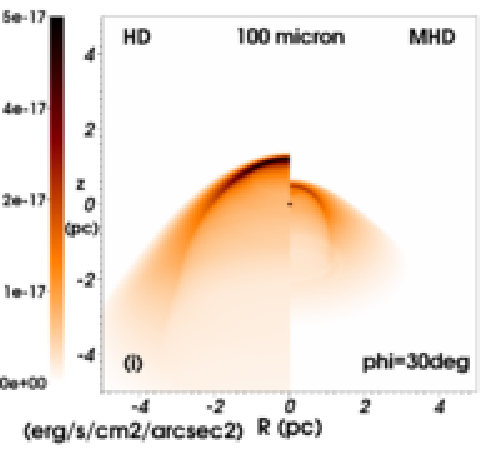}
	\end{minipage}	\\	
	\begin{minipage}[b]{ 0.32\textwidth}
			\centering
	        \includegraphics[width=1.0\textwidth]{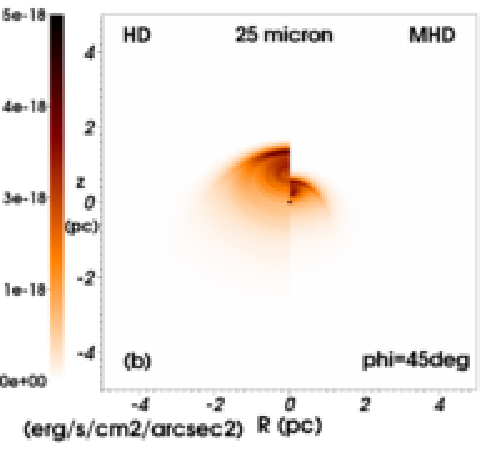}
	\end{minipage}
	\begin{minipage}[b]{ 0.32\textwidth}
			\centering
	        \includegraphics[width=1.0\textwidth]{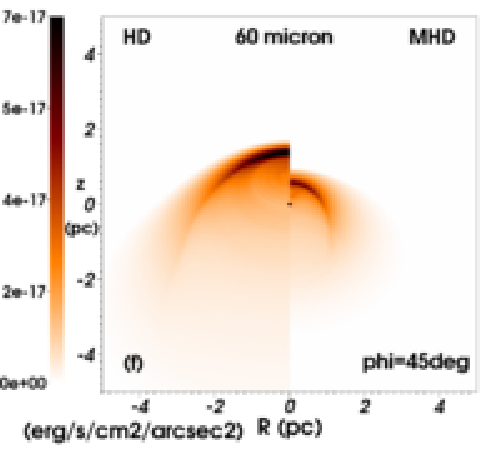}
	\end{minipage}		
	\begin{minipage}[b]{ 0.32\textwidth}
			\centering
	        \includegraphics[width=1.0\textwidth]{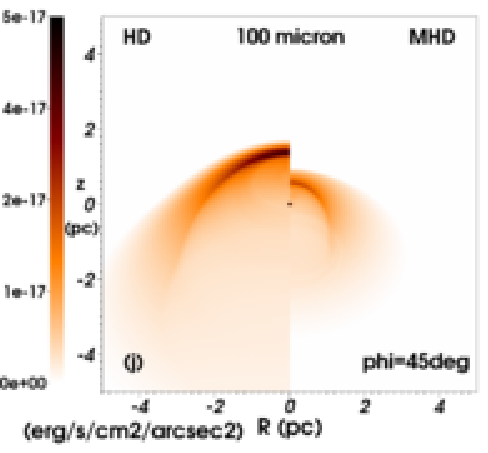}
	\end{minipage}	\\
	\begin{minipage}[b]{ 0.32\textwidth}
			\centering
	        \includegraphics[width=1.0\textwidth]{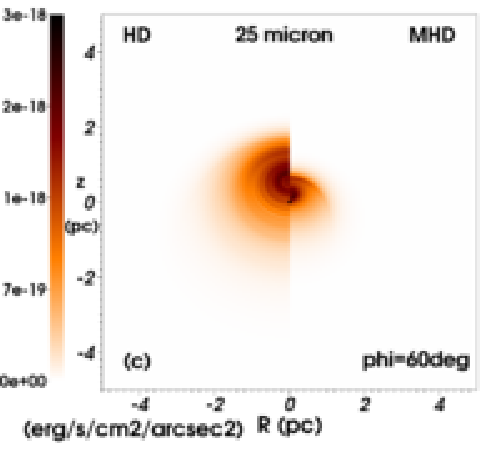}
	\end{minipage}
	\begin{minipage}[b]{ 0.32\textwidth}
			\centering
	        \includegraphics[width=1.0\textwidth]{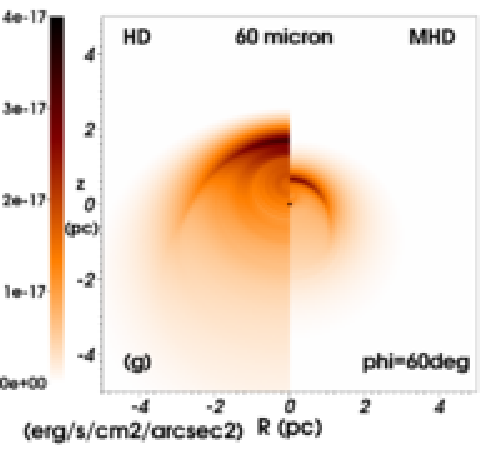}
	\end{minipage}	
	\begin{minipage}[b]{ 0.32\textwidth}
	      \centering
	        \includegraphics[width=1.0\textwidth]{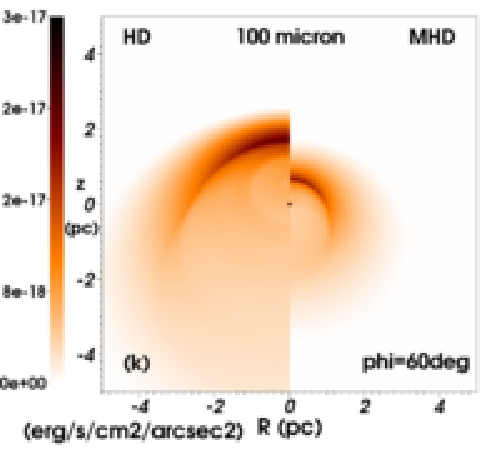}
	\end{minipage}	\\
	\begin{minipage}[b]{ 0.32\textwidth}
			\centering
	        \includegraphics[width=1.0\textwidth]{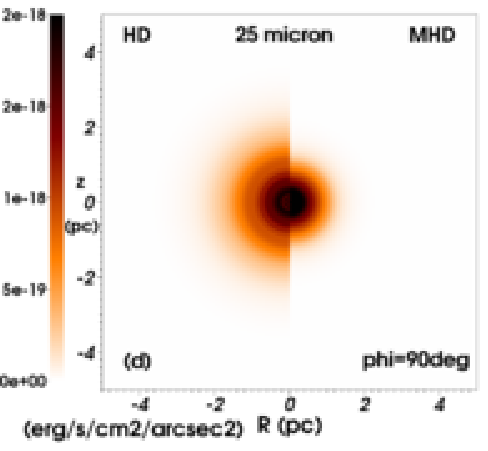}
	\end{minipage}
	\begin{minipage}[b]{ 0.32\textwidth}
			\centering
	        \includegraphics[width=1.0\textwidth]{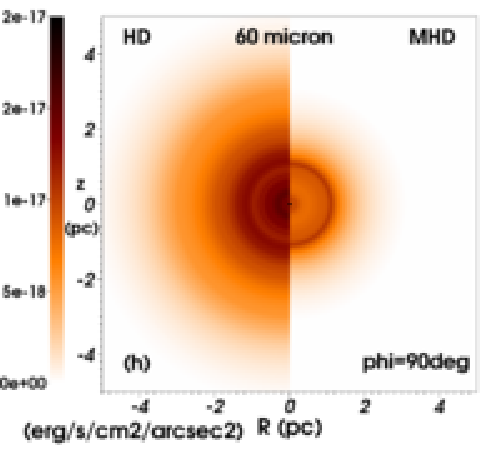}
	\end{minipage}	
	\begin{minipage}[b]{ 0.32\textwidth}
	      \centering
	        \includegraphics[width=1.0\textwidth]{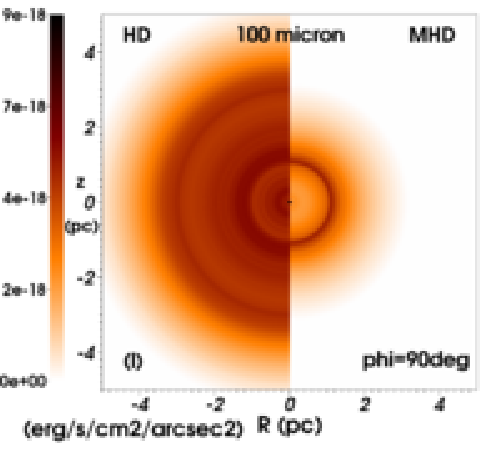}
	\end{minipage}		
	\caption{
	         \textcolor{black}{
	         Isophotal infrared emission maps of our bow shock models MS2040 and MHD2040AllB7. 
	         It represents our initially $20\, \rm M_{\odot}$ star moving with velocity $40\, \rm km\, \rm
s^{-1}$ as seen at wavelengths $\lambda=25$ (a-d), $60\, \rm \mu m$ (e-h) and $100\, \rm \mu m$ (i-l). The projected
flux is in units of $\rm erg\, \rm s^{-1}\, \rm cm^{-2}\, \rm arcsec^{-2}$. 
The maps are generated with an inclination 
angle of $\phi=30$ (a,e,i), $45$ (b,f,j), $60$ (c,g,k) and $90\degree$ (d,h,l) with respect to 
the plane of the sky. For each panel, the surface brightness is plotted in the linear scale and 
its maximum corresponds to the maximum of the hydrodynamical (left) and 
magneto-hydrodynamical bow shock models (right). 
		 } 
		 }
	\label{fig:maps1}  
\end{figure*}


The ratio of our bow shocks models' maximum [O{\sc iii}] and H$\alpha$ maximum
surface brightness increases \textcolor{black}{in} the presence of the magnetic field, e.g. the
hydrodynamical model MS2040 has $\Sigma_{[\rm O{\sc III}]]}^{\rm
max}/\Sigma_{[\rm H\alpha]}^{\rm max}\approx2.1$ whereas our model MHD2040AllB7 has
$\Sigma_{[\rm O{\sc III}]]}^{\rm max}/\Sigma_{[\rm H\alpha]}^{\rm
max}\approx5.5$ if $B_{\rm ISM}=7\, \mu \rm G$. We notice that the spectral line
ratio $\Sigma_{[\rm O{\sc III}]]}^{\rm max}/ \Sigma_{[\rm H\alpha]}^{\rm max}$
augments with the increasing space velocity of the star, e.g. our models
MHD2020B7, MHD2040AllB7 and MHD2070B7 have $\Sigma_{[\rm O{\sc III}]]}^{\rm
max}/\Sigma_{[\rm H\alpha]}^{\rm max}\approx4.0$, $5.5$ and $25.0$,
respectively. This difference between [O{\sc iii}] $\lambda \, 5007$ and 
H$\alpha$ emission is more pronounced in our magneto-hydrodynamical simulations.
As for our hydrodynamical study, the region of maximum emission peaks close to
the contact discontinuity in the layer of shocked ISM material, in the region of
the stagnation shock (Paper~I, see also Figs.~\ref{fig:maps}a,b).

\begin{figure}
	\begin{minipage}[b]{ 0.48\textwidth}
		\includegraphics[width=1.0\textwidth]{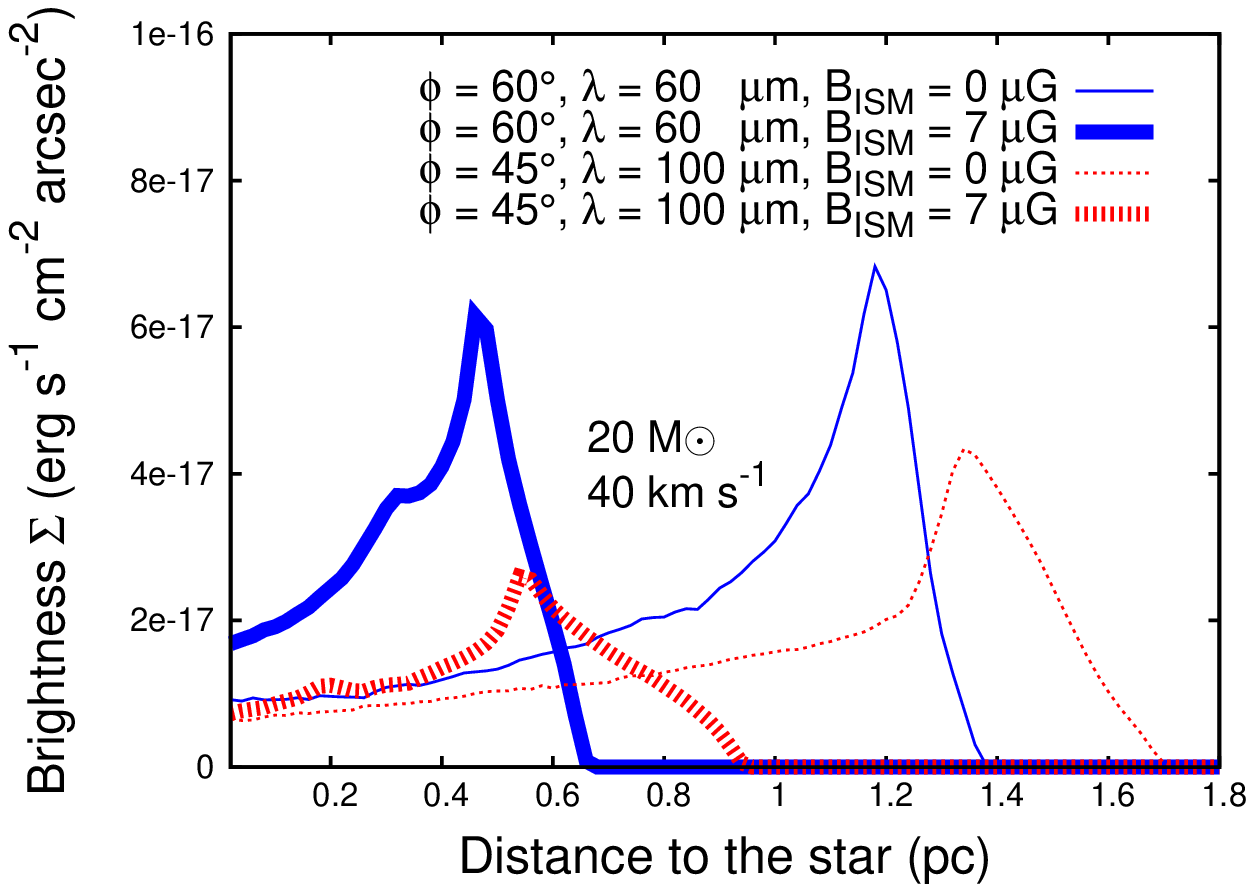}
	\end{minipage}  	
	\caption{
		\textcolor{black}{
		Cross-sections taken through the direction of motion of the bow shock 
of our initially $20\, \rm M_{\odot}$ star moving in a medium 
with velocity $v_{\star}=40\, \rm km\, \rm s^{-1}$, both in a medium with 
$B_{\rm ISM}=0$ and $7\, \mu\rm G$. The emission are shown for a viewing angle 
of $\phi=45\degree$ and at the waveband $\lambda=60\, \mu\rm m$ (dotted red curves)
and for $\phi=60\degree$ at $\lambda=100\, \mu\rm m$ (solid blue curves).  
The surface brightness (in $\rm erg\, \rm s^{-1}\, \rm cm^{-2}\, \rm arcsec^{-2}$) 
is plotted as a function of the distance to the star (in $\rm pc$). 
The position of the star is located at the origin.} 
		 }
	\label{fig:cuts_mhd}  
\end{figure}

The ISM magnetic field does not change the order relations we previously
established with hydrodynamical bow shocks generated by main-sequence stars 
(Fig.~13a in Paper~I), i.e. $L_{\rm wind} < L_{\rm H\alpha} < L_{\rm total} <
L_{\rm IR}$ (see orange dots, blue crosses of Saint-Andrew, dark green triangles
and black squares in Fig.~\ref{fig:feedback}a, respectively). Additionally, as
discussed above in the context of projected emission maps, we find that the 
optical spectral line emission that we consider are such that $L_{[\rm O{\sc III}]}
> L_{\rm H\alpha}$. This confirms and extend to magneto-hydrodynamical bow shocks 
a result previously obtained \textcolor{black}{by} integrating the optically-thin emission in the range
$8000 \le T \leq 10^{6}\, \rm K$ (Paper~I). Our magneto-hydrodynamical bow shock
models have H$\alpha$ and $[\rm O{\sc III}]$ emission originating from the shocked 
ISM gas and their emission from the wind material is negligible ($L_{\rm total}/L_{\rm
wind}\approx10^{-6}$). Moreover, we find that the bow shocks X-rays emission are
very small in all our simulations ($L_{\rm X}/L_{\rm wind}\approx10^{-1}$, see
black crosses in Fig.~\ref{fig:feedback}a).

\subsection{Results: dust continuum infrared emission}
\label{sect:infrared}


\subsubsection{Spectral energy distribution}
\label{sect:infrared_seds}

\textcolor{black}{
Fig.~\ref{fig:dust}b plots a comparison betwenn the SEDs of two bow shock models 
generated by our $20\, \rm M_{\odot}$ star moving with velocity $40\, \rm km\, 
\rm s^{-1}$,  either through an unmagnetized ISM (model 
MS2040, solid blue line) or in a medium with $B_{\rm ISM}=7\, \mu\rm G$ (model MHD2040AllB7, 
dotted red line) for a viewing angle of the nebulae of $\phi=0\degree$. The 
figure represents the flux density $F_{\lambda}$ (in $\rm Jy$) as a function of 
the wavelength $\lambda$ (in $\mu \rm m$) for the waveband including the 
$0.01\le \lambda \le 2000 \, \mu \rm m$. The star is 
responsible for the component in the range $0.01\le \lambda \le 10 \, \mu \rm m$ 
that corresponds to a black body spectrum of temperature $T_{\rm eff}=33900\, 
\rm K$ (see Table~\ref{tab:wind_para}) while the circumstellar dust produces 
the feature in the waveband $10\le \lambda \le 2000 \, \mu \rm m$.  
The bow shock's component is in the waveband including the wavelengths at which 
stellar wind bow shock are typically recorded, e.g. at $60\, \mu \rm m$~\citet{buren_apj_329_1988,
vanburen_aj_110_1995,noriegacrespo_aj_113_1997}. 
}

\textcolor{black}{
The SED of the magnetized bow shock has a slightly larger flux than the 
SED of the hydrodynamical bow shock in the waveband $10\le \lambda 
\le 30\, \mu \rm m$, because its smaller size makes the shell of dense ISM 
gas closer to the star, increasing therefore the dust temperature (Fig.~\ref{fig:dust}b). 
At $\lambda \approx 30\, \mu \rm m$, the hydrodynamical bow shock emits by slightly 
more than half an order of magnitude than the magnetized nebula, 
e.g. at $\lambda \approx 60\, \mu \rm m$ our model MS2040 has a density flux 
$F_{\lambda} \approx 3\times 10^{-1}\, \rm Jy$ whereas our model MHD2040AllB7 shows  
$F_{\lambda} \approx 1\times 10^{-1}\, \rm Jy$, respectively. This is consistent 
with the previously discussed reduction of the projected optical emission of 
our bow shocks. This relates to the changes in size of the nebulae induced by 
the inclusion of the magnetic field in our simulations, which reduces the mass of dust in 
the structure responsible for the reprocessing of the starlight, e.g. 
our models MS2040 and MHD2040AllB7 contain about 
$M_{\rm d}\approx3\times 10^{-2}\, \rm M_{\odot}$ and 
$M_{\rm d}\approx2\times 10^{-3}\, \rm M_{\odot}$, respectively, where 
$M_{\rm d}$ is the dust mass trapped into the nebulae. The reduced mass    
of dust into the magnetized bow shock absorbs a lesser amount of the stellar 
radiation and therefore re-emits a smaller quantity of energy,  
reducing $F_{\lambda}$ in the waveband $\lambda\ge\, 30\, \mu$m 
(Fig.~\ref{fig:dust}b). 
Note that the infrared 
surface brightness of a bow shock is also sensible to the density of its ambient 
medium, i.e. $F_{\lambda}$ is much larger in the situation of a runaway star moving 
in a medium with $n_{\rm ISM}\simeq 1000\, \rm cm^{-3}$~\citep{acreman_mnras_456_2016}.  
%
}

\subsubsection{Synthetic infrared emission maps}
\label{sect:infrared_maps}

\textcolor{black}{
Our Fig.~\ref{fig:maps1} plots a series of synthetic infrared emission maps of our bow 
shock models produced by an initially $20\, \rm M_{\odot}$ star moving with 
velocity $40\, \rm km\, \rm s^{-1}$ in its purely hydrodynamical (MS2040) or 
magneto-hydrodynamical configuration (MHD2040AllB7) at the wavelengths 
corresponding to the central wavelengths of the {\it IRAS} facility's main 
broadband images~\citep{vanburen_apj_329_1988}, i.e. $\lambda=25\, \rm \mu m$ (left column 
of panels), $60\, \rm \mu m$ (middle column of panels) and $100\, \rm \mu m$ 
(right column of panels). The maps are represented with an inclination angle of 
$\phi=30\degree$ (Fig.~\ref{fig:maps1}a,e,i), $45\degree$ 
(Fig.~\ref{fig:maps1}b,f,j), $60\degree$ (Fig.~\ref{fig:maps1}c,g,k) and 
$90\degree$ (Fig.~\ref{fig:maps1}d,h,l) with respect to the plane of the 
sky and the projected flux is plotted in units of $\rm erg\, \rm s^{-1}\, \rm 
cm^{-2}\, \rm arcsec^{-2}$. As in the context of their optical emission 
(Fig.~\ref{fig:maps}), the overall size of the infrared magnetized bow shocks is 
smaller than in the hydrodynamical case because of the reduction of their 
stand-off distance $R(0)$, see, e.g. Fig.~\ref{fig:maps1}a,e,j. 
The global morphology of our infrared bow shock 
nebulae does not change significantly. It remains a single, bright arc at the 
front of an ovoid structure that is symmetric with respect to the direction of 
motion of the runaway star and extended to the trail ($z \le 0$) of the bow shocks 
due to the supersonic motion of the star~\citet{acreman_mnras_456_2016}. In the 
hydrodynamical case, the region of maximum emission is the region containing the 
ISM dust which temperature is smaller than a few $10^{4}\, \rm K$, i.e. 
between the contact discontinuity and the forward shock of the bow 
shock~\citep[Paper~I,][]{acreman_mnras_456_2016} whereas in the magnetized case, 
the maximum emission is reduced to a thin region close to the discontinuity 
between hot stellar wind and colder ISM. Both the shocked stellar wind and the 
shocked ISM of the bow shock do not contributes to these emission 
because the material is too hot. 
}

\textcolor{black}{
Fig.~\ref{fig:cuts_mhd} reports cross-sections taken along the direction of 
motion of the bow shock and comparing their surface brightesses at several 
wavebands $\lambda$ and viewing angles $\phi$. It illustrates that, as in the 
case of the optical emission, the presence of the ISM magnetic field makes the 
bow shocks slightly dimmer, e.g. for $\phi=45\degree$ our model has a maximal 
surface brighness of $\Sigma^{\rm max}_{100\,\mu\rm m} \approx 4.3\times 
10^{-17}\, \rm erg\, \rm s^{-1}\, \rm cm^{-2}\, \rm arcsec^{-2}$ whereas 
$\Sigma^{\rm max}_{100\,\mu\rm m} \approx 2.6\times 10^{-17}\, \rm erg\, \rm 
s^{-1}\, \rm cm^{-2}\, \rm arcsec^{-2}$ for $B_{\rm ISM}=0$ and $7\, \mu\rm G$, 
respectively. Fig.~\ref{fig:cuts} shows different cross-sections of the 
projected infrared emission the magnetized bow shock of our initially $20\, \rm 
M_{\odot}$ star moving with velocity $v_{\star}=40\, \rm km\, \rm s^{-1}$. The 
emission at $\lambda=60\, \mu\rm m$ is more important that at $\lambda=25\, 
\mu\rm m$ and at $\lambda=100\, \mu\rm m$, e.g. it peaks at $\Sigma^{\rm 
max}_{60\,\mu\rm m} \approx 8.2 \times 10^{-17}\, \rm erg\, \rm s^{-1}\, \rm 
cm^{-2}\, \rm arcsec^{-2}$ whereas $\Sigma^{\rm max}_{25\,\mu\rm m} \approx 2.6 
\times 10^{-17}\, \rm erg\, \rm s^{-1}\, \rm cm^{-2}\, \rm arcsec^{-2}$ and 
$\Sigma^{\rm max}_{100\,\mu\rm m} \approx 3.0 \times 10^{-17}\, \rm erg\, \rm 
s^{-1}\, \rm cm^{-2}\, \rm arcsec^{-2}$, respectively, at a distance of $0.55\, 
\rm pc$ from the star and assuming an inclination angle of the bow shock of 
$\phi=45\degree$ (Fig.~\ref{fig:cuts}a). All our models have similar behaviour 
of their infrared surface brightness as a function of $\lambda$ and $\phi$. Note 
also that the evolution of the position of the stand-off distance of the bow 
shock is consistent with the study of~\citet{acreman_mnras_456_2016} in the 
sense that it increases at larger $\phi$ (Fig.~\ref{fig:cuts}b). 
}


\begin{figure*}
	\begin{minipage}[b]{ 0.48\textwidth}
		\includegraphics[width=1.0\textwidth]{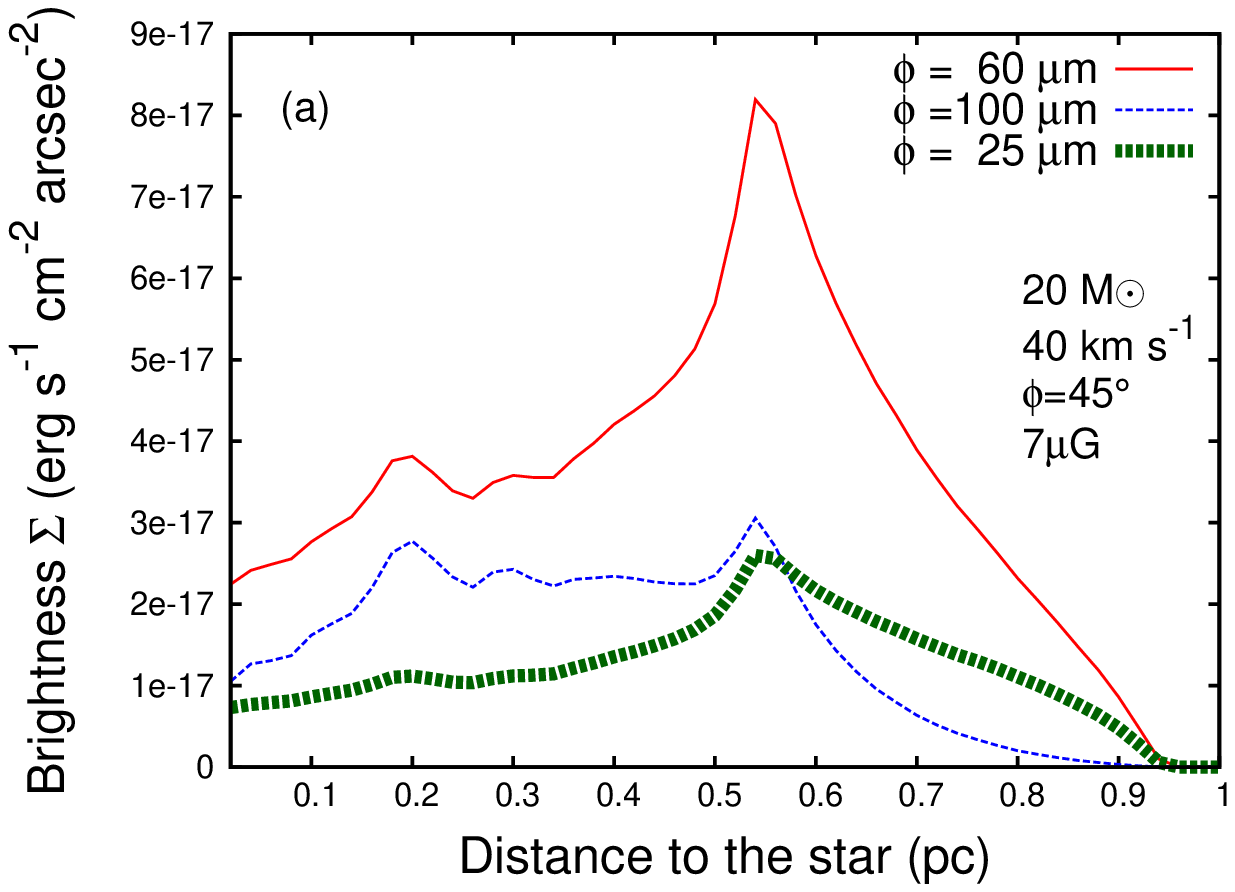}
	\end{minipage} 
	\begin{minipage}[b]{ 0.48\textwidth}
		\includegraphics[width=1.0\textwidth]{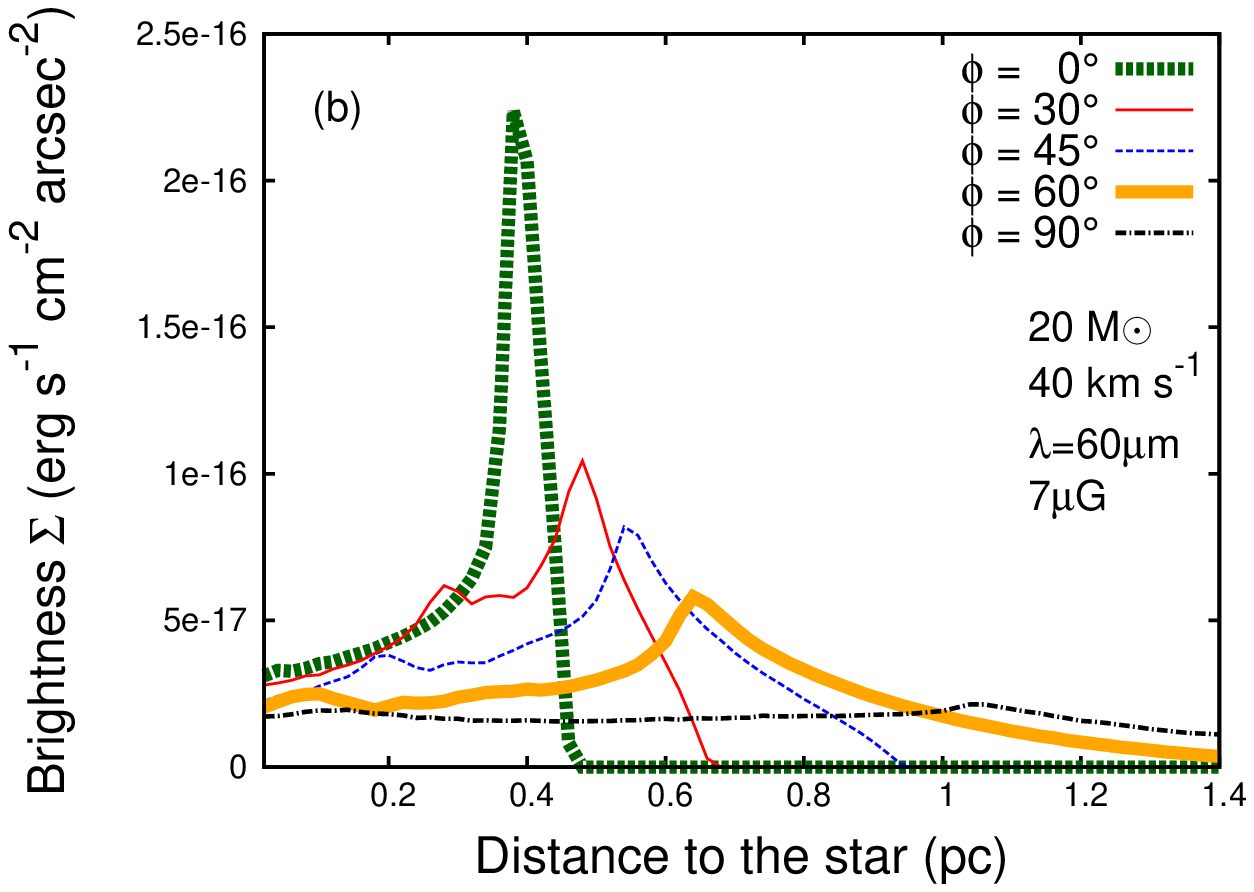}
	\end{minipage}  	
	\caption{
		\textcolor{black}{Cross-sections taken through the direction of motion of the bow shock 
of our initially $20\, \rm M_{\odot}$ star moving in a medium with $B_{\rm ISM}=7\, \mu\rm G$ 
with velocity $v_{\star}=40\, \rm km\, \rm s^{-1}$. The emission are shown for the principal 
broadband images of the $\it IRAS$ telescope for a viewing angle $\phi=45\degree$ (a) and for 
the wavelength $\lambda=60\, \mu\rm m$ as a function of different viewing angle $\phi$ (b).  
The surface brightness (in $\rm erg\, \rm s^{-1}\, \rm cm^{-2}\, \rm arcsec^{-2}$) is plotted as a 
function of the distance to the star (in $\rm pc$). The position of the star is located at the origin.} 
		 }
	\label{fig:cuts}  
\end{figure*}


\subsection{Implications of our results and discussion}
\label{sect:implication}

\subsubsection{Bow shocks H$\alpha$ and [OIII] observability}
\label{sect:observability}

The surface brightnesses at H$\alpha$ and $[\rm O{\sc III}]$ $\lambda \, 5007$
spectral line emission of our stellar wind bow shocks reported in
Table~\ref{sect:obs} indicate that (i) the presence of the ISM magnetic field makes their
projected emission $\Sigma_{\rm H \alpha}$ and $\Sigma_{[\rm OIII]}$ fainter by
two and 1$-$2 orders of magnitude and (ii) that this reduction of the nebulae's
emission is more important as the strength of the B field is larger.
Consequently, the emission signature of a purely hydrodynamical bow shock model
that is above the the diffuse emission sensitivity threshold of, e.g. the
SuperCOSMOS H-Alpha Survey (SHS) of $\Sigma_{\rm SHS}\approx1.1$-$2.8\times
10^{-17}\, \rm erg\, \rm s^{-1}\, \rm cm^{-2}\, \rm arcsec^{-2}$ \textcolor{black}{can drop down below it} 
once the ISM magnetic field is switched-on. As an example, our
hydrodynamical model of a $20\, M_{\odot}$ star moving with velocity
$v_{\star}=70\, \rm km\, \rm s^{-1}$ (Paper~I) could be observed since it has
$\Sigma_{\rm H \alpha}\approx 1.5\times 10^{-16} \ge \Sigma_{\rm SHS}$ whereas
our magneto-hydrodynamical model of the same runaway star has $\Sigma_{\rm H
\alpha}\approx 8.0\times 10^{-18} \ll \Sigma_{\rm SHS}$ and would be invisible
regarding to the SHS facility (our Table~\ref{tab:sigma}).

This may explain why not so many stellar wind bow shocks are discovered at 
H$\alpha$ around isolated, hot massive stars, despite of the fact the ionisation 
of their circumstellar medium must produce such 
emission~\citep{brown_aa_439_2005}. Since $\Sigma_{\rm H \alpha}\propto n^{2}$ 
(see Appendix A of Paper~I), it implies that the more diluted the ISM 
constituting the surrounding of an exiled star, i.e. the higher the runaway 
star's Galactic latitude, the smaller the probability to observe its bow shock 
at H$\alpha$. In other words, the search for bow shocks at this wavelength may 
work well within the Galactic plane or in relatively dense regions of the ISM. 
Note also that in the presence of the magnetic field, all models have 
$\Sigma_{[\rm OIII]}/\Sigma_{\rm H \alpha}>1$, which is consistent with the 
discovery of the first bow-shock-producing massive stars $\zeta$ Ophiuchi in 
$[\rm OIII]$ $\lambda \, 5007$ emission.

\subsubsection{Surrounding \hii region and dust composition}
\label{sect:hii_region}

\textcolor{black}{
Massive stars release huge amount of ultraviolet 
photons~\citep{diazmiller_apj_501_1998} that ionize the hydrogen constituting  
their surroundings~\citep{dyson_ass_35_1975}, giving birth to an \hii region 
overwhelming the stellar wind bubble of the 
star~\citep{weaver_apj_218_1977,vanmarle_phd_2006}. In the case of a runaway 
star, the stellar motion produces a bow shock surrounded by a cometary \hii 
region~\citep{raga_apj_300_1086,maclow_apj_369_1991,raga_rmxaa_33_1997, 
arthur_apjs_165_2006,zhu_apj_812_2015}, which presence in our study is simply 
taken into account assuming that the ambient medium of the star is fully 
ionized, however, we neglect its turbulent internal structure. The gas that is 
between the forward shock of the bow shock and the outer part of the \hii region 
is filled by ISM dust that emits infrared thermal emission by efficiently 
reprocessing the stellar radiation, i.e. it is brighter that the emission by 
gas cooling (Paper~I). 
}

\textcolor{black}{
While our study shows that our nebulae are brighter at $60\, \mu \rm m$ 
(Fig.~\ref{fig:cuts}), i.e. at the waveband at which catalogues of bow shocks 
from OB stars have been 
compiled~\citep{buren_apj_329_1988,vanburen_aj_110_1995,
noriegacrespo_aj_113_1997}, the study of~\citet{mackey_aa_586_2016} compared the 
respective brightnesses of the front of a distorted circumstellar bubble with 
the outer edge of its surrounding \hii region and find the $24\, \mu \rm m$ 
waveband to be ideal to observe the structure generated by the stellar wind. 
However, the presence of the ISM background magnetic field makes our infrared 
arc  smaller and slightly dimmer, i.e. more difficult to detect in the case of a 
distant runaway star which could explain why a large proportion of observed \hii 
regions do not contain dust-free cavities encircled with bright mid-infrared 
arcs~\citep{sharpless_apj_1959,churchwell_apj_649_2006,watcher_aj_139_2010,
simpson_mnras_424_2012}. Further radiation magneto-hydrodynamics simulations are 
required to fully assess the question of the infrared screening of stellar wind bow shocks by 
their own \hii regions, particularly for an ambient medium corresponding to the 
Galactic plane ($n_{\rm ISM}\simeq1\, \rm cm^{-3}$). 
}

\textcolor{black}{ 
Following~\citet{pavlyuchenkov_rep_57_2013}, we consider that the dust filling 
the \hii region and penetrating into the bow shock is similar of that of the 
ISM. Our radiative transfer calculations nevertheless suffer from uncertainties 
regarding to the composition of this ISM dust. Our mixture is made of 
Silicates~\citep{draine_apj_285_1984} which could be modified, e.g. changing the slope 
of the dust size distribution. Particularly, the inclusion of very small grains such as polycyclic 
aromatic hydrocarbon~\citep[PAHs, see][]{wood_apj_688_2008} may be an 
appropriate update of the dust mixture, as it have been shown to be necessary 
to fit observations of mid-infrared bow shocks around O stars in dense medium in 
M17 and RCW 49~\citep{povich_apj_689_2008}. Enlarging our work in a wider study, 
e.g. scanning the parameter space of the quantities governing the formation of 
Galactic stellar wind bow shocks ($v_{\star}$, $n_{\rm ISM}$, $M_{\star}$) in 
order to discuss both their SEDs and infrared images will be considered in a 
follow-up paper, e.g. performing a systematic post-processing of the grid of bow 
shock simulations of~\citet{meyer_mnras_459_2016} with {\sc radmc-3d}. Then, 
thorough comparison of numerical simulations with, e.g. the {\it IRAS} 
observations 
of~\citet{buren_apj_329_1988,vanburen_aj_110_1995,noriegacrespo_aj_113_1997}  
would be achievable.  
}

\subsubsection{Shaping of the circumstellar medium of runaway massive stars at the pre-supernova phase}
\label{sect:presn}

It has been shown in the context of Galactic, high-mass runaway stars, that the 
pre-shaped circumstellar medium in which these stars die and explode as a type 
II supernova is principally constituted of its own main-sequence wind bubble, 
distorted by the stellar motion. Further evolutionary phase(s) produce 
additional bubble(s) and/or shell(s) whose evolution is contained inside the 
initial bow shock~\citep{brighenti_mnras_270_1994, brighenti_mnras_277_1995}. 
The expansion   of the subsequent supernova shock wave is strongly impacted by 
the progenitor's pre-shaped circumstellar medium inside which it develops 
initially~\citep[see, e.g.][]{cox_mnras_250_1991}. Particularly, the more 
well-defined and stable the walls of the tunnel formed by the reverse shock of 
the bow shock are, the easier the channeling the supernovae ejecta inside 
it~\citep[see in particular Appendix A of][and references 
therein]{meyer_mnras_450_2015}.

Our study shows that the presence of background ISM magnetic field aligned with
the direction of motion of a main-sequence runaway star inhibits the growth of
both shear instabilities that typically affect these
circumstellar structures (Fig.~\ref{fig:physics}).
Consequently, a planar-aligned magnetic field would further shape the reverse
shock of moving stars' bow shocks as a smooth tube in which shocks waves could
be channeled as a jet-like extension, e.g. as in~\citet{cox_mnras_250_1991}.
Additionally, the shock wave outflowing out of the forward shock of circumstellar 
structures of runaway stars that are sufficiently dense to make their
subsequent supernova remnant asymmetric~\citep{meyer_mnras_450_2015} would be
more collimated along the direction of motion of its progenitor and/or ambient magnetic
field. \textcolor{black}{This may produce additional asymmetries to the elongated shape 
of supernovae remnants exploding in a magnetized ISM~\citep{rozyczka_274_MNRAS_1995}.}

\subsubsection{The case of the hot runaway star $\zeta$ Ophiuchi}
\label{sect:obs}

\begin{figure*}
	\begin{minipage}[b]{ 0.76\textwidth}
		\includegraphics[width=1.0\textwidth]{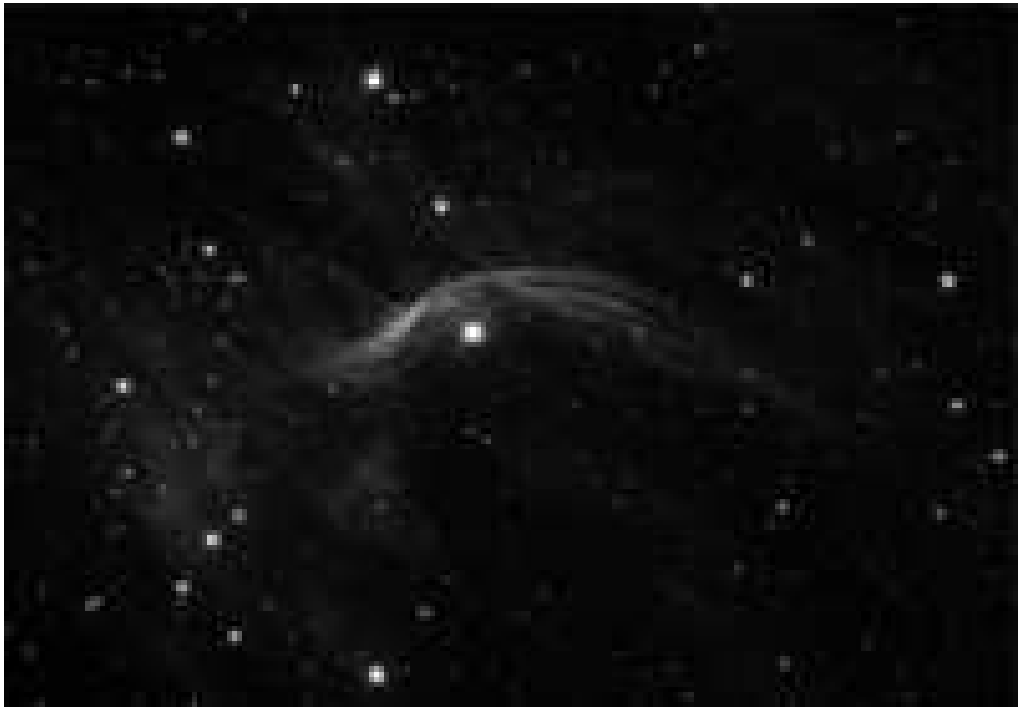}
	\end{minipage} 
	\caption{
	         ${\it WISE}$ $3.4\, \mu m$~\citep[band W1, ][]{wright_aj_140_2010} 
	         observation of the stellar wind bow shock surrounding the massive 
	         runaway O9.5 V star $\zeta$ Ophiuchi. \textcolor{black}{The image 
	         represents about $35\, \rm arcmin$ in the horizontal direction, 
	         which at a distance of $112\, \rm pc$ 
	         corresponds to about $1.12\, \rm pc$.} 
		 }
	\label{fig:obs}  
\end{figure*}

The O9.5 V star $\zeta$ Ophiuchi is the Earth's closest massive, main-sequence 
runaway star. Infrared observations, e.g. with the ${\it WISE}$ $3.4\, \mu m$ 
facility~\citep[band W1, ][see Fig.~\ref{fig:obs}\footnote{http://wise2.ipac.caltech.edu/docs/release/allsky/}]{wright_aj_140_2010} highlighted the complex topology of its stellar wind bow 
shock, originally discovered in [O{\sc iii}] $\lambda \, 5007$ spectral 
line~\citep{gull_apj_230_1979} and further observed in the infrared 
waveband~\citep{vanburen_apj_329_1988}. The properties of the particular, 
\textcolor{black}{non-axisymmetric} shape of its circumstellar nebula which moves 
in the \hii region Sh 2-27~\citep{sharpless_apj_1959} is studied in a relatively 
large literature~\citep[see][and references therein]{mackey_mnras_2013}. The 
mass-loss of $\zeta$ Ophiuchi has been estimated in the range $\dot{M}_{\zeta} 
\approx 1.58 \times 10^{-9}$$-$ $1.43 \times 10^{-7}\, \rm M_{\odot}\, \rm 
yr^{-1}$~\citep{gvaramadze_mnras_427_2012}, which, according to 
Eq.~(\ref{eq:Ro}), taking $R(0)\approx\, 0.16\, \rm 
pc$~\citep{gvaramadze_mnras_427_2012}, adopting $v_{\star} \approx 26.5\, \rm 
km\, \rm s^{-1}$ and considering a typical OB star wind velocity of $v_{\rm 
w}\approx1500\, \rm km\, \rm s^{-1}$, constrains its ambient medium density to 
$n_{\zeta}\approx 3$-$4\, \rm cm^{-3}$~\citep[cf.][]{gull_apj_230_1979}.

\textcolor{black}{
Assuming (i) the magnetisation of the close surrounding of $\zeta$ Ophiuchi to be 
$B_{\rm ISM}=7\, \mu \rm G$~\citep{mackey_mnras_2013}, (ii) that the 
conditions for switch-on shocks to be permitted are fulfilled, i.e. plasma and Alfv\' enic 
velocities are normal to the shock, and (iii)} considering that its
ISM properties are, in addition to the above presented quantities, such that
$\gamma=1.67$, $T_{\rm ISM}=8000\, \rm  K$, it comes that \textcolor{black}{$\beta \ge 2/\gamma$} 
and $M_{\rm A}<1$. \textcolor{black}{This indictes that, under our hypothesises, 
the ambient medium of $\zeta$ Ophiuchi  
does not allow the existence of switch-on shocks. Consequently, the imperfect 
shape of its bow shock (Fig.~\ref{fig:obs}) may not be explained invoking the
particular double-front topology of bow shocks that can be produced in such regime, but 
rather by the presence of a background ISM magnetic field whose direction is
not aligned with respect to the motion of the star.} Further
tri-dimensional \textcolor{black}{magneto-hydrodynamical models are needed in order to
assess the question} of $\zeta$ Ophiuchi's background ISM magnetic field direction, the
position of its contact discontinuity and a more precise estimate of its stellar
wind mass-loss.

\subsubsection{The case of runaway cool stars}
\label{sect:coolstars}

Our results apply to bow shocks generated by hot, main-sequence OB stars that 
move through the hot ionized gas of their own \hii 
region~\citep{raga_rmxaa_33_1997} and archetype of which is the nebulae 
surrounding $\zeta$ Ophiuchi (see above discussion). Externally-photoionized 
cool runaway stars that move rapidly in the \hii region produced by an other 
source of ionizing radiation have particularly bright optical emission, see e.g. 
the cases of the red supergiant Betelgeuse~\citep{mohamed_aa_541_2012,2014Natur.512..282M} and 
IRC$-$10414~\citep{meyer_mnras_439_2014}. These circumstellar structures are 
themselves \textcolor{black}{sensitive} to the presence of even a weak ISM 
background magnetic field of a few $\mu \rm G$~\citep{vanmarle_aa_561_2014}. 
Consequently, one can expect that the inclusion of such a field in numerical 
models tailored to these objects would affect their associated synthetic 
emission maps and update the current estimate of their driving star's mass loss 
and/or ambient medium density~\citep{meyer_mnras_459_2016}.

\textcolor{black}{
According to the fact that the warm phase of the ISM is typically magnetized, the 
reduction of both optical and infrared surface brightnesses of circumstellar 
structures generated by massive stars should be a rather common phenomenon. In 
particular, it should also concern bow shocks of OB runaway stars once they have 
evolved through the red supergiant phase (Paper~I).
However, the proportion of red supergiant stars amongst the population of all 
runaway massive stars should be similar to the proportion of red supergiant 
with respect to the population of static OB stars, which is, to the best of 
our knowledge, contradicted by observations. The recent study of~\citet{vanmarle_aa_561_2014}
shows that a background ISM magnetic field can inhibits the growth of shear 
instabilities, i.e. forbids the development of potentially bright infrared 
knots, in the bow shock of Betelgeuse, and, this may participate in 
explaining why the scientific literature only reports 4 known runaway red 
superigant stars, amongst which only 3 have a detected bow shock, i.e. 
Betelgeuse~\citep{noriegacrespo_aj_114_1997}, 
IRC$-$10414~\citep{meyer_mnras_439_2014} and $\mu$ 
Cep~\citep{cox_aa_537_2012}. The extragalactic, hyperveloce red supergiant star 
J004330.06+405258.4 in M31 has all kinematic characteristics to generate a bow 
shock but it has not been observed so far~\citep{evans_aj_150_2015}. 
This remark is also valid for bow shocks generated by runaway massive stars 
experiencing other evolutionary stages such as the so-called blue supergiant 
phase~\citep[see, e.g.][]{kaper_apj_475_1997}. 
}

\subsubsection{Comparison with the bow shock around the Sun}
\label{sect:sun}

\textcolor{black}{
The Sun is moving into the warm phase of the ISM~\citep{mccomas_apj_801_2015} 
and the properties of its ambient surrounding, the so-called local interstellar 
medium (LISM) are similar to the ISM in which our runaway stars move, especially 
in terms of Alfv\' enic Mach number and plasma 
$\beta$~\citep{florinski_apj_604_2004,burlaga_apj_804_2015}. The study of the 
interaction between our Sun and the LISM led to a large literature, including, 
amongst other, numerical investigations of the bow shock formed by the solar 
wind~\citep[see, e.g.][and references 
therein]{pogorelov_JGR_103_1998,baranov_jgr_1993,zank_araa_53_2015}. If obvious similitudes between the 
bow shock of the Sun and those of our massive stars indicate that the physical 
processes governing the formation of circumstellar nebulae around OB stars such 
as electronic thermal conduction or the influence of the background local 
magnetic field have to be included in the modelling of those 
structures~\citep{zank_ssrv_146_2009}, nevertheless, the bow shock of the Sun 
is, partly due to the differences in terms of effective temperature and wind 
velocity, on a totally different scale. Further resemblances with bow-like 
nebulae from massive stars are therefore mostly morphological. 
}

\textcolor{black}{ 
As a low-mass star ($<8\, \rm M_{\odot}$), the Sun is much cooler ($T_{\rm eff}\approx 6000\, \rm K$) 
than the runaway OB stars considered in the present work ($T_{\rm eff} > 
20000\, \rm K$) and its mass-loss ($\dot{M}_{\odot}\approx 10^{-14}\, \rm M_{\odot}\, 
\rm yr^{-1}$) is much smaller than that of a main-sequence star with 
$M_{\star} \ge 20\, \rm M_{\odot}$ (our Table~\ref{tab:wind_para}), 
which makes its stellar luminosity fainter by several orders of magnitude 
($L_{\star}/L_{\odot} \ge 10^{3}$). Moreover, the solar wind velocity at 
1 $\rm AU$ is about $350\, \rm km\, \rm s^{-1}$~\citep{golub_1997} 
whereas our OB stars have larger wind velocities ($>1000\, \rm km\, 
\rm s^{-1}$, see Table~\ref{tab:wind_para}). 
Stellar winds from solar-like stars consequently develop a smaller ram pressure 
and expel less linear momentum than massive stars such as our $20\, \rm 
M_{\odot}$ star and their associated corresponding circumstellar structures, i.e. wind 
bubbles or bow shocks are scaled down to a few tens or hundreds of $\rm AU$. 
Note also that the Sun is too cool to produce ionizing radiations and generated 
an \hii region that is susceptible screen its optical/infrared wind bubble. In other 
words, if the numerical methods developed 
to study the bow shock surrounding the Sun are similar to the ones utilised in 
our study, the solar solutions are more appropriated to investigate the surroundings 
of cool, low-mass stars such as, asymptotic giant stars (AGB), see~\citep{wareing_apj_660_2007, 
WareingZijlstraOBrien_2007_MNRAS_382_1233_AGB_bowshocks,raga_apj_680_2008,
esquivel_apj_725_2010,villaver_apj_748_2012,chiotellis_mnras_547_2016}, 
or the trails let by planetesimals moving in stellar systems presenting a common 
envelope~\citep[see][and references therein]{thun_aa_589_2016}. 
}

\textcolor{black}{
Early two-dimensional numerical models of the solar neighbourhood were carried out assuming 
that the respective directions of both the Sun's motion and the LISM magnetic 
field are considered as parallel, as we hereby do with our massive 
stars~\citep{pogorelov_JGR_103_1998}. More sophisticated simulations have produced three-dimensional 
models in which the Sun moves obliquely through the LISM~\citep[see, 
e.g.][]{baranov_astl_22_1996,boley_apj_776_2013}. Such investigation is  observationally motivated 
by the perturbated and non-uniform appearance of the heliopause, e.g. the 
boundary between the interplanetary and interstellar 
medium~\citep{kawamura_2010} which revealed the need for 3D calculations,  able 
to report the non-stationary character of the trail of the bow shock of the 
Sun~\citep{washimi_ssrv_78_1996,linde_jgr_103_1998,ratkiewicz_aa_335_1998}. 
Those models are more complex than our simplistic two-dimensional simulations 
and investigate, e.g. the charges exchanges arising between the stellar wind 
and the LISM~\citep{fitzenreiter_jgr_95_1990}. These studies also 
highlighted the complexity and fragility of such models, e.g. regarding to the 
variety of instable MHD discontinuities that affects shock waves propagating 
through a magnetized flow and differentiating the shocks from purely 
hydrodynamical discontinuities described by the Rankine-Hugoniot~\citep[see 
also][]{sterck_phpl_1998,sterck_aa_343_1999}. Additionally, those solutions are 
affected by the spatial resolution of the calculations and the inclusion of 
numerical viscosity in the models~\citep[][and references 
therein]{lopez_angeo_29_2011,wang_jgra_119_2014}. 
}

\textcolor{black}{
Finally, let mention an other obvious difference between bow shock of the Sun and 
the nebulae generated by the runaway OB stars that we model. The proximity of the Earth 
with the Sun makes it easier to be studied and analysed by means of, e.g. radio 
observations~\citep{baranov_sovastronlett_1_1975} while its innermost 
substructures are directly reachable with spacecrafts such as {\it Voyager 1} and 
{\it Voyager 2}\footnote{http://voyager.jpl.nasa.gov/mission/interstellar.html}. 
Their missions partly consisted in leaving the neighbourhood of our Sun in order 
to explore the heliosheath, i.e. the layer corresponding to the region of shocked solar wind 
that is between the contact discontinuity (the heliopause) and the 
reverse shock of the solar bow shock (the wind termination shock). The {\it Voyager} 
engines crossed the outermost edge of the solar system between 2004 and 2007 at a the expected 
distance of 94 and 84 AU from the Earth~\citep{linde_jgr_103_1998}, giving 
the first experimental data on 
the physics of the interstellar medium~\citep{chalov_mnras_455_2016}. Those 
measures proved the existence of the solar bow shock, but also highlighted the 
particular conditions of the outer space in terms of magnetic 
phenomenon~\citep{richardson_2016} and effects of cosmic 
rays~\citep{webber_2016}. In order to make our models more realistic, those 
physical processes should be taken into account into future simulations of bow 
shocks from runaway high-mass stars.  
}


\section{Conclusion}
\label{section:conclusion}

In this study, we present\textcolor{black}{ed} magneto-hydrodynamical models of the
circumstellar medium of runaway, main-sequence, massive stars moving
supersonically through the plane of the Galaxy. Our two-dimensional simulations first
investigate\textcolor{black}{d} the conjugated effects of optically-thin radiative cooling and
heating together with anisotropic thermal transfers on a field-aligned, 
magneto-hydrodynamical bow shock flow around an OB-type, fast-moving star. We
then explore\textcolor{black}{d} the effects of the stellar motion with respect to the bow shocks, 
focusing on an initially $20\, \rm M_{\odot}$ star moving with velocities
$v_{\star}=20$, $40$ and $70\, \rm km\, \rm s^{-1}$. We present\textcolor{black}{ed} additional
models of an initially $10\, \rm M_{\odot}$~\textcolor{black}{star} moving with velocities
$v_{\star}=40\, \rm km\, \rm s^{-1}$ and of an initially $40\, \rm M_{\odot}$~\textcolor{black}{star} 
moving with velocities $v_{\star}=70\, \rm km\, \rm s^{-1}$. The ISM magnetic
field strength is set to $B_{\rm ISM}=7\, \mu \rm G$. We also
consider\textcolor{black}{ed} bow shock nebulae produced within a weaker ISM magnetic field ($B_{\rm
ISM}=3.5\, \mu \rm G$). The other ISM properties are unchanged for each models.

Our models show that although the magnetization of the ISM does not radically
change the global aspect of our bow shock nebulae, it slightly modifies their
internal organiation. Anisotropic thermal transfers do not split the region of
shocked ISM gas as in our hydrodynamical models (Paper~I), since the presence of
the magnetic field in the regions of shocked material \textcolor{black}{forbids heat 
conduction perpendicular to the magnetic field lines}. 
\textcolor{black}{The field lines, initially parallel to the direction of stellar
motion, are bent round by the bow shock into a sheath around
the fast stellar wind bubble.}
\textcolor{black}{As showed in~\citet{heitsch_apj_665_2007}}, the presence 
of the magnetic field stabilises \textcolor{black}{the} contact
discontinuities inhibiting the growth of the Kelvin-Helmholtz instabilities that typically
occur in pure hydrodynamical models at the interface between shocked ISM and shocked stellar wind. 


As in our previous hydrodynamical study (Paper~I), bow shocks are brighter
in infrared reprocessed starlight. Their emission by optically-thin radiation
mostly originates from the shocked ISM and their [O{\sc iii}] $\lambda \, 5007$
spectral line emission are higher than their H$\alpha$ emission. Notably, their 
X-rays emission are negligible compared to their optical luminosity and
therefore it does not constitute the best waveband to search for hot massive stars'
stellar wind bow shocks. We find that the presence of an ISM background 
magnetic field \textcolor{black}{has the effect of} reducing the optical 
synthetic emission maps of our models, 
making them fainter by one and two orders
of magnitude at [O{\sc iii}] $\lambda \, 5007$ and H$\alpha$, respectively. This
may explain why not so many of them are observed at these spectral lines. We
confirm that, under our assumptions and even in the presence of a magnetic field, 
circumstellar structures produced by high-mass, slowly-moving stars are the 
easiest observable bow shock nebulae in the warm neutral phase of the Milky Way.

\textcolor{black}{
We performed dust continuum radiative transfer calculations of our bow 
shocks models~\citep[cf.][]{acreman_mnras_456_2016} and generated spectral 
energy distributions and isophotal emission maps for different wavelengths $25\, 
\le \lambda\, \le 100\, \mu\rm m$ and viewing angles $0\degree\, \le \phi\, \le 90\degree$. 
Consistently with the observation of~\citet{buren_apj_329_1988,vanburen_aj_110_1995,noriegacrespo_aj_113_1997}, 
the calculations show that our bow shocks are brighter at $60\, \mu\rm m$. 
The projected infrared emission can also be diminished the presence of the ISM magnetic field, 
in particular at wavelengths $\lambda\ge 60\, \mu\rm m$, since the amount of 
dust trapped into the bow shock is smaller. We also notice that 
the change in surface brightness of our emission maps as a function of 
the viewing angle of the bow shock is similar as in the optical waveband, i.e. it 
is brighter if $\phi=0\degree$ and fainter if $\phi=90\degree$~\citep[see][]{meyer_mnras_459_2016}. 
}

In future models, we would like to extend this pioneering study of 
massive stars' bow shocks within the magnetized ISM towards 
three-dimensional models in which the ISM magnetic field is unaligned with
respect to the motion of the star, as it has been done in order to appreciate
its influence on the morphology of the global
heliopause~\citep{pogorelov_JGR_103_1998}. Such simulations will help to
better understand the structure of the circumstellar nebulae forming
around hot, ionising, massive runaway stars and allow us to predict more
accurately, e.g. the optical emission signatures of these bow shocks.
Thorough comparison with particular hot, bow-shock-producing massive stars, 
e.g. $\zeta$ Ophiuchi, might then be feasible.


\section*{Acknowledgements}

The authors thank Tom Hartquist, for his kind help and very helpful advices 
when reviewing this paper and Richard Stancliffe for numerous 
grammatical comments when carefully reading the manuscript.
D.~M.-A.~Meyer gratefully thanks T. Robitaille and C. Dullemond for their 
support with the {\sc hyperion} and the {\sc radmc-3d} raditative 
transfer codes, respectively, as well as D. Thun for fruitful discussion concerning his master thesis.
Are also acknowledged Prof. R. Jalabert and G. Weick from the Institute of 
Physics and Chemistry of Materials of Strasbourg (IPCMS) for their kind help 
concerning the Mie theory. 
This study was in parts conducted within the Emmy Noether research group on
"Accretion Flows and Feedback in Realistic Models of Massive Star Formation" 
funded by the German Research Foundation under grant no. KU 2849/3-1.  
\textcolor{black}{
The authors gratefully acknowledge the computing time provided on the 
supercomputer JUROPA at J\" ulich Supercomputing Centre (JSC) 
and on the bwGrid cluster T\" ubingen. 
}
This research has made use of ''Aladin sky atlas" and ``VizieR catalogue access 
tool" both developed at CDS, Strasbourg Observatory, France.


\bibliographystyle{mn2e}

\footnotesize{
\bibliography{grid}
}


\end{document}